\documentclass[preprint,12pt]{elsarticle}
\journal{European Physical Journal Special Topics}

\usepackage{amsmath}
\usepackage{amssymb}
\usepackage{graphicx}
\usepackage{natbib}
\biboptions{numbers,sort&compress}
\usepackage{csquotes}
\usepackage{siunitx}
\usepackage{hyperref}
\usepackage{enumitem}
\usepackage{physics}
\usepackage{caption}
\usepackage{subcaption}
\usepackage{booktabs}
\usepackage{multirow}
\usepackage{cancel}
\usepackage{bm}
\usepackage[capitalise]{cleveref}
\usepackage{color}
\usepackage[toc,page]{appendix}
\usepackage{cleveref}

\DeclareSIUnit\Molar{\textsc{M}}
\DeclareSIUnit\ergs{ergs}
\DeclareSIUnit\esu{esu}
\DeclareSIUnit\statV{stat\textsc{V}}
\DeclareSIUnit\cm{\centi\metre}

\begin{document}
	\begin{frontmatter}
	
	
\title{Unified Symmetry Breaking in Confined Electrolytes:\\
	\large Charge, Chemical Potential, and the Nonlinear Capacitance of Hollow Nanoparticles}
	
	\author{M. Lozada-Cassou\corref{cor1}}
	\ead{marcelolcmx@ier.unam.mx}
	
	\address{Instituto de Energías Renovables, Universidad Nacional Autónoma de México, Priv. Xochicalco S/N, Temixco, Morelos 62580, México.}
	\cortext[cor1]{Corresponding author}
	
	\begin{abstract}
		\small
		We study the nonlinear electrostatic response of electrolyte-filled, hollow charged nanoparticles, modeled as nanocapacitors with finite wall thickness and curved geometry. Using the linearized Poisson–Boltzmann (LPB) approximation, we derive analytical expressions for the electric double layers (EDLs) and compute the differential capacitance \( C_d \) as a function of topology, wall thickness, and electrolyte concentration.
			
			Remarkably, we identify two forms of symmetry-breaking that yield the same macroscopic capacitance: one due to variations in surface charge density (charge symmetry breaking), and another due to differing bulk chemical potentials (chemical potential symmetry breaking). In the first case, increasing the wall charge alters the internal and external EDL structures without affecting the overall capacitance—provided the Debye length and geometry remain fixed. In the second case, two distinct electrolytes (e.g., 1:1 at concentration \( \rho_0 \) and 2:2 at \( \rho_0/4 \)) yield indistinguishable EDLs and capacitance, despite differing significantly in chemical potential.
			
			These phenomena reflect a deeper electrostatic invariance governed by confinement, topology, and the violation of the local electroneutrality condition (VLEC), rather than by absolute charge or thermodynamic potential. Our findings establish a unified framework for understanding nonlinear capacitance in nanoconfined systems and suggest design principles for energy storage and biological applications where control of electrostatic response is critical.
	\end{abstract}
\end{frontmatter}

\section{Introduction}\label{introduction}
	
Electric double-layer capacitors (EDLCs) have been extensively studied both experimentally~\cite{Chmiola_2006,Kim_2013,Supercapacitors-Book-2013,Beguin_2014,ElKady_2014,Ke_2016} and theoretically~\cite{Lozada_1984,Mier_1988,Vlachy1989,Yeomans1993,Yu_1997,Vlachy2001,Grosse-2002,Henderson2005,Aguilar_2007,Peng2009,Henderson2012,Pizio2012,Lamperski-2014,Henderson2015,Yang2019,biagooi-Nature2020,Enrique-Henry2021,Keshavarzi_2022,Feng-nanopores-topology-2023}, particularly in the context of porous electrodes. Among these systems, charged hollow nanoparticles immersed in electrolyte solutions exhibit a nonlinear capacitance behavior that depends sensitively on both the electrolyte concentration and the geometry of the cavity~\cite{Adrian-JML-2023}. These phenomena have implications for applications in energy storage and biological systems~\cite{Evans-Wennerstrom-1999}, as well as in biomaterials~\cite{Bohinc_2008,Bohinc-2018}, medicine~\cite{Coffey-Biology-2023}, and enhanced oil recovery~\cite{HUANG-oil-1996,Oil-Recovery-book-2019}.

In such systems, the electrolyte confined inside and outside the hollow nanoparticles gives rise to electrical double layers (EDLs) along both surfaces of the shell. The structure and properties of these EDLs are determined by a combination of the surface charge density, electrolyte concentration, cavity radius, and the topology of the nanoparticle geometry.

The simplest model for an electrolyte consists of point ions immersed in a structureless solvent characterized by a dielectric constant $\varepsilon_{\scriptscriptstyle{0}}\varepsilon$. The Poisson--Boltzmann (PB) equation applied to this model yields the Debye--H"uckel theory for bulk electrolytes~\cite{McQuarrie_StatMech}, the Gouy--Chapman model for planar EDLs~\cite{Gouy_1910,Chapman_1913,Hiemenz-book-1977}, and the Verwey--Overbeek theory for colloidal stability~\cite{Verwey_TheoryStabilityLyophobicColloids_1948}. The PB equation is a nonlinear, second-order differential equation that has been analytically solved for planar EDLs and numerically solved in various geometries~\cite{Lozada_1982,Gonzalez_1985,Gonzalez_1989}.

More refined models, such as the restricted primitive model (RPM), consider ions as charged hard spheres. These models have been addressed using statistical mechanical frameworks, including integral equations~\cite{Carnie_1981,Henderson_1992_FIF,Lozada_1992_FIF,Attard_1996,Croxton_1981,Henderson_1982,Bratko_1982,Vlachy-Donnan-1992}, density functional theory~\cite{Patra_1994,Gillespie_2005,Goel_2008,Hartel_2017,Patra_2020}, modified PB equations~\cite{Outhwaite_1986,Bhuiyan_1993_CMT,Bhuiyan_1994}, and Monte Carlo simulations~\cite{Bret_1984,Bratko-Vlachy-osmotic-1985,Card_1970,Torrie_1980,Degreve_1993,Goel_2008,Boda-confinement-2024}. Prior studies have also examined electrolytes confined in cavity shells~\cite{Lozada_1984,Kjellander-Marcelja-Chem-Phys-Letts-1986,Lozada-Cassou-Yamada-1988,Kjellander-two-plates-1988,Kjellander-Akesson-Jonsson-Marcelja-1992,Lozada_1990-I,Lozada_1990-II,Spada_two_plates_2018,Ala-Nisila_2011}, using both point-ion and RPM approaches.

Although the full PB equation cannot generally be solved analytically, its linearized form---the linear Poisson--Boltzmann (LPB) equation---can. This approximation is valid in the low surface charge and low electrolyte concentration regime. The LPB model has been shown to yield good agreement with results from more sophisticated theories and simulations~\cite{Lozada_1982,Degreve_1993,Degreve_1995,Gonzalez_2018,Boda-confinement-2024}, making it a useful benchmark for understanding electrostatics in confined systems.

In this article, we solve the LPB equation to obtain analytical expressions for the electric double layers inside and outside hollow nanoparticles with planar, cylindrical, and spherical geometries, under constant chemical potential. From these, we derive the electric field, electrostatic potential profiles, and differential capacitance.

We demonstrate that, although the capacitance depends explicitly only on the electrolyte concentration, valence, temperature, and geometrical factors, it implicitly exhibits a nontrivial asymmetry with respect to the surface charge. This effect, which we refer to as \emph{charge symmetry-breaking}, arises from the interplay of curvature, confinement, and wall thickness. Despite symmetric boundary conditions, the electrostatic profiles are altered in a way that breaks charge symmetry---not via charge inversion or overcharging~\cite{Lozada-Cassou_JML-2025}---but through the induced variation in the electric field structure, rooted in geometry and confinement. This phenomenon is analogous to spontaneous symmetry breaking in soft matter physics~\cite{chaikin1995principles}, where a system with symmetric governing equations exhibits asymmetric outcomes due to boundary or structural constraints.

In addition, we report another symmetry-breaking effect, here associated with the chemical potential. Specifically, we show that for distinct bulk electrolytes---such as a 1:1 salt at concentration \(\rho_0\) and a 2:2 salt at concentration \(\rho_0/4\)---the charge concentration profile and capacitance remain identical, despite differences in their bulk chemical potentials. This arises because the capacitance depends only on the Debye length and the confinement geometry, not on the absolute value of the chemical potential. Nevertheless, the positive and negative ions' local electrochemical potential and structural profiles vary significantly, reflecting the underlying inhomogeneous electrostatic environment.

This phenomenon reflects a kind of \textit{chemical potential symmetry-breaking}, where global observables such as the capacitance exhibit invariance under transformations that preserve, while local fields---such as the mean electrostatic potential and ionic profiles---do not. Thus, the capacitance is a nonlocal quantity insensitive to the absolute value of the chemical potential, whereas the local electrostatic and ionic environments remain highly sensitive to it.

This paper is organized as follows. In \cref{Theory}, we outline the derivations of the linear Poisson-Boltzmann, the electric and mean electrostatic potential profiles, and the capacitances' formulas for model hollow nanoshells of planar, cylindrical and spherical geometries. In \cref{Res_Disc}, we analyze the capacitance as a function of key parameters and report two novel confinement-induced symmetry breaking: \emph{charge symmetry breaking} and \emph{chemical potential symmetry breaking}. In \cref{Conclusions}, we summarize our findings and outline possible extensions.


\section{Theory}\label{Theory}

We model the system as hollow nanoshells with planar, cylindrical, or spherical geometry, immersed in a bulk electrolyte. The electrolyte consists of symmetric point ions suspended in a continuous dielectric medium with dielectric constant $\varepsilon_{\scriptscriptstyle{0}}\varepsilon$. The nanoparticles are represented as shells with finite wall thickness $d$, such that the inner and outer surfaces of the shell are located at $r=R$ and $r=R+d$, respectively, measured from the geometric center. The dielectric constant of the shell's walls is taken to be equal to that of the bulk electrolyte, to avoid image potentials.

\begin{figure}[htbp]
	\centering
	\begin{subfigure}[b]{0.85\textwidth}
		\centering
		\includegraphics[width=\linewidth]{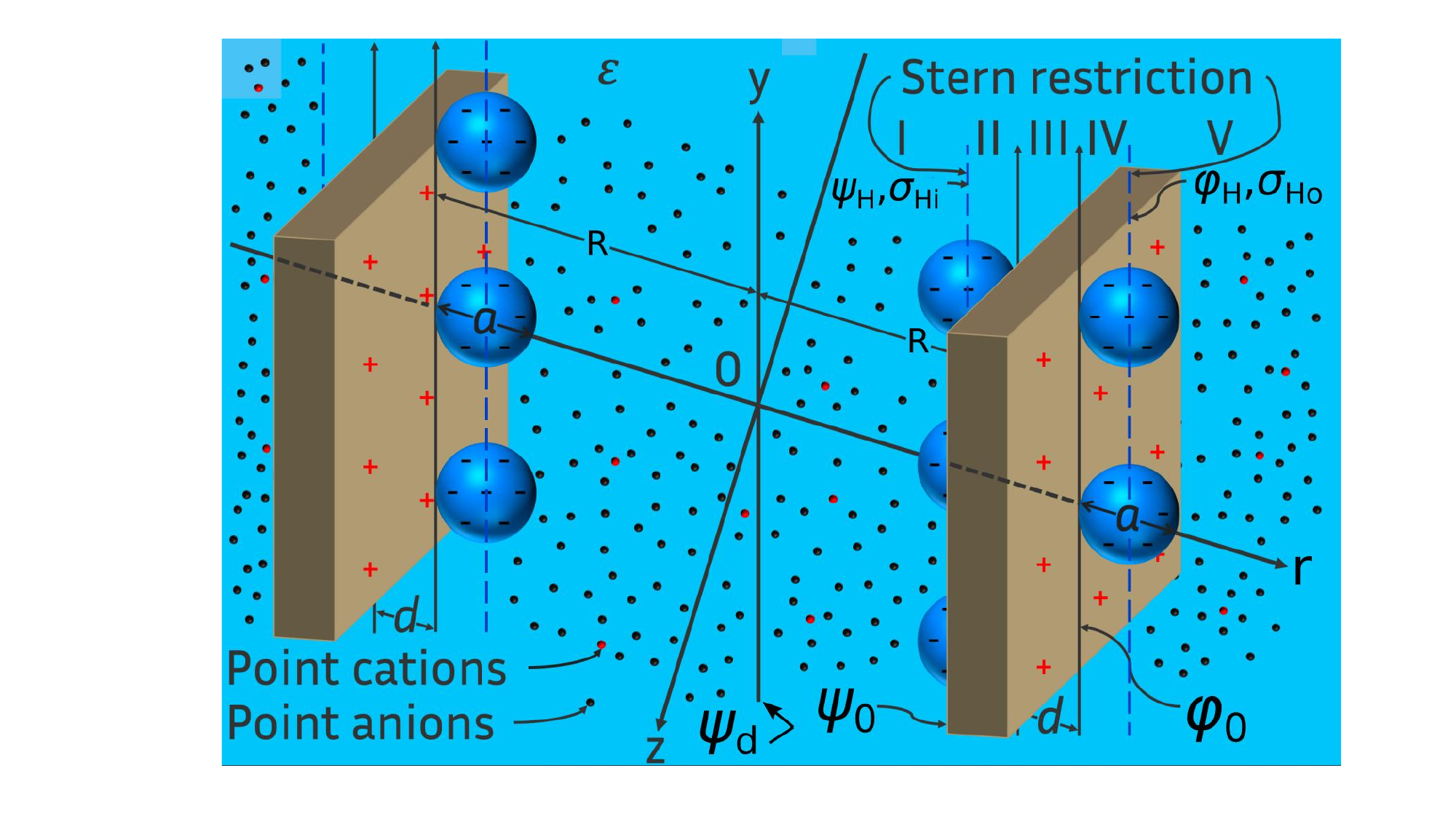}
		\caption{Slit-shell (planar geometry, 2D).}
		\label{Geometry_twoplates}
	\end{subfigure}
	
	\vspace{1.5em} 
	
	\begin{subfigure}[b]{0.48\textwidth}
		\centering
		\includegraphics[width=1.1\linewidth]{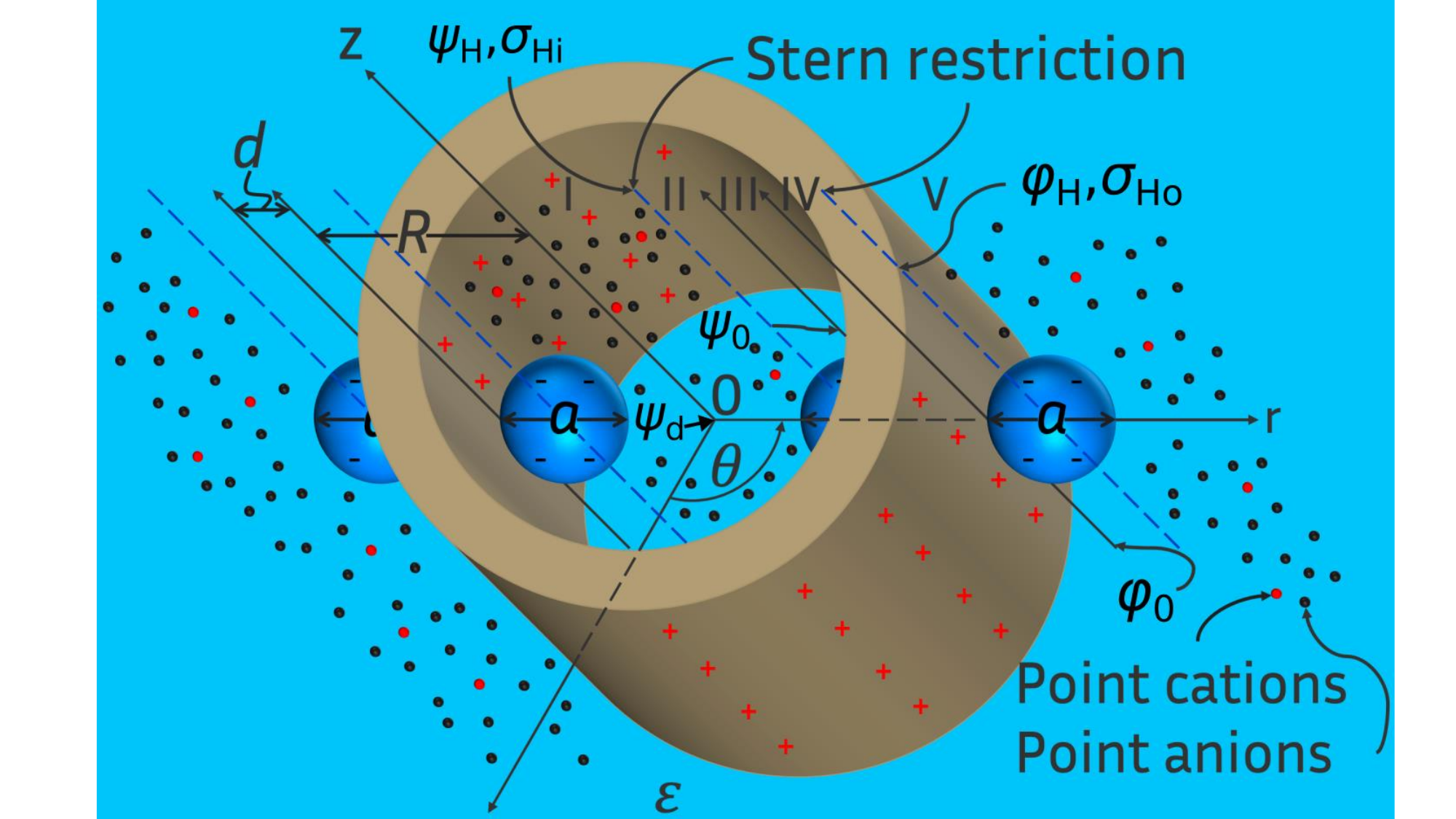}
		\caption{Cylindrical shell (3D).}
		\label{Geometry_cylinder}
	\end{subfigure}
	\hfill
	\begin{subfigure}[b]{0.48\textwidth}
		\centering
		\includegraphics[width=1.1\linewidth]{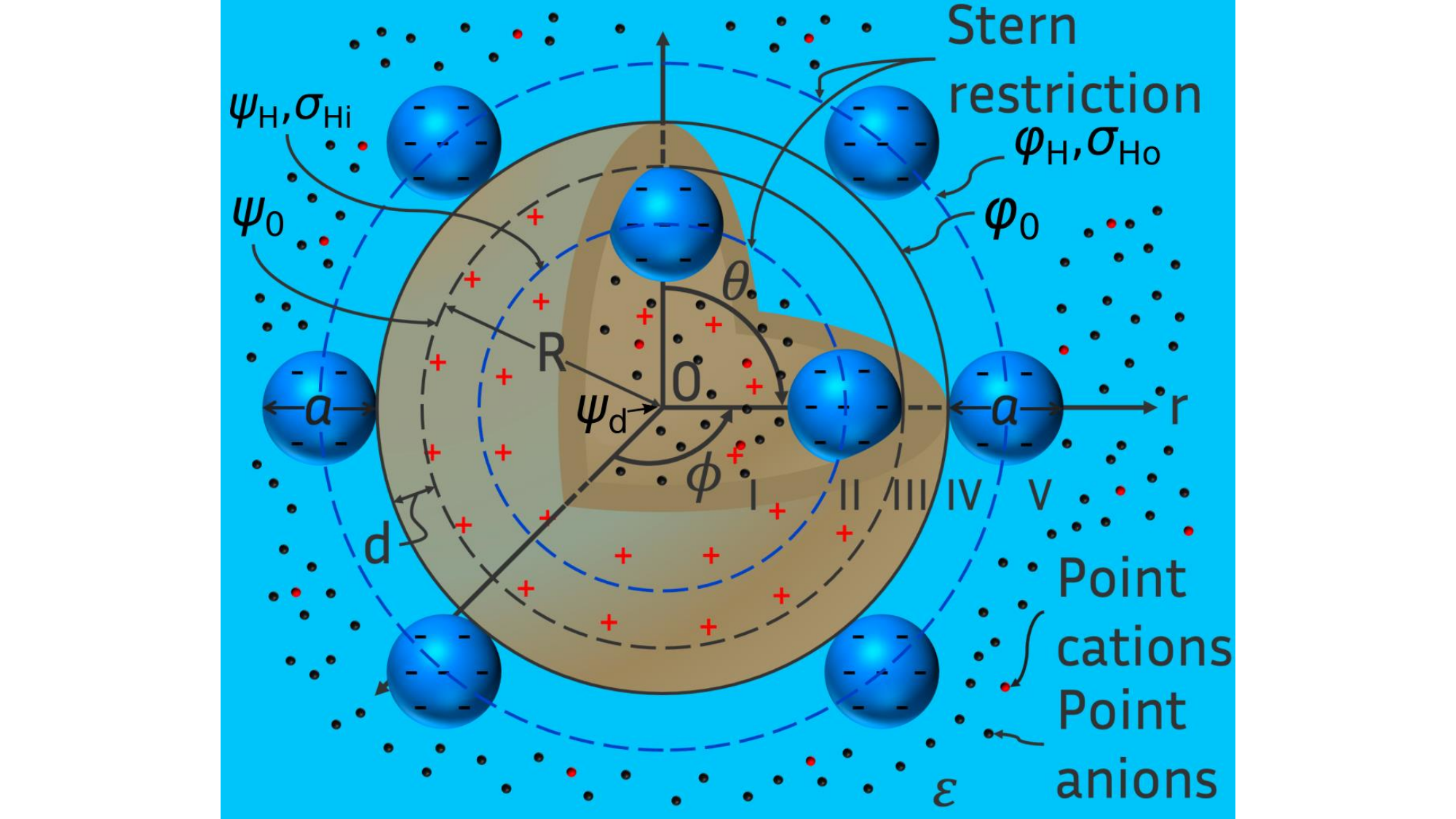}
		\caption{Spherical shell (2D cross-section).}
		\label{Geometry_sphere}
	\end{subfigure}
	\caption{Geometries of the three electrolyte-filled charged shells: planar (slit), cylindrical, and spherical.}
	\label{fig:geometries_all}
\end{figure}

Each shell wall carries a constant surface charge density $\sigma_{\scriptscriptstyle{0}}$ on both inner and outer surfaces. Ion-ion interactions are modeled by the Coulomb potential, while ion-wall interactions are treated using a Stern layer correction~\cite{Verwey_TheoryStabilityLyophobicColloids_1948}, which assigns a finite ionic diameter $a$. Consequently, the electric double layers (EDLs) are restricted to the regions $0 \leq r \leq R - a/2$ and $R + d + a/2 \leq r < \infty$. See \cref{fig:geometries_all} for schematic representation.

We define five spatial regions: $I$, $II$, $III$, $IV$, and $V$, corresponding to $0 \leq r \leq R - a/2$, $R - a/2 \leq r \leq R$, $R \leq r \leq R + d$, $R + d \leq r \leq R + d + a/2$, and $R + d + a/2 \leq r < \infty$, respectively. Regions $I$ and $V$ contain mobile ions and are governed by the Poisson equation,

\begin{equation}
	\nabla^2 \psi (r) = -\frac{1}{\varepsilon_{\scriptscriptstyle{0}}\varepsilon}\rho_{\scriptscriptstyle{el}}(r),
	\label{Poisson-equation}
\end{equation}

\noindent while in regions $II$--$IV$ (devoid of ions), the electrostatic potential satisfies Laplace's equation:

\begin{equation}
	\nabla^2 \psi (r) = 0.
	\label{Laplace-equation}
\end{equation}

In \cref{Poisson-equation}, the local charge density is modeled using the Boltzmann distribution for point ions:

\begin{equation}
	\rho_{\scriptscriptstyle{el}}(r) = \sum_{i=1}^n e z_i \rho_{i0} \exp\left(-\beta e z_i \psi(r)\right),
	\label{Rho_elx2}
\end{equation}

\noindent where $\psi(r)$ is the mean electrostatic potential (MEP), $\rho_{i0}$ is the bulk concentration of ion species $i$, $z_i$ its valence, $e$ the elementary charge, $n$ the number of ionic species, and $\beta = 1/(kT)$. $K$ is the Boltzmann constant and $T$ is the system's temperature.

Substituting \cref{Rho_elx2} into \cref{Poisson-equation} gives the Poisson--Boltzmann (PB) equation. In the linear regime, where $\beta e z_i \psi \ll 1$, we obtain the linearized Poisson--Boltzmann (LPB) equation:

\begin{equation}
	\nabla^2 \psi(r) = \kappa^2 \psi(r),
	\label{LPB-equation}
\end{equation}

\noindent with

\begin{equation}
	\kappa = \sqrt{\frac{2 e^2 z^2 \rho_0}{\varepsilon_{\scriptscriptstyle{0}}\varepsilon k T}},
	\label{Ec.kappa}
\end{equation}

\noindent where $\rho_0$ is the bulk concentration of each species in a symmetric $z:z$ electrolyte, and $\kappa = 1/\lambda_{\scriptscriptstyle{D}}$, with $\lambda_{\scriptscriptstyle{D}}$ denoting the Debye screening length~\cite{Hansen-book-2013}.

The electric field is obtained from the gradient of the potential,

\begin{equation}
	E(r) = -\nabla \psi(r) = \frac{1}{\varepsilon_{\scriptscriptstyle{0}}\varepsilon} \sigma(r),
	\label{electric-field general equation}
\end{equation}

\noindent where vector notation is omitted due to the azimuthal symmetry of the geometries considered (see \cref{fig:geometries_all}). At the characteristic interfaces of the nanoparticle, we define:

\begin{itemize}
	\item $\phi_{\scriptscriptstyle{H}}$ and $\sigma_{\scriptscriptstyle{Ho}}$ at $r = R + d + a/2$,
	\item $\phi_{\scriptscriptstyle{0}}$ and $\sigma_{\scriptscriptstyle{0}}$ at $r = R + d$,
	\item $\psi_{\scriptscriptstyle{0}}$ and $\sigma_{\scriptscriptstyle{0}}$ at $r = R$,
	\item $\psi_{\scriptscriptstyle{Hi}}$ and $\sigma_{\scriptscriptstyle{Hi}}$ at $r = R - a/2$,
	\item $\psi_{\scriptscriptstyle{d}}$ and zero charge at $r = 0$.
\end{itemize}

We also impose the boundary conditions $\lim_{r \to \infty} \psi(r) = 0$ and $\lim_{r \to \infty} E(r) = 0$, which implies $\lim_{r \to \infty} \sigma(r) = 0$. 

These equations, together with the boundary conditions at interfaces of the classical electrodynamics theory~\cite{Jackson_2001}, define the electrostatic problem in the LPB framework, whose analytical solution will be used to compute the mean electrostatic potential, electric field profiles, and capacitance for different nanopore geometries. Please, notice that the mean electrostatic and surface charge densities at all the nanoparticles' boundaries, but at $r\rightarrow \infty$ depend on $R$ and $d$. We will make this dependence explicit, when necessary.

\subsection{The linear Poisson-Boltzmann equation}\label{LPB}

With the Boltzmann distribution, the reduced ions concentration profiles for positive ions, $g_{\scriptscriptstyle{+}}(r)$, and negative ions, $g_{\scriptscriptstyle{-}}(r)$, become

\begin{equation}
	g_{\scriptscriptstyle{+}}(r) = \exp(-\,e\,z\,\beta\psi(r)) \quad r\in[0,(R-a/2)]\cup [(R+d+a/2),\infty),
	\label{Ec.gmas}
\end{equation}

\noindent and

\begin{equation}
	g_{\scriptscriptstyle{-}}(r) =\exp(\,e\,z\,\beta\psi(r)) \quad  r\in[0,(R-a/2)]\cup [(R+d+a/2),\infty).
	\label{Ec.gmenos}
\end{equation}

\noindent Although \cref{Ec.gmas,Ec.gmenos} are symmetrically valid for either positively or negatively charged shells, we will henceforth assume a positively charged shell. In this case, the cations and anions of the electrolyte act as co-ions and counter-ions, respectively. Throughout this work, we use $r$ to denote the distance from the geometrical center of the shell, regardless of its specific geometry.

For the analytical solution of the LPB differential equation, we will use Taylor expansions of \cref{Ec.gmas,Ec.gmenos}. However, when directly evaluating distribution functions—e.g., for plotting $\rho_{\scriptscriptstyle{el}}(r)$—we will employ the full expressions in \cref{Ec.gmas,Ec.gmenos}. This is justified by the fact that, under the low values of $e z_+ \beta \psi_{\scriptscriptstyle{H}}$ and $e z_+ \beta \varphi_{\scriptscriptstyle_{H}}$ used in our calculations, the charge density profiles obtained from the full expressions and those from second-order Taylor expansions are virtually indistinguishable.

In all our calculations, we consider a positively charged shell immersed in an aqueous symmetric electrolyte ($1:1$ or $2:2$), with a relative dielectric constant $\varepsilon = 78.5$ an ion diameter $a = \SI{4.25}{\angstrom}$, and a temperature $T=298.15 K$.

\subsection{Electrostatics}\label{Electrostatics}

Analytical solutions of \cref{Laplace-equation,LPB-equation} for the three hollow nanoparticle geometries have been previously derived~\cite{Adrian-JML-2023}. Here, we present a reformulation of the relevant expressions for the electrostatic potential \( \psi(r) \) and the electric field \( E(r) \), specific to each geometry. We define \( \sigma_{\scriptscriptstyle{Hi}} = \sigma(R - a/2) \) and \( \sigma_{\scriptscriptstyle{Ho}} = \sigma(R + d + a/2) \) as the \textit{effective induced surface charge densities} at the inner and outer electrolyte interfaces, respectively.

These expressions are exact within the LPB framework. The discontinuities in the electric field at \( r = R \) and \( r = R + d \) arise from Maxwell’s boundary conditions at the interfaces between the electrolyte and the charged shell wall.

\subsubsection{Slit-Shell}\label{Slit-shell}

The LPB solution for a symmetric electrolyte in a slit-shell geometry (i.e., two parallel plates separated by \( 2R \)) with a Stern layer correction is given by~\cite{Adrian-JML-2023}:

\paragraph{Mean Electrostatic Potential \(\psi(r)\):}
\begin{equation}
	\psi(r=|x|) =
	\begin{cases}
		\displaystyle \frac{2\sigma_0 - \sigma_{\scriptscriptstyle{Ho}}}{\varepsilon_{\scriptscriptstyle{0}}\varepsilon \kappa} \frac{\cosh[\kappa r]}{\sinh[\kappa(R - a/2)]}, & 0 \leq r \leq R - \frac{a}{2} \\[8pt]
		\displaystyle \psi_{\scriptscriptstyle{H}} + \frac{2\sigma_0 - \sigma_{\scriptscriptstyle{Ho}}}{\varepsilon_{\scriptscriptstyle{0}}\varepsilon} (r - (R - a/2)), & R - \frac{a}{2} \leq r \leq R \\[8pt]
		\displaystyle \varphi_0 - \frac{\sigma_{\scriptscriptstyle{Ho}} - \sigma_0}{\varepsilon_{\scriptscriptstyle{0}}\varepsilon}(r - R - d), & R \leq r \leq R + d \\[8pt]
		\displaystyle \frac{\sigma_{\scriptscriptstyle{Ho}}}{\varepsilon_{\scriptscriptstyle{0}}\varepsilon \kappa}[1 - \kappa(r - R_{\scriptscriptstyle{H}})], & R + d \leq r \leq R_{\scriptscriptstyle{H}} \\[8pt]
		\displaystyle \frac{\sigma_{\scriptscriptstyle{Ho}}}{\varepsilon_{\scriptscriptstyle{0}}\varepsilon \kappa} e^{-\kappa(r - R_{\scriptscriptstyle{H}})}, & R_{\scriptscriptstyle{H}} \leq r
	\end{cases}
	\label{Plates-MEP(r)}
\end{equation}

\paragraph{Electric Field \(E(r)\):}
\begin{equation}
	E(r) =
	\begin{cases}
		\displaystyle \frac{\sigma_{\scriptscriptstyle{Ho}} - 2\sigma_0}{\varepsilon_{\scriptscriptstyle{0}}\varepsilon} \frac{\sinh[\kappa r]}{\sinh[\kappa(R - a/2)]}, & 0 \leq r \leq R - \frac{a}{2} \\[8pt]
		\displaystyle \frac{\sigma_{\scriptscriptstyle{Ho}} - 2\sigma_0}{\varepsilon_{\scriptscriptstyle{0}}\varepsilon}, & R - \frac{a}{2} \leq r \leq R \\[8pt]
		\displaystyle \frac{\sigma_{\scriptscriptstyle{Ho}} - \sigma_0}{\varepsilon_{\scriptscriptstyle{0}}\varepsilon}, & R < r < R + d \\[8pt]
		\displaystyle \frac{\sigma_{\scriptscriptstyle{Ho}}}{\varepsilon_{\scriptscriptstyle{0}}\varepsilon}, & R + d \leq r \leq R_{\scriptscriptstyle{H}} \\[8pt]
		\displaystyle \frac{\sigma_{\scriptscriptstyle{Ho}}}{\varepsilon_{\scriptscriptstyle{0}}\varepsilon} e^{-\kappa(r - R_{\scriptscriptstyle{H}})}, & R_{\scriptscriptstyle{H}} \leq r
	\end{cases}
	\label{Plates-E(r)}
\end{equation}

Here, \( R_{\scriptscriptstyle{H}} = R + d + a/2 \) is the outermost contact point of the shell and electrolyte. The surface potential values are:
\begin{align}
	\varphi_{\scriptscriptstyle{H}} &= \psi(R_{\scriptscriptstyle{H}}) = \frac{\sigma_{\scriptscriptstyle{Ho}}}{\varepsilon_{\scriptscriptstyle{0}}\varepsilon \kappa}, \label{Plates-phi-H} \\
	\varphi_0 &= \psi(R + d) = \varphi_{\scriptscriptstyle{H}} + \frac{a}{2} \frac{\sigma_{\scriptscriptstyle{Ho}}}{\varepsilon_{\scriptscriptstyle{0}}\varepsilon}, \label{Plates-phi-0} \\
	\psi_0 &= \psi(R) = \varphi_0 + d\left( \frac{\sigma_{\scriptscriptstyle{Ho}} - \sigma_0}{\varepsilon_{\scriptscriptstyle{0}}\varepsilon} \right), \label{Plates-psi-0} \\
	\psi_{\scriptscriptstyle{H}} &= \psi(R - a/2) = \psi_0 + \frac{a}{2} \left( \frac{\sigma_{\scriptscriptstyle{Ho}} - 2\sigma_0}{\varepsilon_{\scriptscriptstyle{0}}\varepsilon} \right), \label{Plates-psi-H} \\
		\psi_{\scriptscriptstyle{d}}&\equiv\psi(r=0)=\frac{\sigma_{\scriptscriptstyle{Hi}}}{\varepsilon_{\scriptscriptstyle{0}}\varepsilon\kappa}csch[\kappa (R-a/2)].\label{Plates-psi-d}
\end{align}

\paragraph{Effective Surface Charge Density:}

The effective outer surface charge density \( \sigma_{\scriptscriptstyle{Ho}} \) is given by:
\begin{equation}
	\sigma_{\scriptscriptstyle{Ho}} = \left[ \frac{\kappa(a + d) + 2 \coth[\kappa(R - a/2)]}{1 + \kappa(a + d) + \coth[\kappa(R - a/2)]} \right] \sigma_0.
	\label{Plates_sigmaHO}
\end{equation}


\subsubsection{Cylindrical Shell}\label{Cylindrical-shell-EDL}

The LPB solutions for a cylindrical shell geometry (infinitely long hollow cylinder) immersed in a symmetric electrolyte, with Stern correction, are given by~\cite{Adrian-JML-2023}.

\paragraph{Mean Electrostatic Potential \(\psi(r)\):}
\begin{equation}
	\psi(r) = 
	\begin{cases}
		\displaystyle -\left( \frac{R_{\scriptscriptstyle{H}} \sigma_{\scriptscriptstyle{Ho}} - (2R + d)\sigma_0}{\varepsilon_{\scriptscriptstyle{0}}\varepsilon} \right) \frac{I_0(\kappa r)}{\kappa (R - \frac{a}{2}) I_1[\kappa(R - \frac{a}{2})]}, & 0 \leq r \leq R - \frac{a}{2} \\[8pt]
		\displaystyle \psi_{\scriptscriptstyle{H}} + \left( \frac{R_{\scriptscriptstyle{H}} \sigma_{\scriptscriptstyle{Ho}} - (2R + d)\sigma_0}{\varepsilon_{\scriptscriptstyle{0}}\varepsilon} \right) \ln\left( \frac{R - \frac{a}{2}}{r} \right), & R - \frac{a}{2} \leq r \leq R \\[8pt]
		\displaystyle \psi_0 + \left( \frac{R_{\scriptscriptstyle{H}} \sigma_{\scriptscriptstyle{Ho}} - (R + d)\sigma_0}{\varepsilon_{\scriptscriptstyle{0}}\varepsilon} \right) \ln\left( \frac{R}{r} \right), & R \leq r \leq R + d \\[8pt]
		\displaystyle \varphi_0 + \frac{\sigma_{\scriptscriptstyle{Ho}} R_{\scriptscriptstyle{H}}}{\varepsilon_{\scriptscriptstyle{0}}\varepsilon} \ln\left( \frac{R + d}{r} \right), & R + d \leq r \leq R_{\scriptscriptstyle{H}} \\[8pt]
		\displaystyle \frac{\sigma_{\scriptscriptstyle{Ho}}}{\varepsilon_{\scriptscriptstyle{0}}\varepsilon \kappa} \frac{K_0(\kappa r)}{K_1(\kappa R_{\scriptscriptstyle{H}})}, & R_{\scriptscriptstyle{H}} \leq r
	\end{cases}
	\label{Cylinder-MEP}
\end{equation}

\paragraph{Electric Field \(E(r)\):}
\begin{equation}
	E(r) =
	\begin{cases}
		\displaystyle \left( \frac{R_{\scriptscriptstyle{H}} \sigma_{\scriptscriptstyle{Ho}} - (2R + d)\sigma_0}{\varepsilon_{\scriptscriptstyle{0}}\varepsilon} \right) \frac{I_1(\kappa r)}{(R - \frac{a}{2}) I_1[\kappa(R - \frac{a}{2})]}, & 0 \leq r \leq R - \frac{a}{2} \\[8pt]
		\displaystyle \left( \frac{R_{\scriptscriptstyle{H}} \sigma_{\scriptscriptstyle{Ho}} - (2R + d)\sigma_0}{\varepsilon_{\scriptscriptstyle{0}}\varepsilon} \right) \frac{1}{r}, & R - \frac{a}{2} \leq r \leq R \\[8pt]
		\displaystyle \left( \frac{R_{\scriptscriptstyle{H}} \sigma_{\scriptscriptstyle{Ho}} - (R + d)\sigma_0}{\varepsilon_{\scriptscriptstyle{0}}\varepsilon} \right) \frac{1}{r}, & R < r < R + d \\[8pt]
		\displaystyle \frac{\sigma_{\scriptscriptstyle{Ho}} R_{\scriptscriptstyle{H}}}{\varepsilon_{\scriptscriptstyle{0}}\varepsilon r}, & R + d \leq r \leq R_{\scriptscriptstyle{H}} \\[8pt]
		\displaystyle \frac{\sigma_{\scriptscriptstyle{Ho}}}{\varepsilon_{\scriptscriptstyle{0}}\varepsilon K_1(\kappa R_{\scriptscriptstyle{H}})} K_1(\kappa r), & R_{\scriptscriptstyle{H}} \leq r
	\end{cases}
	\label{Cylinder-E(r)}
\end{equation}

\paragraph{Surface Potential Values:}
\begin{align}
	\varphi_{\scriptscriptstyle{H}} &= \frac{K_0(\kappa R_{\scriptscriptstyle{H}}) \sigma_{\scriptscriptstyle{Ho}}}{\varepsilon_{\scriptscriptstyle{0}}\varepsilon \kappa K_1(\kappa R_{\scriptscriptstyle{H}})}, \label{Cyl-phi-H} \\
	\varphi_0 &= \varphi_{\scriptscriptstyle{H}} + \frac{R_{\scriptscriptstyle{H}} \sigma_{\scriptscriptstyle{Ho}}}{\varepsilon_{\scriptscriptstyle{0}}\varepsilon} \ln\left( \frac{R_{\scriptscriptstyle{H}}}{R + d} \right), \label{Cyl-phi-0} \\
	\psi_0 &= \varphi_0 + \left( \frac{R_{\scriptscriptstyle{H}} \sigma_{\scriptscriptstyle{Ho}} - (R + d)\sigma_0}{\varepsilon_{\scriptscriptstyle{0}}\varepsilon} \right) \ln\left( \frac{R + d}{R} \right), \label{Cyl-psi-0} \\
	\psi_{\scriptscriptstyle{H}} &= \psi_0 + \left( \frac{R_{\scriptscriptstyle{H}} \sigma_{\scriptscriptstyle{Ho}} - (2R + d)\sigma_0}{\varepsilon_{\scriptscriptstyle{0}}\varepsilon} \right) \ln\left( \frac{R}{R - \frac{a}{2}} \right), \label{Cyl-psi-H} \\
	 \\ 	\psi_{\scriptscriptstyle{d}} &= \psi_{\scriptscriptstyle{H}} + \frac{\left[R_{\scriptscriptstyle_{H}}\sigma_{\scriptscriptstyle{Ho}}-(2R+d)\sigma_{o}\right](I_{\scriptscriptstyle{0}}[\kappa(R-a/2)]-1)}{\varepsilon_{\scriptscriptstyle{0}}\varepsilon\kappa(R-a/2)I_{\scriptscriptstyle{1}}[\kappa(R-a/2)]}. \label{Cyl-psi-d}
\end{align}

\paragraph{Effective Surface Charge:}
\begin{equation}
	\sigma_{\scriptscriptstyle{Ho}} = \frac{L_2}{L_1} \sigma_0
	\label{Cylinder-sigmaHo}
\end{equation}
where:
\begin{align*}
	L_1 &= \frac{R_{\scriptscriptstyle{H}} I_0[\kappa(R - \frac{a}{2})]}{\kappa(R - \frac{a}{2}) I_1[\kappa(R - \frac{a}{2})]} + R_{\scriptscriptstyle{H}} \ln\left( \frac{R_{\scriptscriptstyle{H}}}{R - \frac{a}{2}} \right) + \frac{K_0(\kappa R_{\scriptscriptstyle{H}})}{\kappa K_1(\kappa R_{\scriptscriptstyle{H}})} \\
	L_2 &= \frac{(2R + d) I_0[\kappa(R - \frac{a}{2})]}{\kappa(R - \frac{a}{2}) I_1[\kappa(R - \frac{a}{2})]} + (2R + d) \ln\left( \frac{R}{R - \frac{a}{2}} \right) + (R + d) \ln\left( \frac{R + d}{R} \right)
\end{align*}

\subsubsection{Spherical Shell}\label{Spherical-shell-EDL}

The LPB analytical solution for a spherical shell immersed in a symmetric electrolyte, with Stern correction, is given by~\cite{Adrian-JML-2023}.

\paragraph{Mean Electrostatic Potential \(\psi(r)\):}
\begin{equation}
	\psi(r) = 
	\begin{cases}
		\displaystyle \frac{R_{\scriptscriptstyle{H}}^2 \sigma_{\scriptscriptstyle{Ho}} - [(R+d)^2 + R^2]\sigma_0}{\varepsilon_{\scriptscriptstyle{0}}\varepsilon D(R)} \cdot \frac{\sinh(\kappa r)}{r}, & 0 \leq r \leq R - \frac{a}{2} \\[8pt]
		\displaystyle \psi_{\scriptscriptstyle{H}} + \frac{R_{\scriptscriptstyle{H}}^2 \sigma_{\scriptscriptstyle{Ho}} - [(R+d)^2 + R^2]\sigma_0}{\varepsilon_{\scriptscriptstyle{0}}\varepsilon (R - \frac{a}{2})} \left( \frac{R - \frac{a}{2}}{r} - 1 \right), & R - \frac{a}{2} \leq r \leq R \\[8pt]
		\displaystyle \psi_0 + \frac{R_{\scriptscriptstyle{H}}^2 \sigma_{\scriptscriptstyle{Ho}} - (R + d)^2 \sigma_0}{\varepsilon_{\scriptscriptstyle{0}}\varepsilon R} \left( \frac{R}{r} - 1 \right), & R \leq r \leq R + d \\[8pt]
		\displaystyle \varphi_0 + \frac{R_{\scriptscriptstyle{H}}^2 \sigma_{\scriptscriptstyle{Ho}}}{\varepsilon_{\scriptscriptstyle{0}}\varepsilon (R + d)} \left( \frac{R + d}{r} - 1 \right), & R + d \leq r \leq R_{\scriptscriptstyle{H}} \\[8pt]
		\displaystyle \frac{R_{\scriptscriptstyle{H}}^2 \sigma_{\scriptscriptstyle{Ho}}}{\varepsilon_{\scriptscriptstyle{0}}\varepsilon (1 + \kappa R_{\scriptscriptstyle{H}})} \cdot \frac{e^{-\kappa (r - R_{\scriptscriptstyle{H}})}}{r}, & R_{\scriptscriptstyle{H}} \leq r
	\end{cases}
	\label{Sphere-MEP}
\end{equation}

Here, the denominator
\[
D(R) = \sinh[\kappa (R - \frac{a}{2})] - \kappa (R - \frac{a}{2}) \cosh[\kappa (R - \frac{a}{2})]
\]

\paragraph{Electric Field \(E(r)\):}
\begin{equation}
	E(r) = 
	\begin{cases}
		\displaystyle \frac{R_{\scriptscriptstyle{H}}^2 \sigma_{\scriptscriptstyle{Ho}} - [(R+d)^2 + R^2]\sigma_0}{\varepsilon_{\scriptscriptstyle{0}}\varepsilon D(R)} \cdot \frac{\sinh(\kappa r) - \kappa r \cosh(\kappa r)}{r^2}, & 0 \leq r \leq R - \frac{a}{2} \\[8pt]
		\displaystyle \frac{R_{\scriptscriptstyle{H}}^2 \sigma_{\scriptscriptstyle{Ho}} - [(R+d)^2 + R^2]\sigma_0}{\varepsilon_{\scriptscriptstyle{0}}\varepsilon r^2}, & R - \frac{a}{2} \leq r \leq R \\[8pt]
		\displaystyle \frac{R_{\scriptscriptstyle{H}}^2 \sigma_{\scriptscriptstyle{Ho}} - (R + d)^2 \sigma_0}{\varepsilon_{\scriptscriptstyle{0}}\varepsilon r^2}, & R < r < R + d \\[8pt]
		\displaystyle \frac{R_{\scriptscriptstyle{H}}^2 \sigma_{\scriptscriptstyle{Ho}}}{\varepsilon_{\scriptscriptstyle{0}}\varepsilon r^2}, & R + d \leq r \leq R_{\scriptscriptstyle{H}} \\[8pt]
		\displaystyle \frac{R_{\scriptscriptstyle{H}}^2 (1 + \kappa r) \sigma_{\scriptscriptstyle{Ho}}}{\varepsilon_{\scriptscriptstyle{0}}\varepsilon (1 + \kappa R_{\scriptscriptstyle{H}})} \cdot \frac{e^{-\kappa(r - R_{\scriptscriptstyle{H}})}}{r^2}, & R_{\scriptscriptstyle{H}} \leq r
	\end{cases}
	\label{Sphere-E(r)}
\end{equation}

\paragraph{Surface Potential Values:}
\begin{align}
	\varphi_{\scriptscriptstyle{H}} &= \frac{R_{\scriptscriptstyle{H}} \sigma_{\scriptscriptstyle{Ho}}}{\varepsilon_{\scriptscriptstyle{0}}\varepsilon (1 + \kappa R_{\scriptscriptstyle{H}})}, \label{Sph-phi-H} \\
	\varphi_0 &= \varphi_{\scriptscriptstyle{H}} + \frac{R_{\scriptscriptstyle{H}} \sigma_{\scriptscriptstyle{Ho}}}{\varepsilon_{\scriptscriptstyle{0}}\varepsilon (R + d)} \cdot \frac{a}{2}, \label{Sph-phi-0} \\
	\psi_0 &= \varphi_0 + \frac{[R_{\scriptscriptstyle{H}}^2 \sigma_{\scriptscriptstyle{Ho}} - (R + d)^2 \sigma_0] d}{\varepsilon_{\scriptscriptstyle{0}}\varepsilon R (R + d)}, \label{Sph-psi-0} \\
	\psi_{\scriptscriptstyle{H}} &= \psi_0 + \frac{[R_{\scriptscriptstyle{H}}^2 \sigma_{\scriptscriptstyle{Ho}} - (R+d)^2 - R^2]\sigma_0 \cdot \frac{a}{2}}{\varepsilon_{\scriptscriptstyle{0}}\varepsilon R(R - \frac{a}{2})}, \label{Sph-psi-H} \\ \psi_{\scriptscriptstyle_{d}}& = \frac{\kappa}{\varepsilon_{\scriptscriptstyle{0}}\varepsilon} \left\{ \frac{[R_{\scriptscriptstyle_{H}}^2\sigma_{\scriptscriptstyle{Ho}}-\left[ R^2 + (R+d)^2 \right]\sigma_{\scriptscriptstyle{0}}]}{\sinh[\kappa (R-a/2)]-\kappa (R-a/2)\cosh[\kappa(R-a/2)]}\right\}.\label{Sph-psi-d}	
\end{align}

\paragraph{Effective Surface Charge:}
\begin{equation}
	\sigma_{\scriptscriptstyle{Ho}} = \frac{L_2}{L_1} \sigma_0
	\label{Sphere-sigmaHo}
\end{equation}

where:
\begin{align*}
	L_1 &= \frac{R_{\scriptscriptstyle{H}}}{1 + \kappa R_{\scriptscriptstyle{H}}} + \frac{\frac{a}{2} R_{\scriptscriptstyle{H}}}{R + d} + \frac{d R_{\scriptscriptstyle{H}}^2}{R(R + d)} + \frac{\frac{a}{2} R_{\scriptscriptstyle{H}}^2}{R(R - \frac{a}{2})} \\
	&\quad - \frac{\sinh[\kappa(R - \frac{a}{2})] R_{\scriptscriptstyle{H}}^2}{(R - \frac{a}{2}) D(R)} \\[5pt]
	L_2 &= \frac{d(R + d)}{R} + \frac{\frac{a}{2} [(R + d)^2 + R^2]}{R(R - \frac{a}{2})} \\
	&\quad - \frac{\sinh[\kappa(R - \frac{a}{2})] [(R + d)^2 + R^2]}{(R - \frac{a}{2}) D(R)}
\end{align*}

\subsection{Electroneutrality Condition}\label{electroneutrality condition}

For all three hollow nanoparticle geometries, the total charge balance is given by:
\begin{equation}\label{Electroneutrality-condition-general}
	Q_{\scriptscriptstyle{0}}(R) + Q_{\scriptscriptstyle{0}}(R+d) + Q_{\scriptscriptstyle{Hi}}(R - a/2) + Q_{\scriptscriptstyle{Ho}}(R + d + a/2) = 0,
\end{equation}
\noindent where \( Q_{\scriptscriptstyle{0}}(R) \) and \( Q_{\scriptscriptstyle{0}}(R+d) \) are the fixed surface charges located at \( A_\gamma(R) \) and \( A_\gamma(R+d) \), respectively, and \( Q_{\scriptscriptstyle{Hi}} \) and \( Q_{\scriptscriptstyle{Ho}} \) represent the induced charges in the inner and outer electric double layers (EDLs). These are defined as:
\begin{align}
	Q_{\scriptscriptstyle{Hi}}(R - a/2) &= \int\limits_{\boldsymbol{\omega_{\scriptscriptstyle{in}}}} \rho_{\text{el}}(r) \, d\mathbf{V} = f_{\scriptscriptstyle{\gamma}} \int_{0}^{R - a/2} r^\gamma \rho_{\text{el}}(r) \, dr, \label{Induced-charge-in}\\
	Q_{\scriptscriptstyle{Ho}}(R + d + a/2) &= \int\limits_{\boldsymbol{\omega_{\scriptscriptstyle{out}}}} \rho_{\text{el}}(r) \, d\mathbf{V} = f_{\scriptscriptstyle{\gamma}} \int_{R + d + a/2}^{\infty} r^\gamma \rho_{\text{el}}(r) \, dr. \label{Induced-charge-out}
\end{align}

\noindent The integrals in \cref{Induced-charge-in,Induced-charge-out} are taken over the internal and external electrolyte volumes \( \boldsymbol{\omega_{\scriptscriptstyle{in}}} \) and \( \boldsymbol{\omega_{\scriptscriptstyle{out}}} \), respectively. Here, \( f_{\scriptscriptstyle{\gamma}} \) is a geometry-dependent constant, with \( \gamma = 0, 1, 2 \) for planar, cylindrical, and spherical geometries.

From these expressions, the induced surface charge density profiles for \( r \leq R - a/2 \) and \( r \geq R + d + a/2 \) are given by:
\begin{align}
	\sigma_{\scriptscriptstyle{\gamma}}(r) &= \frac{1}{r^\gamma} \int_0^r r^\gamma \rho_{\text{el}}(r) \, dr, \label{Induced-charge-density-in}\\
	\sigma_{\scriptscriptstyle{\gamma}}(r) &= -\frac{1}{r^\gamma} \int_r^\infty r^\gamma \rho_{\text{el}}(r) \, dr. \label{Induced-charge-density-out}
\end{align}

Substituting these into \cref{Electroneutrality-condition-general} yields the general electroneutrality condition:
\begin{equation}\label{Electroneutrality-condition-general2}
	\begin{split}
		R^\gamma \sigma_{\scriptscriptstyle{0}} + (R + d)^\gamma \sigma_{\scriptscriptstyle{0}} + (R - a/2)^\gamma \sigma_{\scriptscriptstyle{\gamma}}(R - a/2)
		= (R + d + a/2)^\gamma \sigma_{\scriptscriptstyle{\gamma}}(R + d + a/2).
	\end{split}
\end{equation}

\noindent This equation applies to all three geometries presented in \cref{Slit-shell,Cylindrical-shell-EDL,Spherical-shell-EDL} and allows for the analytical determination of \( \sigma_{\scriptscriptstyle{Hi}} \) when \( \sigma_{\scriptscriptstyle{Ho}} \) is known.

The terms \( \sigma_{\scriptscriptstyle{\gamma}}(R - a/2)/(\varepsilon_{\scriptscriptstyle{0}}\varepsilon) \) and \( \sigma_{\scriptscriptstyle{\gamma}}(R + d + a/2)/(\varepsilon_{\scriptscriptstyle{0}}\varepsilon) \) correspond to the effective electric fields at \( r = R - a/2 \) and \( r = R + d + a/2 \), respectively. Thus, \cref{Electroneutrality-condition-general2} implies an electric field balance:
\begin{equation}\label{Electrical-field-balance}
	\begin{split}
		R^\gamma E_\gamma(R) \mathbf{e_r} + (R - a/2)^\gamma E_\gamma(R - a/2) \mathbf{e_r} + (R + d)^\gamma E_\gamma(R + d) \mathbf{e_r} \\
		= (R + d + a/2)^\gamma E_\gamma(R + d + a/2) \mathbf{e_r},
	\end{split}
\end{equation}

\noindent where \( \mathbf{e_r} \) is the unit radial vector. Since \( \sigma_{\scriptscriptstyle{0}} > 0 \), the electric field is positive at \( r = R \), \( R + d \), and \( R + d + a/2 \), but negative at \( r = R - a/2 \). According to \cref{Induced-charge-density-out}, \( \sigma_{\scriptscriptstyle{\gamma}}(R + d + a/2) \) equals the negative of the induced surface charge density outside the shell, while \cref{Induced-charge-density-in} defines the total induced charge inside.

In Ref.~\cite{Adrian-JML-2023}, a different sign convention was adopted for \( \sigma_{\scriptscriptstyle{\gamma}}(R - a/2) \), which does not affect the physical interpretation but leads to a reversed sign for \( E(r) \) in the inner region. In this work, we adopt the convention consistent with \cref{Electrical-field-balance}, and define:
\[
\sigma_{\scriptscriptstyle{Hi}} \equiv \sigma_{\scriptscriptstyle{\gamma}}(R - a/2), \quad \sigma_{\scriptscriptstyle{Ho}} \equiv \sigma_{\scriptscriptstyle{\gamma}}(R + d + a/2),
\]
with the understanding that both quantities depend on \( R \), \( d \), \( \rho_0 \), and \( \sigma_{\scriptscriptstyle{0}} \). We will occasionally write \( \sigma_{\scriptscriptstyle{Hi}}(R) \) and \( \sigma_{\scriptscriptstyle{Ho}}(R) \) for clarity, omitting \( d \) when unambiguous.

It can be shown that:
\begin{equation}\label{Eq.Violation_Local_Elec_Cond}
	E_\gamma(R - a/2)\mathbf{e_r} + E_\gamma(R)\mathbf{e_r} \neq \vec{0}, \quad \text{for all finite  R}.
\end{equation}

Given that \( E_\gamma(R) = \sigma_0 / (\varepsilon_0 \varepsilon) \), this indicates a violation of the local electroneutrality condition (VLEC) within the shell. The VLEC originates from the nanoparticles' topology and ion-ion correlations across their shells~\cite{Lozada-Cassou-PRE1997}. This effect has been predicted by PB and integral equation theories~\cite{Lozada_1984,Lozada1996,Lozada-Cassou-PRL1996,Yu_1997,Aguilar_2007,Levin_electroneutrality-2016} and corroborated by density functional theory, simulations~\cite{Levy-electroneutrality-2020,Levy-electroneutrality-PRE-2021,Keshavarzi_2020}, and experiments~\cite{Cuvillier-Rondelez-1998,Luo-electroneutrality-nature-2015}.

Nonetheless, global electroneutrality is preserved via \cref{Electrical-field-balance}. While slit-shells may approach local neutrality for moderate \( R \), spherical and cylindrical shells require much larger \( R \) to do so. However, in the limit \( R \to \infty \),
\[
\lim_{R \to \infty} [E_\gamma(R - a/2) + E_\gamma(R)] = 0, \quad
\lim_{R \to \infty} [E_\gamma(R + d + a/2) - E_\gamma(R + d)] = 0.
\]
Thus, the local electroneutrality condition is recovered asymptotically. In this limit, \cref{Electrical-field-balance} reduces to:
\[
R^\gamma \sigma_0 + (R + d)^\gamma \sigma_0 + (R - a/2)^\gamma \sigma_0 = (R + d + a/2)^\gamma \sigma_0,
\]
implying that in this limit \textit{local electroneutrality is independently satisfied inside and outside the shell}.

\subsection{Hollow nanoparticles capacitance}\label{Hollow-nanoparticles-capacitance}

As shown above, a charged nanopore immersed in an electrolyte induces electric double layers (EDLs) both inside and outside its walls. The self-capacitance of such a nanopore electrode corresponds to the charge transported from a reference point—taken here as infinity—to its walls, producing an electrostatic potential difference. The voltage drops across the five electrostatic regions are illustrated in \cref{fig:geometries_all}.

Accordingly, the total capacitance of the nanopore is that of five capacitive regions connected in series:

\begin{equation}
	C_d = \left( \frac{1}{C_1} + \frac{1}{C_2} + \frac{1}{C_3} + \frac{1}{C_4} + \frac{1}{C_5} \right)^{-1}, \label{Ec.CT}
\end{equation}

\noindent where $C_j$ denotes the capacitance associated with the $j$-th region. The \textit{capacitance per unit area} in each region is defined as

\begin{gather}
	c_j = \frac{C_j}{A_j(r_i)} = \frac{Q_j(r_i)/A_j(r_i)}{\Delta\psi_j(r_i)} = \frac{\sigma_j(r_i)}{\Delta\psi_j(r_i)},
	\label{Ec.Cs.ind-definition}
\end{gather}

\noindent where $Q_j(r_i)$ is the charge, $A_j(r_i)$ the area, and $\Delta \psi_j(r_i)$ the voltage drop across the $j$-th region, evaluated at location $r_i$. The total \textit{specific capacitance}, i.e., capacitance per unit area, is thus

\begin{equation}
	c_{\scriptscriptstyle{d}} = \left( \sum_{j=1}^5 \frac{1}{c_j} \right)^{-1}. \label{Ec.cT}
\end{equation}

\noindent Analytical expressions for $\sigma_j(r_i)$ and $\Delta \psi_j(r_i)$ for each of the five regions and all three nanopore topologies have been presented in the previous sections. For the slit nanopore, using \cref{Plates-E(r),Plates-phi-H,Plates-phi-0,Plates-psi-0,Plates-psi-H,Plates-psi-d}, we obtain:

\begin{gather}
	c_1 = \frac{\sigma_{1}(R - a/2)}{\psi_d - \psi_H} = \frac{\varepsilon_0 \varepsilon \kappa \sinh[\kappa(R - a/2)]}{\cosh[\kappa(R - a/2)] - 1}, \label{Equ.Cap-Plate1} \\
	c_2 = \frac{\sigma_2(R)}{\psi_H - \psi_0} = \frac{\varepsilon_0 \varepsilon}{a/2}, \label{Equ.Cap-Plate2} \\
	c_3 = \frac{\sigma_3(R)}{\psi_0 - \varphi_0} = \frac{\varepsilon_0 \varepsilon}{d}, \label{Equ.Cap-Plate3} \\
	c_4 = \frac{\sigma_4(R + d)}{\varphi_0 - \varphi_H} = \frac{\varepsilon_0 \varepsilon}{a/2}, \label{Equ.Cap-Plate4} \\
	c_5 = \frac{\sigma_{\scriptscriptstyle{Ho}}}{\varphi_H} = \varepsilon_0 \varepsilon \kappa. \label{Equ.Cap-Plate5}
\end{gather}

For the cylindrical nanopore, using \cref{Cylinder-E(r),Cyl-phi-H,Cyl-phi-0,Cyl-psi-0,Cyl-psi-H,Cyl-psi-d}, we have:

\begin{gather}
	c_1 = \frac{\sigma_1(R - a/2)}{\psi_d - \psi_H} = \frac{\varepsilon_0 \varepsilon \kappa I_1[\kappa(R - a/2)]}{I_0[\kappa(R - a/2)] - 1}, \label{Equ.Cap-Cyl1} \\
	c_2 = \frac{\sigma_2(R)}{\psi_H - \psi_0} = \frac{\varepsilon_0 \varepsilon}{R \ln(R/(R - a/2))}, \label{Equ.Cap-Cyl2} \\
	c_3 = \frac{\sigma_3(R)}{\psi_0 - \varphi_0} = \frac{\varepsilon_0 \varepsilon}{R \ln((R + d)/R)}, \label{Equ.Cap-Cyl3} \\
	c_4 = \frac{\sigma_4(R + d)}{\varphi_0 - \varphi_H} = \frac{\varepsilon_0 \varepsilon}{(R + d) \ln(R_H/(R + d))}, \label{Equ.Cap-Cyl4} \\
	c_5 = \frac{\sigma_{\scriptscriptstyle{Ho}}}{\varphi_H} = \frac{\varepsilon_0 \varepsilon \kappa K_1[\kappa R_H]}{K_0[\kappa R_H]}. \label{Equ.Cap-Cyl5}
\end{gather}

For the spherical nanopore, using \cref{Sphere-E(r),Sph-phi-H,Sph-phi-0,Sph-psi-0,Sph-psi-H,Sph-psi-d}, we obtain:

\begin{gather}
	c_1 = \frac{\sigma_1(R - a/2)}{\psi_d - \psi_H} = \frac{\varepsilon_0 \varepsilon \left[ \sinh(\kappa(R - a/2)) - \kappa(R - a/2)\cosh(\kappa(R - a/2)) \right]}{(R - a/2)\left[ \kappa(R - a/2) - \sinh(\kappa(R - a/2)) \right]}, \label{Equ.Cap-Sphe1} \\
	c_2 = \frac{\sigma_2(R)}{\psi_H - \psi_0} = \frac{\varepsilon_0 \varepsilon(R - a/2)}{R(a/2)}, \label{Equ.Cap-Sphe2} \\
	c_3 = \frac{\sigma_3(R)}{\psi_0 - \varphi_0} = \frac{\varepsilon_0 \varepsilon(R + d)}{R d}, \label{Equ.Cap-Sphe3} \\
	c_4 = \frac{\sigma_4(R + d)}{\varphi_0 - \varphi_H} = \frac{\varepsilon_0 \varepsilon R_H}{(R + d)(a/2)}, \label{Equ.Cap-Sphe4} \\
	c_5 = \frac{\sigma_{\scriptscriptstyle{Ho}}}{\varphi_H} = \frac{\varepsilon_0 \varepsilon (1 + \kappa R_H)}{R_H}. \label{Equ.Cap-Sphe5}
\end{gather}

\noindent Full derivations of these expressions are provided in Ref.~\cite{Adrian-JML-2023}.

From \cref{Ec.cT}, the total specific capacitance $c_{\scriptscriptstyle{d}}$ can be readily obtained using the above expressions for $c_j$, for each of the three geometries.

Notice that the capacitance is defined as $C = Q/V$, while the differential capacitance is defined as $C_d = dQ/dV$. However, since we are employing a linear theory, all capacitances are independent of surface charge densities, and the integral and differential capacitances coincide. This is also the case in density functional studies of RPM electrolytes confined in a slit pore~\cite{Pizio2012}. Henceforth, we will refer to the capacitance and the specific capacitance as $C$ and $c$, respectively .

Although this study is limited to symmetric electrolytes within the linear Poisson–Boltzmann approximation, our methodology can be extended to asymmetric systems in ion size and valence, as well as constant charge, and asymmetrically charged hollow nanoparticles' walls, which would be relevant for the cylindrical and spherical geometries. The analysis of hollow nanoparticle capacitance may have implications not only for energy storage device design, but also for understanding micelles, nanopores, and confined charged systems in biology, medicine, drug delivery, and the chemical industry.

\section{Results and discussion}\label{Res_Disc}

In this section we present results for the differential capacitance of hollow nanoshells. We consider three different nanoparticles' topologies: planar, cylindrical, and spherical. The nanoparticles are at infinite dilutions, hence no nanoparticle-nanoparticle interaction potentials are considered. The electrolyte solution can freely enter into the nanoparticles' cavities. Thus, in accordance with the principle of equivalence between particles and fields~\cite{Aguilar_2007,Odriozola-Fortschritte-der-Physik-2017}, the statistical mechanics Poisson-Boltzmann equation for these systems, by construction, guaranties that the equilibrium chemical potential for the fluid inside and outside the hollow nanoparticles, $\mu_{\scriptscriptstyle_{0}}$, are equal to the that of the bulk solution, $\mu_{\scriptscriptstyle_{bulk}}$. This is an important symmetry condition of these systems. \textit{That is, independently of the hollow nanoparticles' topology, their cavity size, shell-walls' thickness and charge, $\mu_{\scriptscriptstyle_{0}}=\mu_{\scriptscriptstyle_{bulk}}$}. We will come back to this point further below.
 
 \subsection{The linear Poisson-Boltzmann approximation}\label{The LPB approximation}
 
\begin{figure}[!htb]
	\begin{subfigure}{.5\textwidth}
		\centering
		\includegraphics[width=1.1\linewidth]{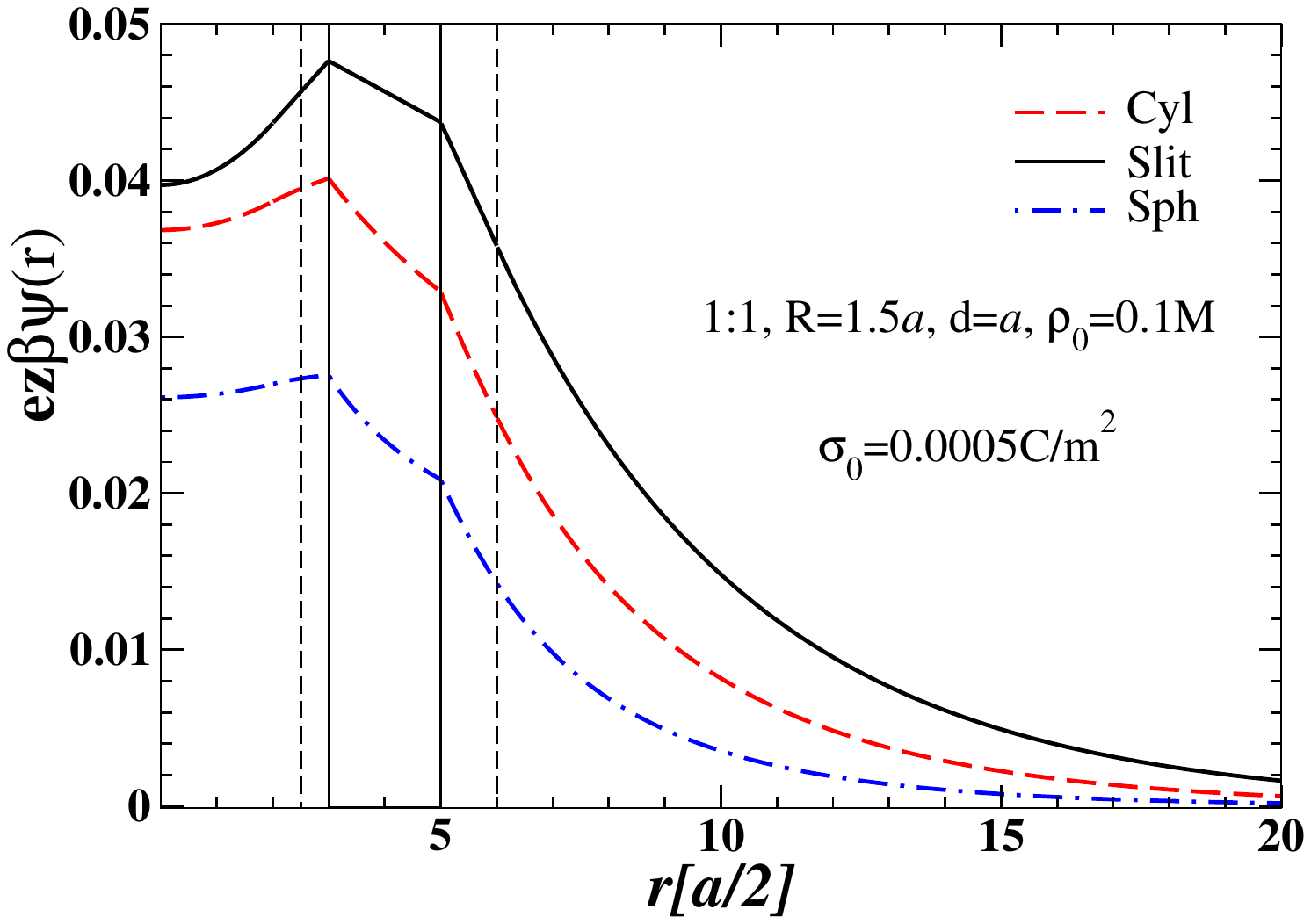}
		\caption{Small cavity radius.}
		\label{Fig.Cd_vs_r-R1.5}
	\end{subfigure}
	\begin{subfigure}{.5\textwidth}
		\centering
		\includegraphics[width=1.1\linewidth]{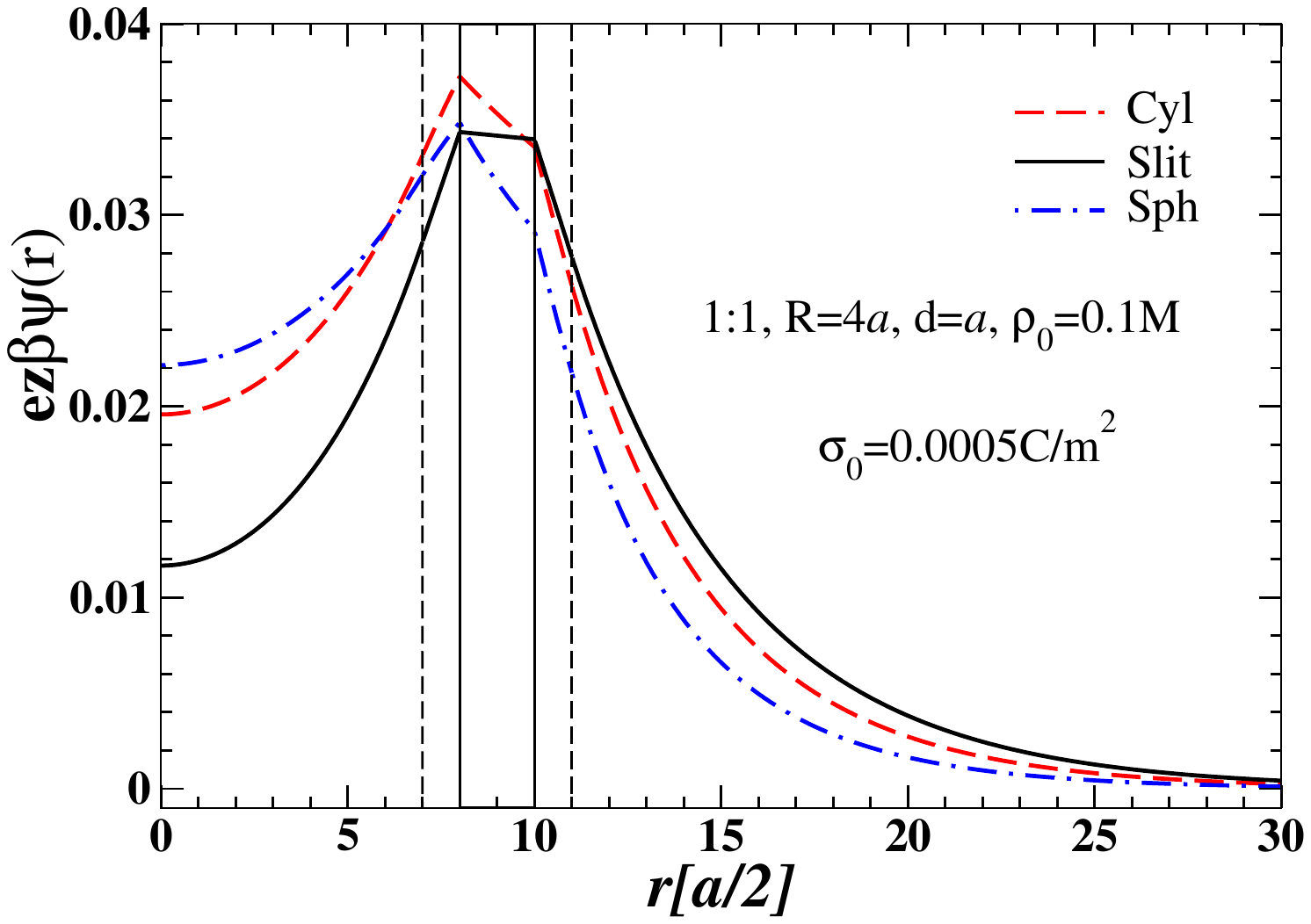}
		\caption{Less small cavity radius.}
		\label{Fig.Cd_vs_r-R4}
	\end{subfigure}
	\begin{subfigure}{.5\textwidth}
		\centering
		\includegraphics[width=1.1\linewidth]{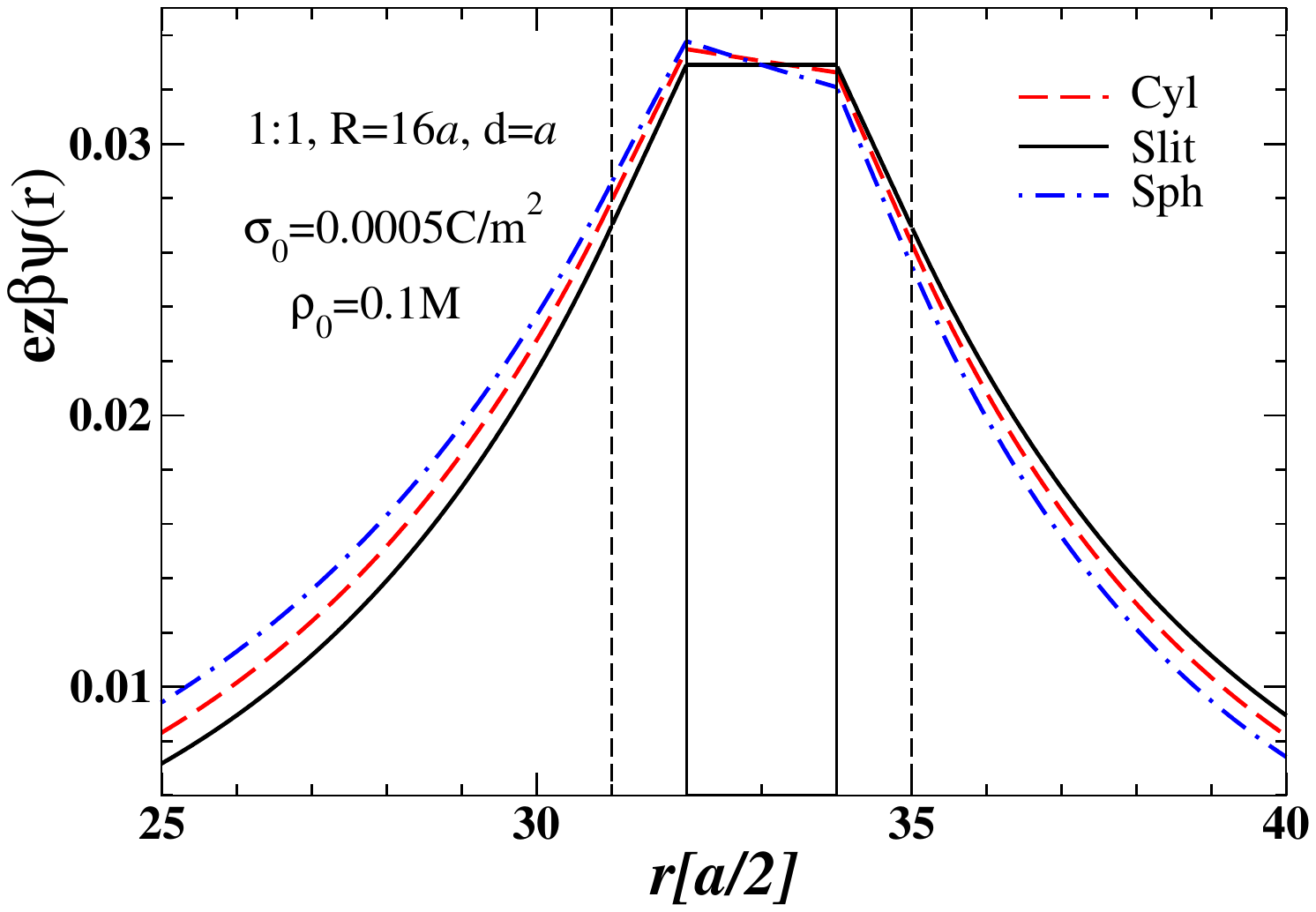}
		\caption{Large cavity radius.}
		\label{Fig.Cd_vs_r-R16}
	\end{subfigure}
	\begin{subfigure}{.5\textwidth}
		\centering
		\includegraphics[width=1.1\linewidth]{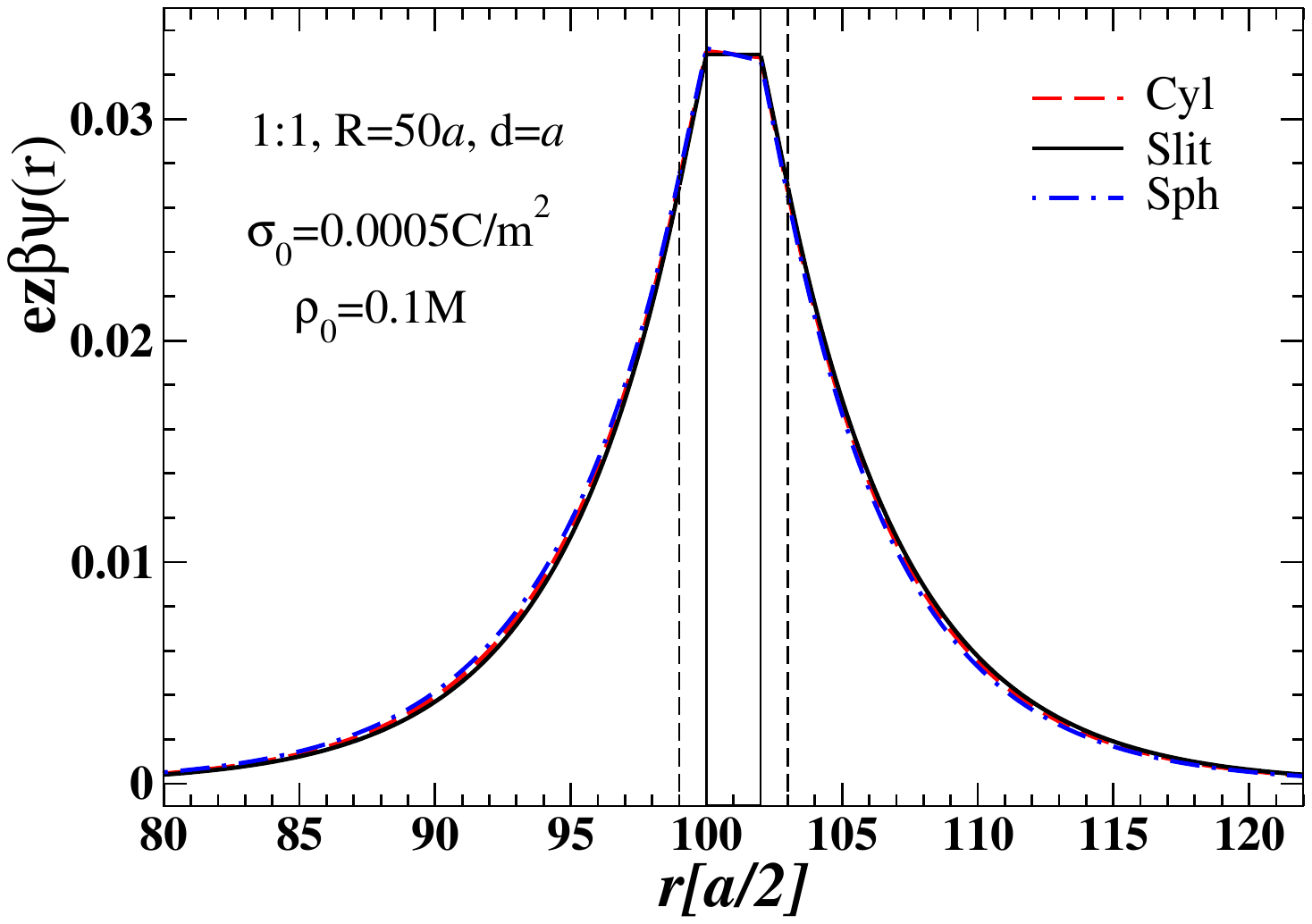}
		\caption{Very large cavity radius.}
		\label{Fig.Cd_vs_r-R50}
	\end{subfigure}
	\caption{Reduced mean electrostatic potential (MEP), \( ez_+\beta\psi(r) \), for hollow nanoparticles with planar, cylindrical, and spherical geometries. The MEP is evaluated at key positions: \( r = 0 \), \( r = R - a/2 \), \( r = R \), \( r = R + d \), and \( r = R + d + a/2 \), corresponding to \( ez_+\beta\psi_{\scriptscriptstyle{d}} \), \( ez_+\beta\psi_{\scriptscriptstyle{H}} \), \( ez_+\beta\psi_{\scriptscriptstyle{0}} \), \( ez_+\beta\varphi_{\scriptscriptstyle{0}} \), and \( ez_+\beta\varphi_{\scriptscriptstyle{H}} \), respectively. The system contains a symmetric 1:1 electrolyte at bulk concentration \( \rho_{\scriptscriptstyle{0}} = 0.1\,\mathrm{M} \), and the surface charge density on both inner and outer walls of the shell is \( \sigma_{\scriptscriptstyle{0}} = 0.0005\,\mathrm{C/m^2} \). The shell thickness is fixed at \( d = a \). Panels (a)--(d) show profiles for cavity radii: (a) \( R = 1.5a \) (\cref{Fig.Cd_vs_r-R1.5}), (b) \( R = 4a \) (\cref{Fig.Cd_vs_r-R4}), (c) \( R = 16a \) (\cref{Fig.Cd_vs_r-R16}), and (d) \( R = 50a \) (\cref{Fig.Cd_vs_r-R50}). Note the non-monotonic and nonlinear behavior of \( ez_+\beta\psi_{\scriptscriptstyle{H}} \) and \( ez_+\beta\varphi_{\scriptscriptstyle{H}} \) as functions of cavity radius in the cylindrical and spherical geometries.}
	\label{Fig.Cd_vs_r}
\end{figure}

In our calculations, we restrict ourselves to charge densities on the nanoparticles' walls and electrolyte concentrations and valence such that $ez\beta \psi (r)<1$, $\forall 0\leq r < \infty$, to be within the mathematical validity of the LPB equation (see \cref{Ec.gmas,Ec.gmenos}). In \cref{Fig.Cd_vs_r} we show the dimensionless mean electrostatic potential (MEP) profile, $ez\beta \psi (r)$, for a $z_1:z_2=1:1$, $\rho_{\scriptscriptstyle{0}}=0.1M$ electrolyte, while the surface charge density on both sides of the nanoparticles wall is $\sigma_{\scriptscriptstyle{0}}=0.0005C/m^2$. Four cavity radii are considered, $R=1.5a, 4a, 16a$ and $5a$. In general, increasing the cavity size implies a lower MEP. We explored the mathematical validity of the LPB equation as a function of all the model parameters.

The MEP decreases with increasing $R$, and increases with increasing electrolyte valence $z_1:z_2$, cavity walls' surface charge density, $\sigma_{\scriptscriptstyle{0}}$ and thickness $d$, although, the walls' thickness increases only slightly the MEP. Notwithstanding, for small cavity sizes,  $ez\beta \psi_{\scriptscriptstyle_{H}} (R)$ and $ez\beta \varphi_{\scriptscriptstyle_{H}} (R)$ of the non-planar geometries exhibit a non-monotonic nonlinear behavior as a function of $R$ and $d$, while the for the slit $ez\beta \psi_{\scriptscriptstyle_{H}} (R)$ and $ez\beta \varphi_{\scriptscriptstyle_{H}} (R)$ are monotonically, descending non-linear functions of $R$ and $d$. In addition, depending on the values of $\rho_{\scriptscriptstyle{0}}$, $R$ and $d$, there are crossovers among the $\psi (r)$ MEP' functions of the different geometries. These nonlinearities of the MEP at the boundaries of the hollow nanoparticles have some consequences on their corresponding effective electrical field and, hence, in their differential capacitances (see \cref{Hollow-nanoparticles-capacitance})~\cite{Lozada-Cassou_JML-2025}. We will differ this discussion to \cref{The electric field}. Of course, the $\lim_{R \to \infty}[ez\beta \Psi (r)]\rightarrow [ez\beta \Psi_{\scriptscriptstyle_{\infty}} (r)]$, where $\Psi_{\scriptscriptstyle_{\infty}} (r)$ is finite $\forall r$. In this limit the MEP and electric field profiles and their capacitances become equal to those of a planar electrode of thickness $d$~\cite{Adrian-JML-2023}. 

 All the calculations were made with temperature, $T=298.15 K$, relative dielectric constant $\epsilon =78.5$, thus $\varepsilon_{\scriptscriptstyle_{0}} \varepsilon= 8.854e-12*78.5 F/m$, and ionic diameter, $a=4.25 \textup{\r{A}}$.

In \cref{Fig.Cd_vs_r}, at $r=0$, $r=R-a/2$, $r=R$, $r=R+d$ and $r=R+d+a/2$, the MEP, $ez_+\beta\psi(r)$,  is denoted as $ez_+\beta\psi_{\scriptscriptstyle{d}}, ez_+\beta\psi_{\scriptscriptstyle{H}}, ez_+\beta\psi_{\scriptscriptstyle{0}}, ez_+\beta\varphi_{\scriptscriptstyle{0}}$, and $ez_+\beta\varphi_{\scriptscriptstyle{H}}$, respectively. These quantities are functions of $R$ and $d$. In general we will not make explicit this dependence of the MEP for simplicity of the notation.

 In each of these nanoparticles' geometries, the maximum of their MEP is at $r=R-a/2$, i.e., for $\psi_{\scriptscriptstyle{H}}$. For large cavity radii, the slit's MEP is lower than that for the cylindrical nanoparticle, and this is lower than that for the spherical hollow nanoparticle, for $r\leq R$, but the opposite occurs for $r\geq (R+d)$.


Interestingly, the LPB approximation remains mathematically valid at molar concentrations as high as $2\,\mathrm{M}$, provided $\sigma_{\scriptscriptstyle{0}}$ is low. This is possible due to the point-ion nature of our model. Moreover, despite its simplicity, the LPB solution yields good agreement with results from integral equations for slit shells in a restricted primitive model (RPM) electrolyte~\cite{Lozada_1990-I,Lozada_1990-II}, and even for properties like the $\zeta$-potential of nano-electrodes, for $2{:}2$ electrolytes at sufficiently low $\rho_{\scriptscriptstyle{0}}$ or $\sigma_{\scriptscriptstyle{0}}$~\cite{McQuarrie_StatMech,Degreve_1993,Degreve_1995}.

In general, thinner external electric double layers (EDLs)—resulting from higher salt concentration, higher valence, or increased surface charge density of the nanocapacitor—enhance the dimensionless electrostatic potential, \( ez_+ \beta \psi(r) \), across the entire spatial domain \( 0 \leq r < \infty \) for all three nanoparticle geometries. Conversely, larger cavity radii reduce the magnitude of \( ez_+ \beta \psi(r) \).

An interesting exception arises in the case of the shell wall thickness \( d \). For planar and cylindrical geometries, increasing \( d \) amplifies the dimensionless electrostatic potential \( ez_+ \beta \psi(r) \) within the internal region \( 0 \leq r \leq R \), while attenuating it in the external region \( r \geq R + d \), due to a reduction in the effective electric field at \( r = R + d \). This behavior is more pronounced in the cylindrical geometry, where the internal enhancement is stronger and the external attenuation is weaker. In contrast, for spherical nanoparticles, increasing the wall thickness enhances \( ez_+ \beta \psi(r) \) both inside and outside the cavity, underscoring the crucial influence of geometry on electrostatic behavior in hollow nanoparticle systems. All of these behaviors are systematically explored in our parameter scans.

In our calculations, $\sigma_{\scriptscriptstyle{0}}$ was typically remain below $0.005\,\mathrm{C/m^2}$ for accuracy. Higher $\rho_{\scriptscriptstyle{0}}$, higher $T$, and/or thicker shells improve the LPB’s applicability. For $2{:}2$ salts, the LPB is valid for $\sigma_{\scriptscriptstyle{0}} \lesssim 0.0005\,\mathrm{C/m^2}$; at higher $\rho_{\scriptscriptstyle{0}}$, this threshold rises to $\sim 0.005\,\mathrm{C/m^2}$. Additionally, the LPB becomes more accurate with increasing $R$, as pointed out above.

\subsection{The capacitance as a function of the model's parameters}\label{Capacitance_concentration}

\begin{figure}[!htb]
	\begin{subfigure}{.5\textwidth}
		\centering
		\includegraphics[width=1.1\linewidth]{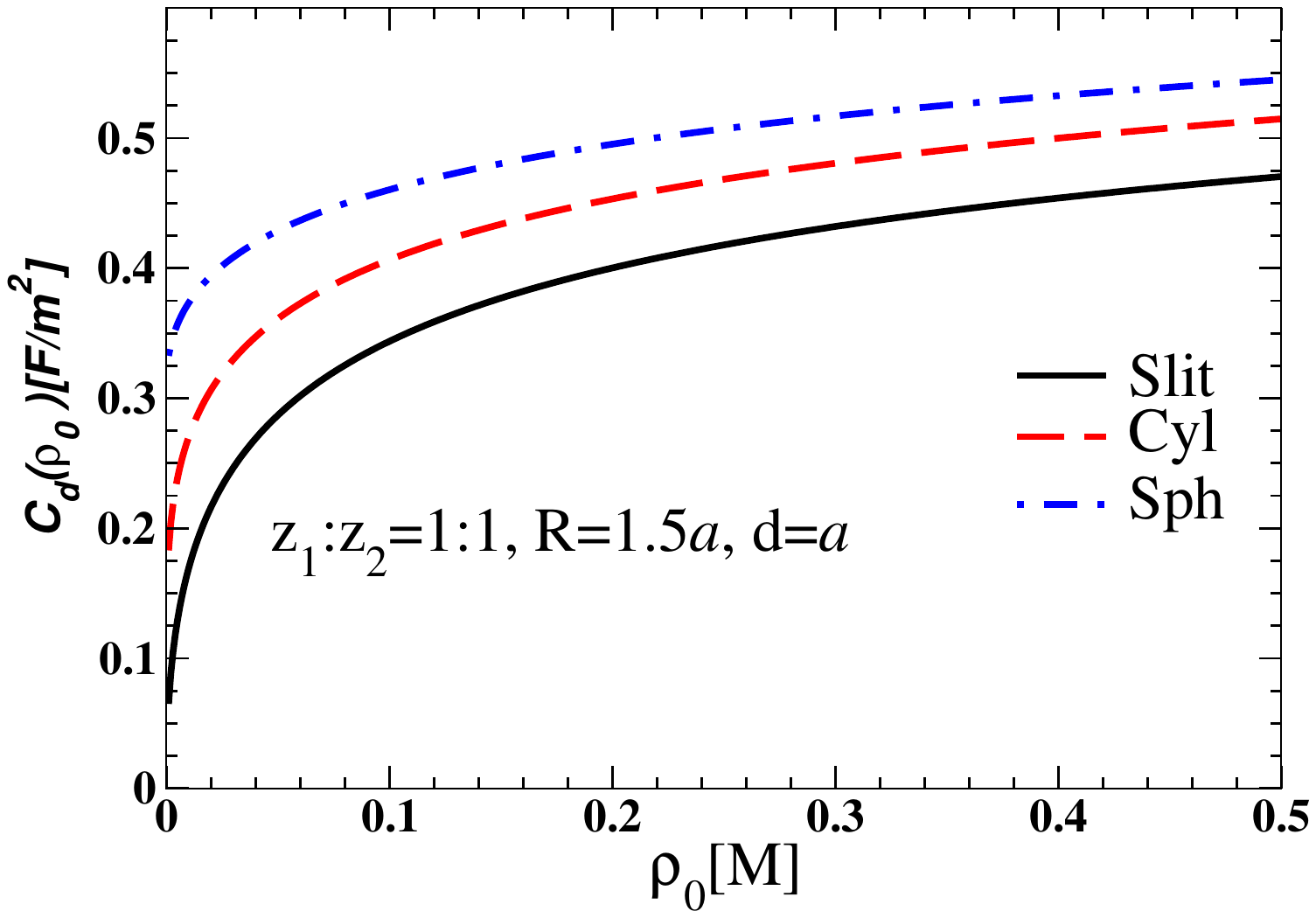}
		\caption{Capacitance as a function of the electrolyte \\ bulk concentration, for narrow cavities and thin\\ walls.}
		\label{Cd_vs_rho_small_R_small_d}
	\end{subfigure}%
	\begin{subfigure}{.5\textwidth}
		\centering
		\includegraphics[width=1.1\linewidth]{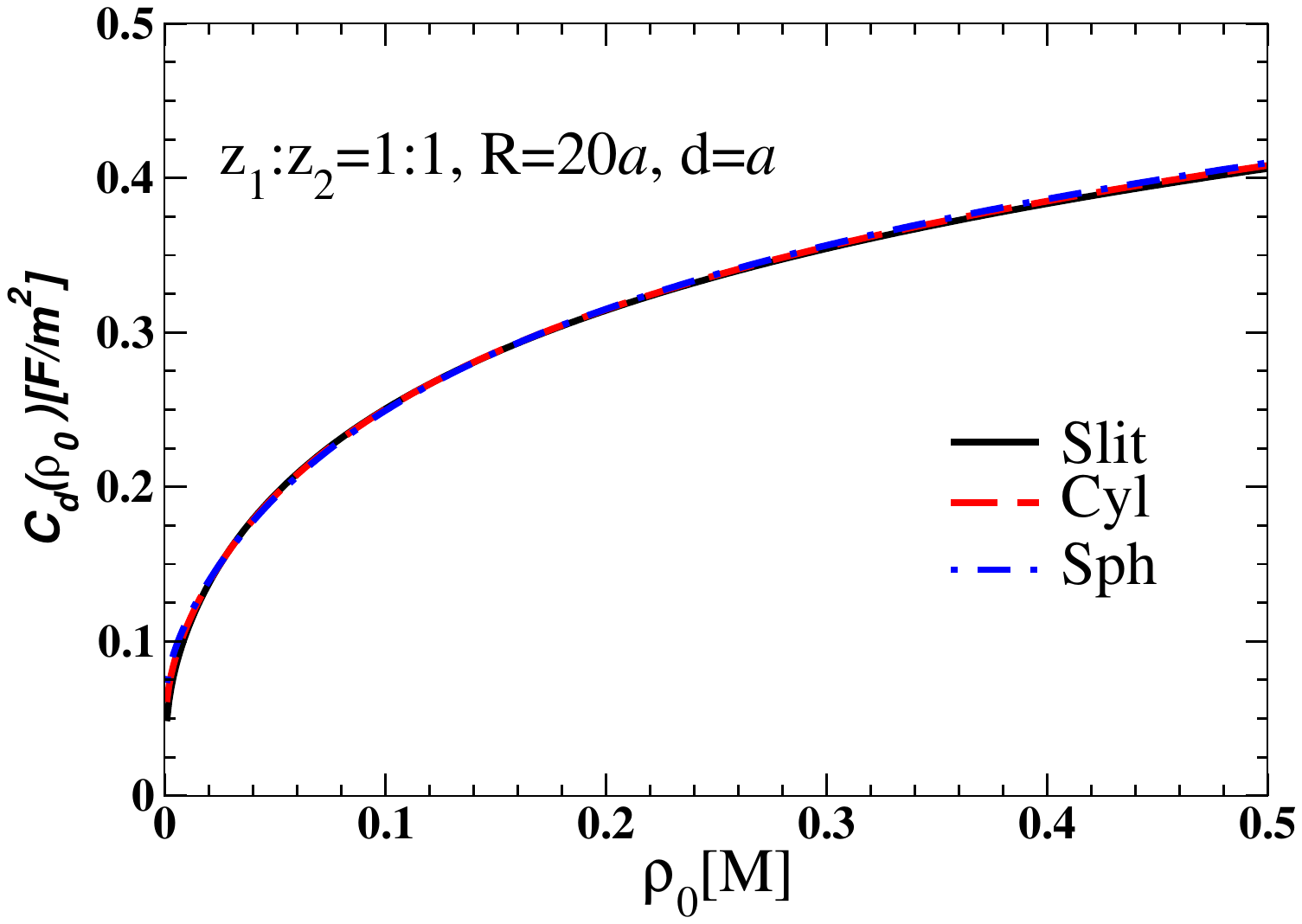}
		\caption{Capacitance as a function of the electrolyte bulk concentration, for large cavities and thin walls.\\}
		\label{Cd_vs_rho_large_R_small_d}
	\end{subfigure}
	\begin{subfigure}{.5\textwidth}
		\centering
		\includegraphics[width=1.1\linewidth]{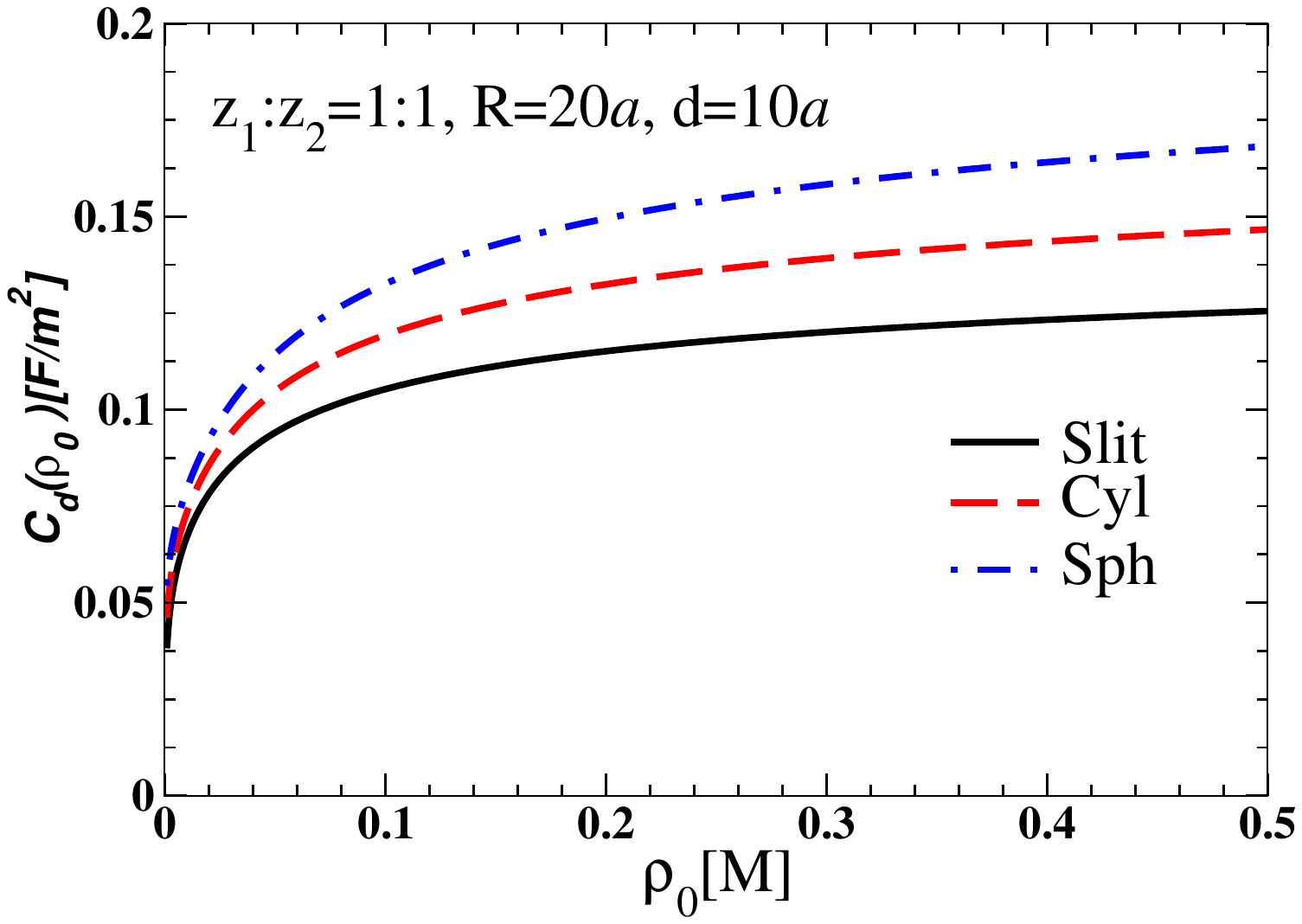}
		\caption{Capacitance as a function of the electrolyte\\ bulk concentration, for large cavities and thick\\ walls.}
		\label{Cd_vs_rho_large_R_large_d}
	\end{subfigure}%
	\begin{subfigure}{.5\textwidth}
		\centering
		\includegraphics[width=1.1\linewidth]{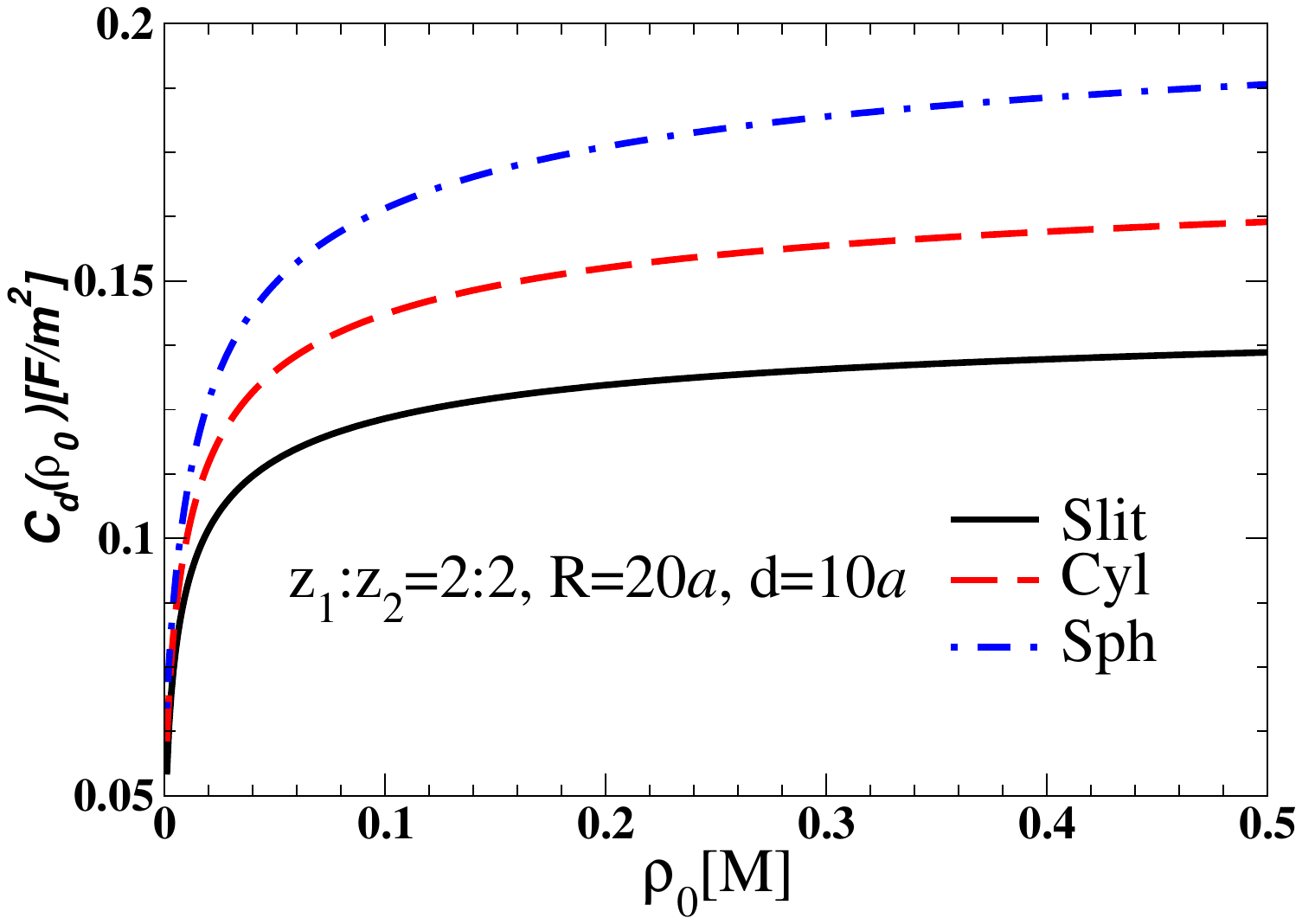}
		\caption{Capacitance of a divalent electrolyte as a function of its bulk concentration, for large cavities and thick walls.}
		\label{Cd_vs_rho_2_2_large_R_large_d}
	\end{subfigure}
	\caption{Specific differential capacitance of hollow nanoparticles with planar, cylindrical, and spherical geometries, plotted as a function of the bulk electrolyte concentration \( \rho_{\scriptscriptstyle{0}} \).}
	\label{Cd_vs_rho_all}
\end{figure}

In \cref{Cd_vs_rho_all}, we present the differential capacitance, \( C_d(\rho_{\scriptscriptstyle{0}}) \), of planar, cylindrical, and spherical hollow nanoparticles as a function of the bulk electrolyte concentration, \( \rho_{\scriptscriptstyle{0}} \). \Cref{Cd_vs_rho_small_R_small_d} shows this dependence for nanoparticles with a small cavity radius, \( R = 1.5a \), and thin walls, \( d = a \), immersed in a symmetric, monovalent electrolyte. In this regime, the capacitance increases nonlinearly and monotonically with the electrolyte concentration. Among the three geometries, the spherical cavity exhibits the highest capacitance, while the planar (slit-like) cavity shows the lowest.

In \cref{Cd_vs_rho_large_R_small_d}, the cavity radius is increased to \( R = 20a \). In this case, the capacitances for the three geometries converge toward a common, lower value, indicating that increasing the cavity size reduces the capacitance and that spherical and cylindrical geometries begin to resemble the planar case. Although not easily distinguished in the figure, the spherical shell still exhibits the highest capacitance, followed by the cylindrical and planar cases.

Increasing the wall thickness to \( d = 10a \) further reduces the capacitance and enhances the differences between geometries, even for large cavities, as shown in \cref{Cd_vs_rho_large_R_large_d}. Conversely, increasing the electrolyte valence to \( z_1\!:\!z_2 = 2\!:\!2 \) significantly raises the capacitance for all geometries, as illustrated in \cref{Cd_vs_rho_2_2_large_R_large_d}.

\begin{figure}[!htb]
	\begin{subfigure}{.5\textwidth}
		\centering
		\includegraphics[width=1.1\linewidth]{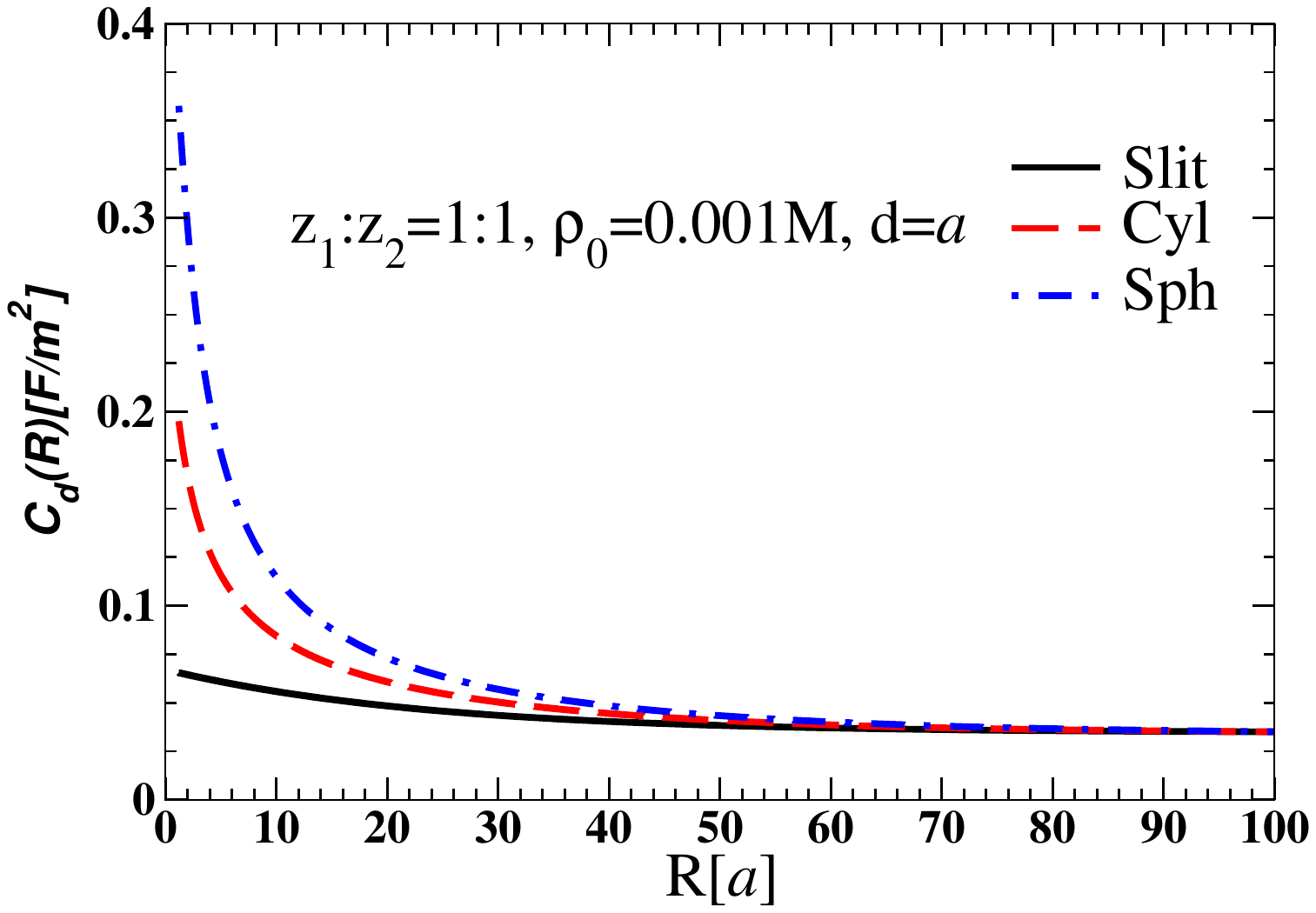}
		\caption{Low bulk concentration}
		\label{Cd_vs_R_low_bulk_concentration}
	\end{subfigure}
	\begin{subfigure}{.5\textwidth}
		\centering
		\includegraphics[width=1.1\linewidth]{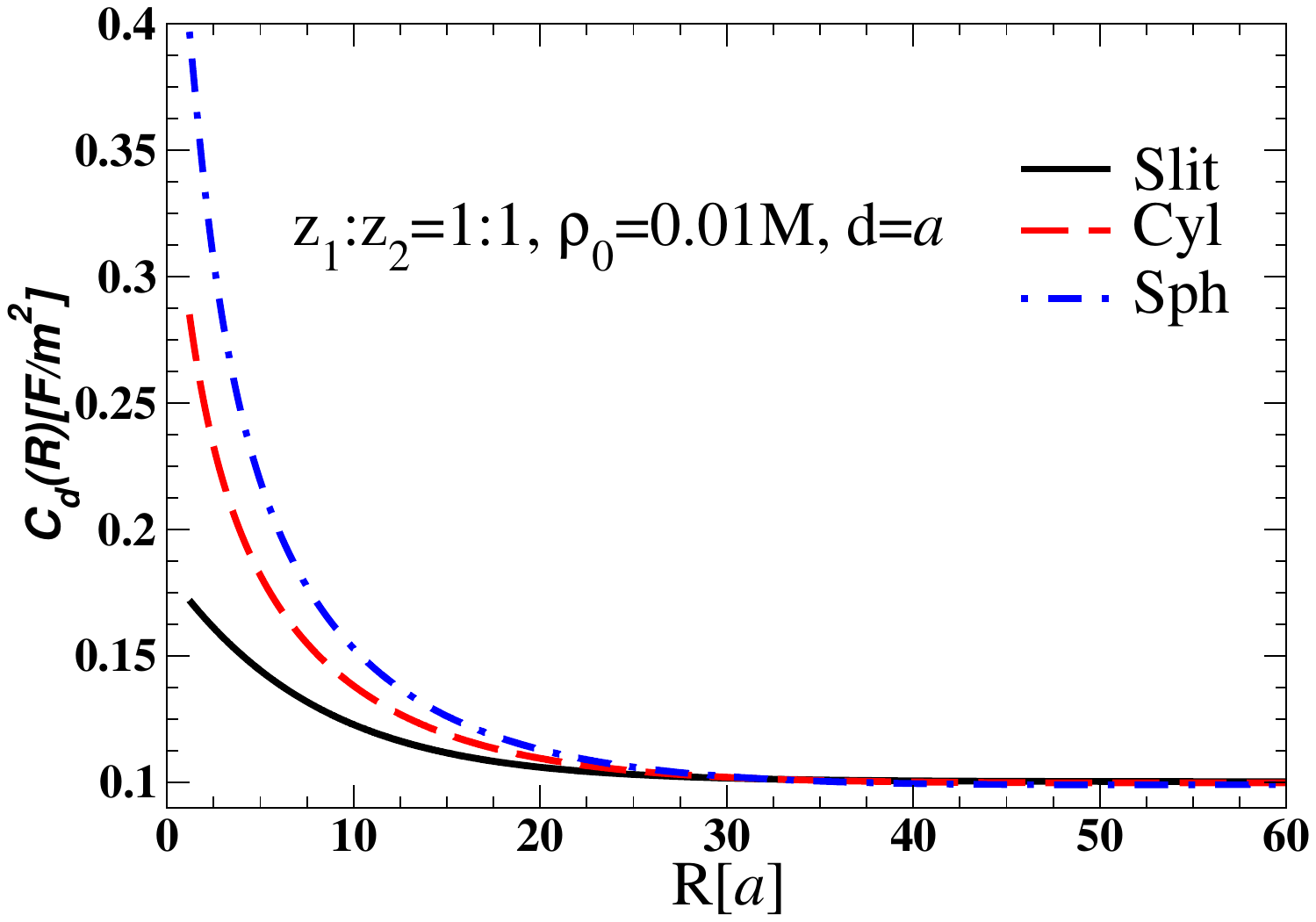}
		\caption{Higher bulk concentration.}
		\label{Cd_vs_R_higher_bulk_concentration}
	\end{subfigure}
	\caption{Capacitance of hollow nanoparticles of planar, cylindrical and spherical geometry, as a function of their cavity radius.}
	\label{Cd_vs_R_all_bulk_concentrations}
\end{figure}

In \cref{Cd_vs_rho_all}, we observe that the capacitance depends sensitively on both the cavity radius and the shell thickness. \Cref{Cd_vs_R_all_bulk_concentrations} shows the variation of capacitance with cavity size for the three nanoparticle geometries. In all cases, the capacitance appears as a nonlinear, monotonically decreasing function of the cavity radius. As expected, the spherical geometry yields the highest capacitance, while the slit geometry yields the lowest.

\begin{figure}[!hbt]
	\begin{subfigure}{.5\textwidth}
		\centering
		\includegraphics[width=1.1\linewidth]{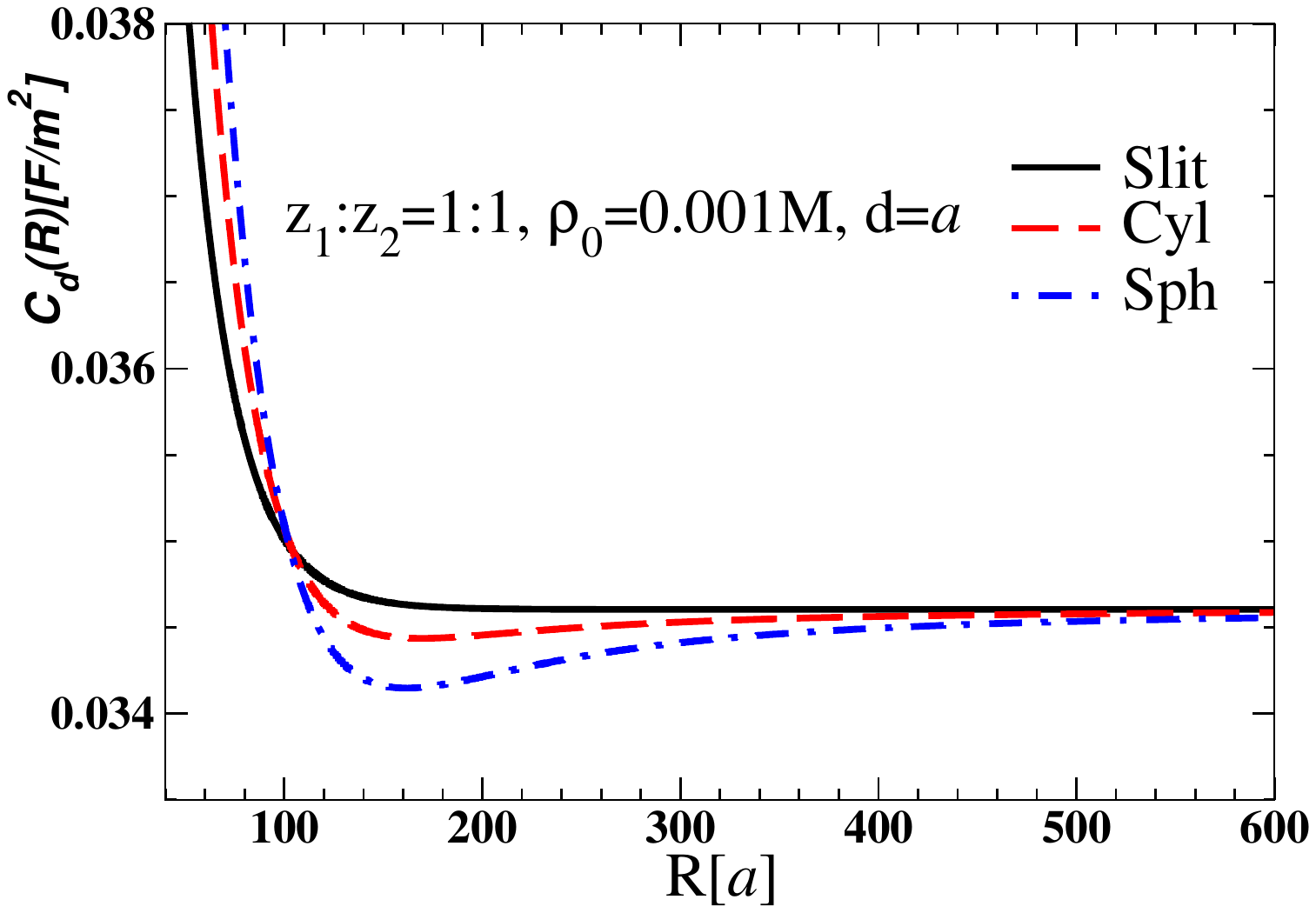}
		\caption{Very low monovalent electrolyte concentration,\\thin nanoparticles' walls.}
		\label{Cd_vs_R_a4p25_Rho0p001_z1_d1p0}
	\end{subfigure}
	\begin{subfigure}{.5\textwidth}
		\centering
		\includegraphics[width=1.1\linewidth]{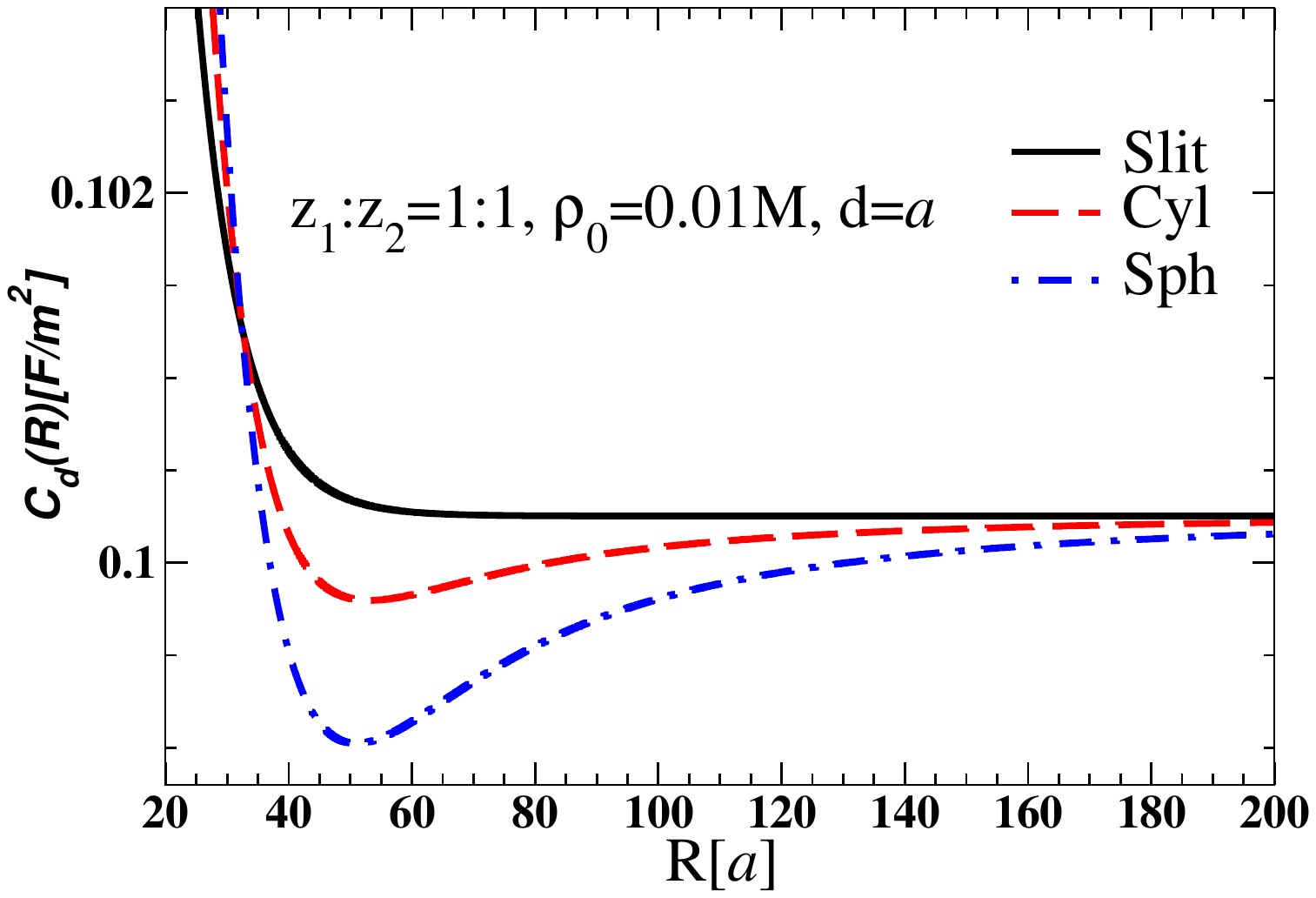}
		\caption{Low monovalent electrolyte concentration,\\thin nanoparticles' walls.}
		\label{Cd_vs_R_a4p25_Rho0p1_z1_d1p0}
	\end{subfigure}\\
	\begin{subfigure}{.5\textwidth}
		\centering
		\includegraphics[width=1.1\linewidth]{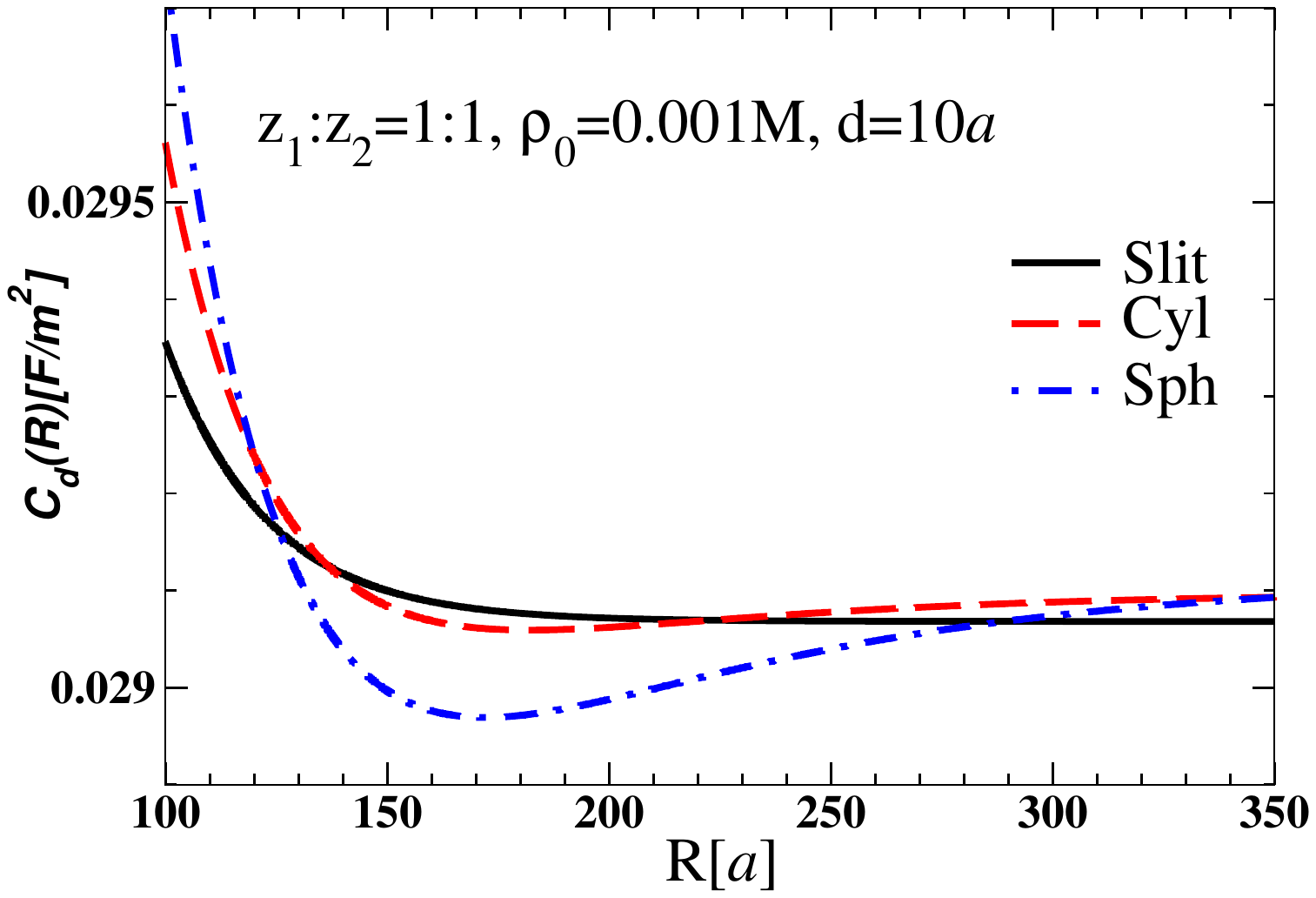}
		\caption{Very low monovalent electrolyte concentration,\\thick nanoparticles' walls.}
		\label{Cd_vs_R_a4p25_Rho0p001_z1_d10p0}
	\end{subfigure}
	\begin{subfigure}{.5\textwidth}
		\centering
		\includegraphics[width=1.1\linewidth]{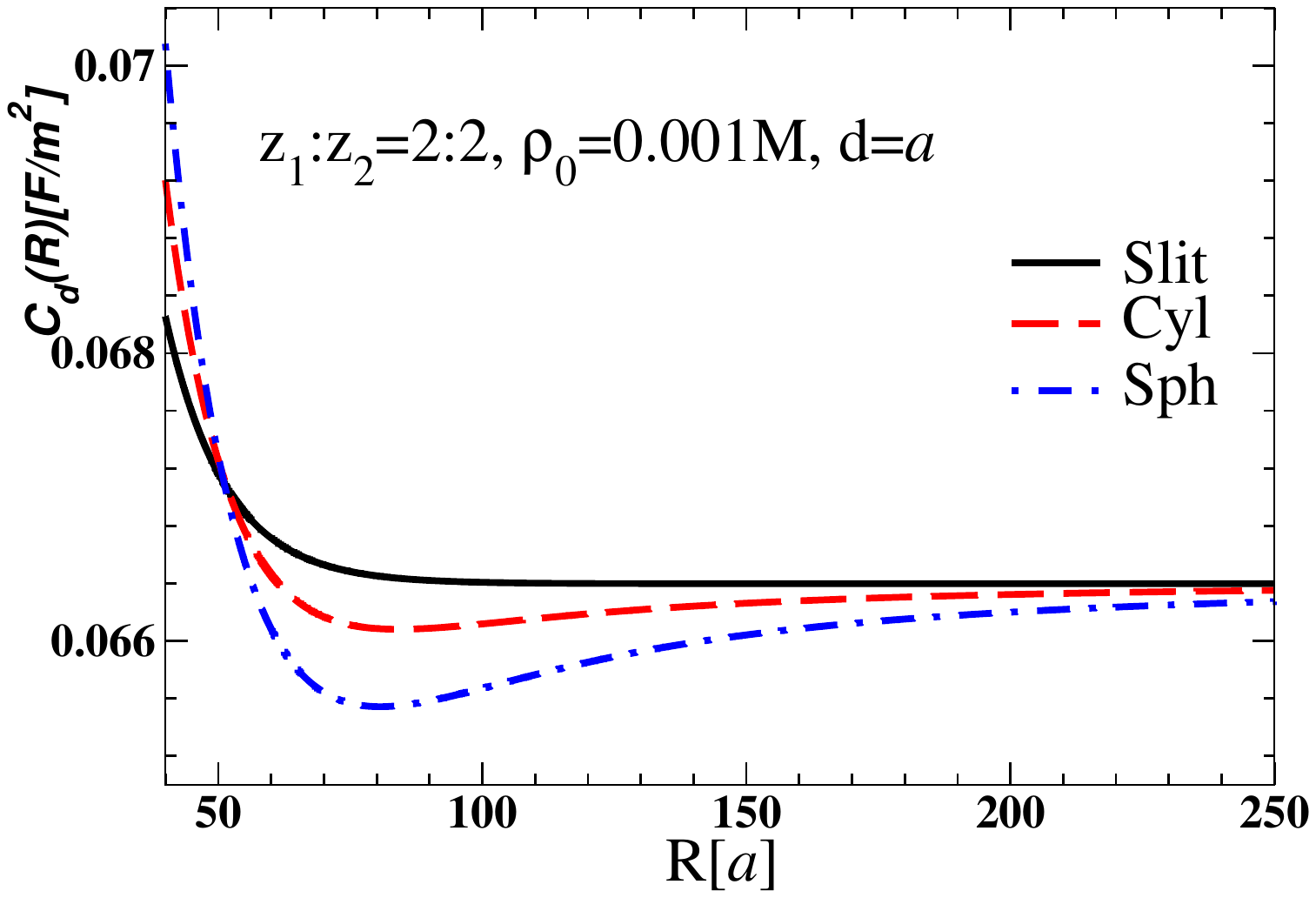}
		\caption{Very low divalent electrolyte concentration,\\thin nanoparticles' walls.}
		\label{Cd_vs_R_a4p25_Rho0p001_z2_d1p0}
	\end{subfigure}
	\caption{Differential capacitance of hollow nanoparticles with planar, cylindrical, and spherical geometry as a function of their cavity radius \( R \) in the large-radius regime.}
	\label{Cd_vs_R_all}
\end{figure}

A closer examination in \cref{Cd_vs_R_all} reveals that for larger cavity sizes, the capacitance curves of the cylindrical and spherical geometries exhibit pronounced nonlinearities, whereas the slit geometry maintains its monotonic, decreasing trend.

For relatively large values of \( R \), the capacitances of the curved geometries (cylindrical and spherical) exhibit crossovers with each other and with that of the slit at specific values of \( R \), followed by a minimum and a subsequent local maximum (see \cref{Cd_vs_R_a4p25_Rho0p001_z1_d1p0}). These minima shift to smaller \( R \) as the electrolyte concentration or ionic valence increases (see \cref{Cd_vs_R_a4p25_Rho0p1_z1_d1p0,Cd_vs_R_a4p25_Rho0p001_z2_d1p0}), and to larger \( R \) values as the wall thickness increases (see \cref{Cd_vs_R_a4p25_Rho0p001_z1_d10p0}).

In the limit \( R \to \infty \), the capacitances for the slit, cylindrical, and spherical geometries converge to a common value:
\begin{equation}
	\lim_{R \to \infty} C_d(R) = \frac{\varepsilon_{\scriptscriptstyle{0}} \varepsilon \kappa}{2 + \kappa (d + a)},
	\label{capacitance-limit-of-R}
\end{equation}
which corresponds to the capacitance of a flat capacitor of thickness \( d \)~\cite{Adrian-JML-2023}, as expected, and above already pointed out (see \cref{The LPB approximation}).

In summary, the capacitance of hollow-shell nanoparticles increases with increasing bulk electrolyte concentration or ionic valence, and with decreasing temperature—that is, with decreasing Debye screening length \( \lambda_{\scriptscriptstyle{D}} \)—and depends nonlinearly on the degree of confinement. Notably, the capacitance expressions do not depend explicitly on the surface charge or potential, but only on the electrolyte parameters and the system geometry.

The oscillatory behavior of the capacitance in cylindrical and spherical geometries as a function of cavity radius appears to originate from geometric factors alone, since a careful analysis of the structure of EDL inside and outside the hollow particles does show a continuous, smooth variation along these oscillations. However, changes in \( \lambda_{\scriptscriptstyle{D}} \) and confinement also modify the internal and external electric double layer (EDL) structures. What role do these EDL structures play in determining the electrostatic response and capacitance of the system? In the following subsections, we address this question in the context of the confinement-induced symmetry-breaking phenomena.

\subsection{The electric field profile}\label{The electric field}

\begin{figure}[!hbt]
	\begin{subfigure}{.5\textwidth}
		\centering
		\includegraphics[width=1.1\linewidth]{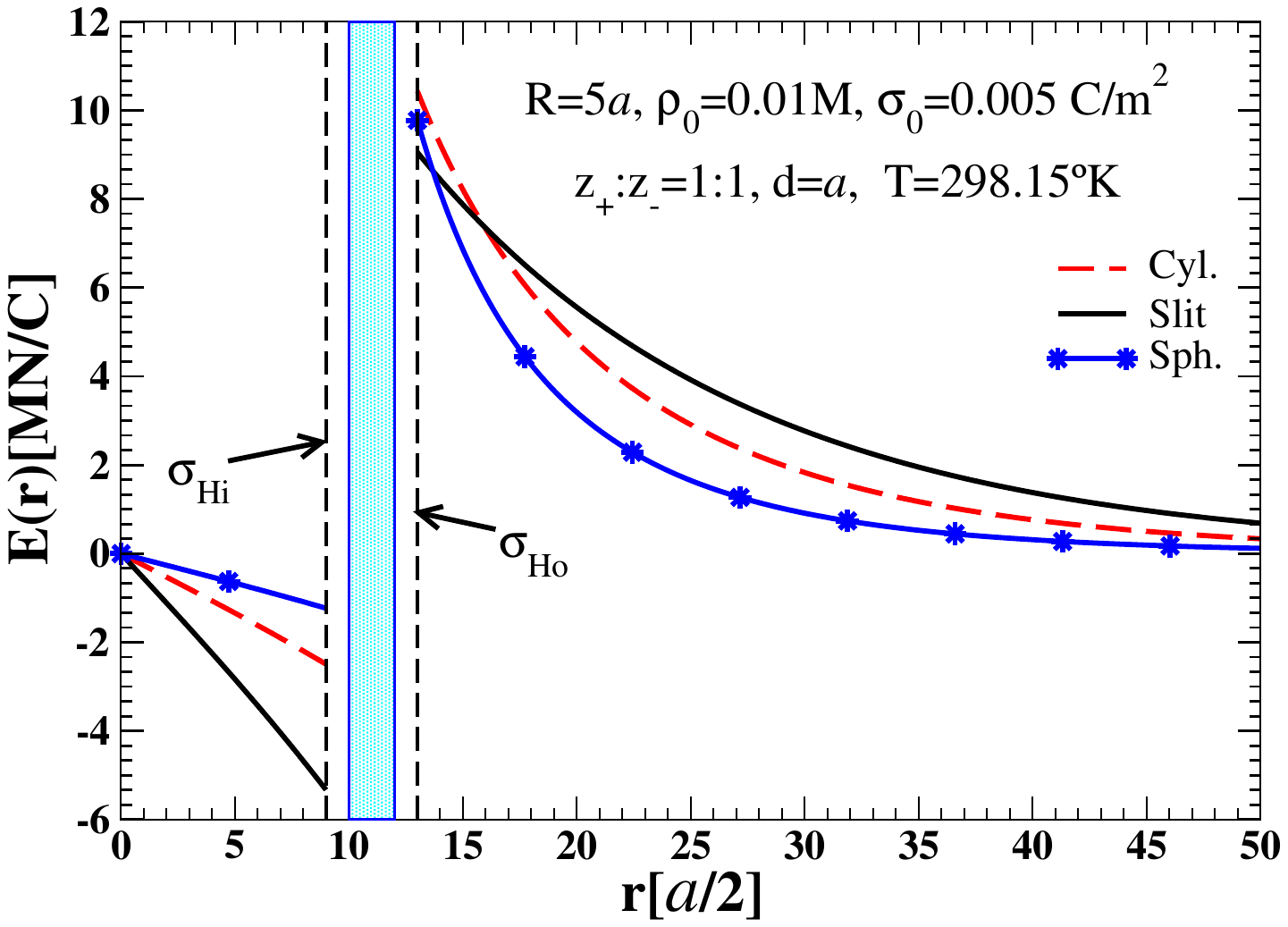}
		\caption{Electric field profile, $E(r)$, for planar,\\ cylindrical, and spherical hollow nanoparticles\\ with cavity radius $R = 5a$, immersed in a $1{:}1$\\ electrolyte.}
		\label{Fig.E(r)_R5a_d1a_rho0p01_s0p005_z1}
	\end{subfigure}
	\begin{subfigure}{.5\textwidth}
		\centering
		\includegraphics[width=1.1\linewidth]{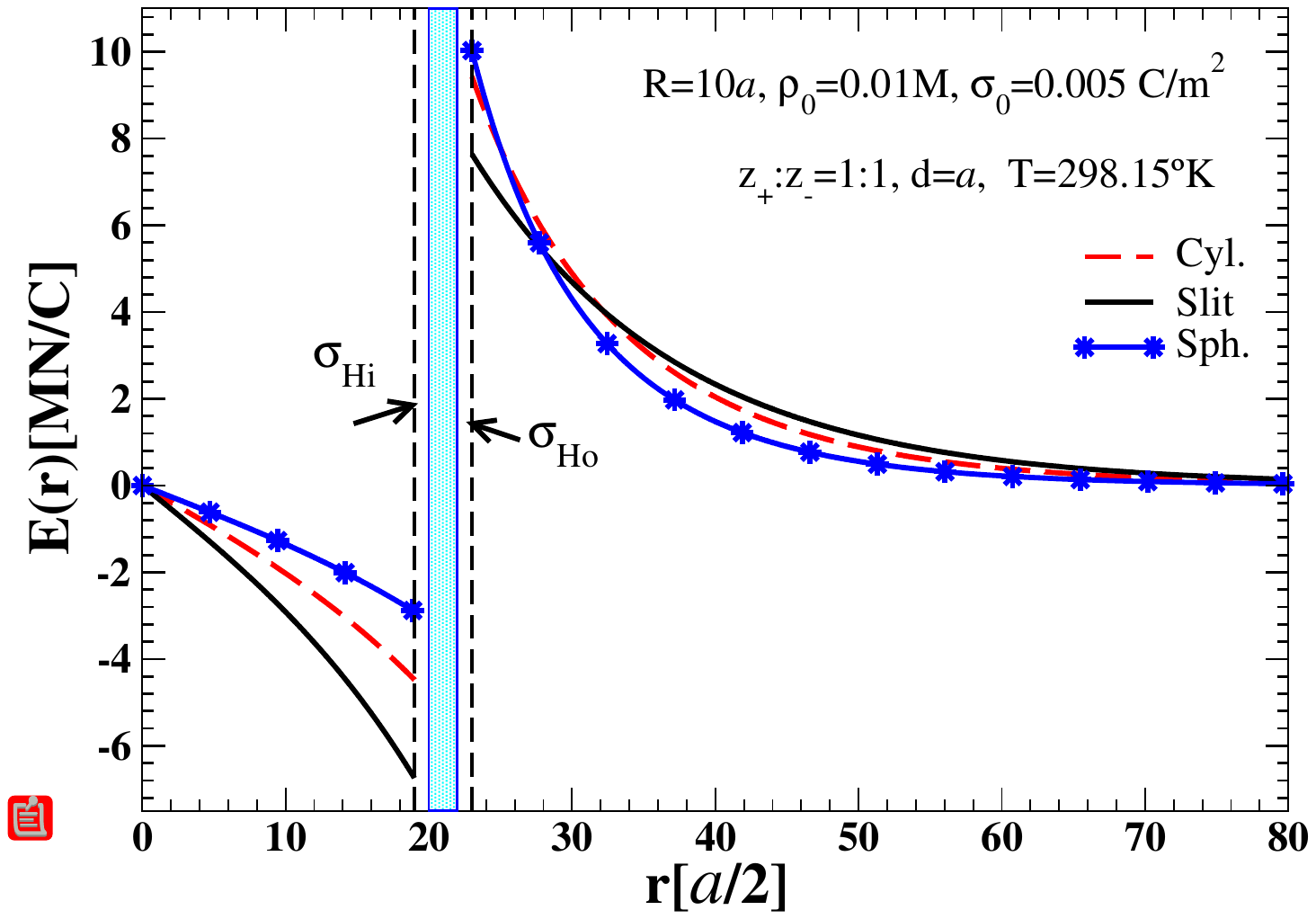}
		\caption{Electric field profile, $E(r)$, for planar,\\ cylindrical, and spherical hollow nanoparticles\\ with a larger cavity radius $R = 10a$, immersed in a $1{:}1$ electrolyte.}
		\label{Fig.E(r)_R10a_d1a_rho0p01_s0p005_z1}
	\end{subfigure}
	\begin{subfigure}{.5\textwidth}
		\centering
		\includegraphics[width=1.1\linewidth]{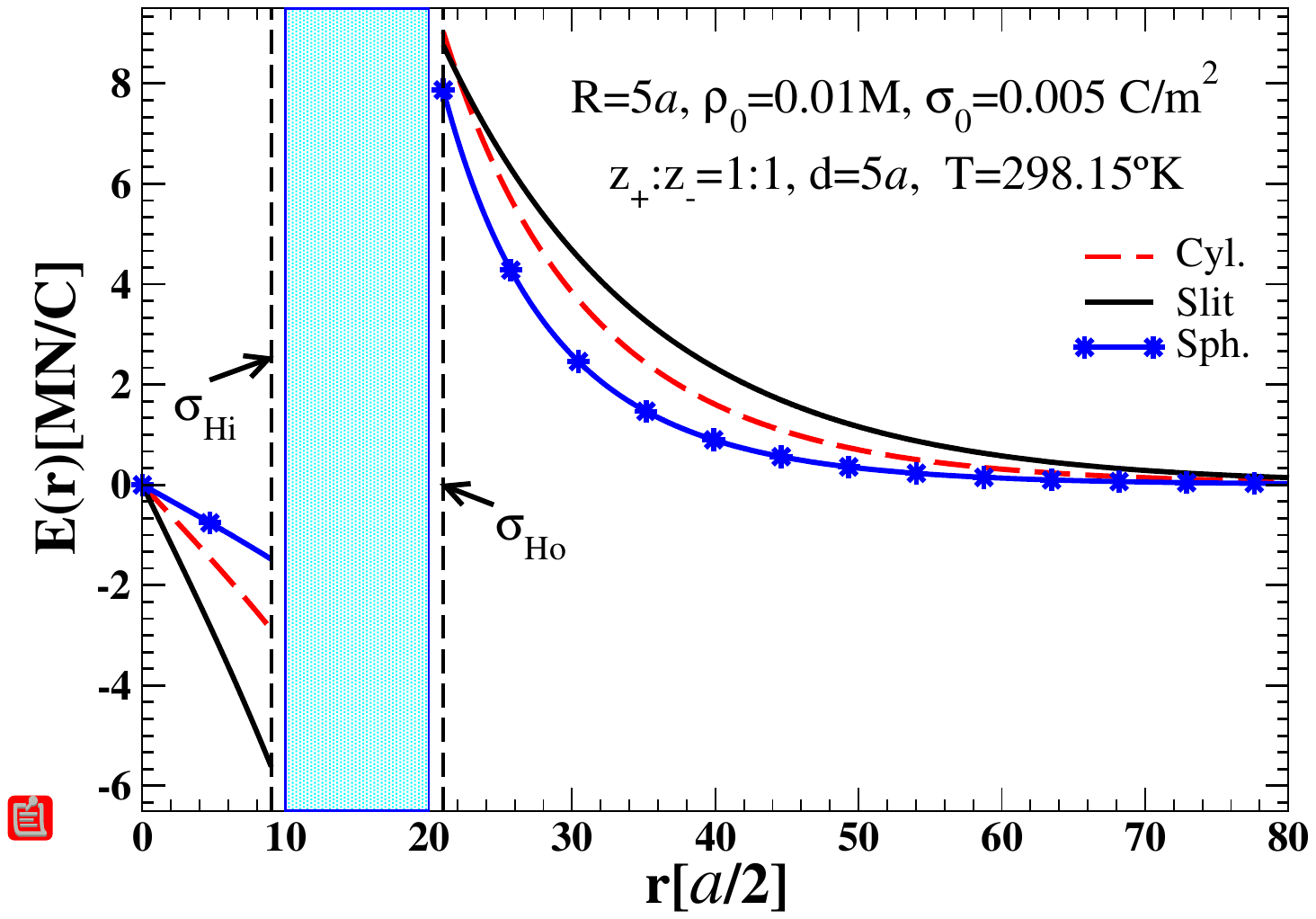}
		\caption{Electric field profile, $E(r)$, for planar,\\ cylindrical, and spherical hollow nanoparticles\\ with thicker walls $d = 5a$, immersed in a $1{:}1$\\ electrolyte.}
		\label{Fig.E(r)_R5_d5a_rho0p01_s0p005_z1}
	\end{subfigure}
	\begin{subfigure}{.5\textwidth}
		\centering
		\includegraphics[width=1.1\linewidth]{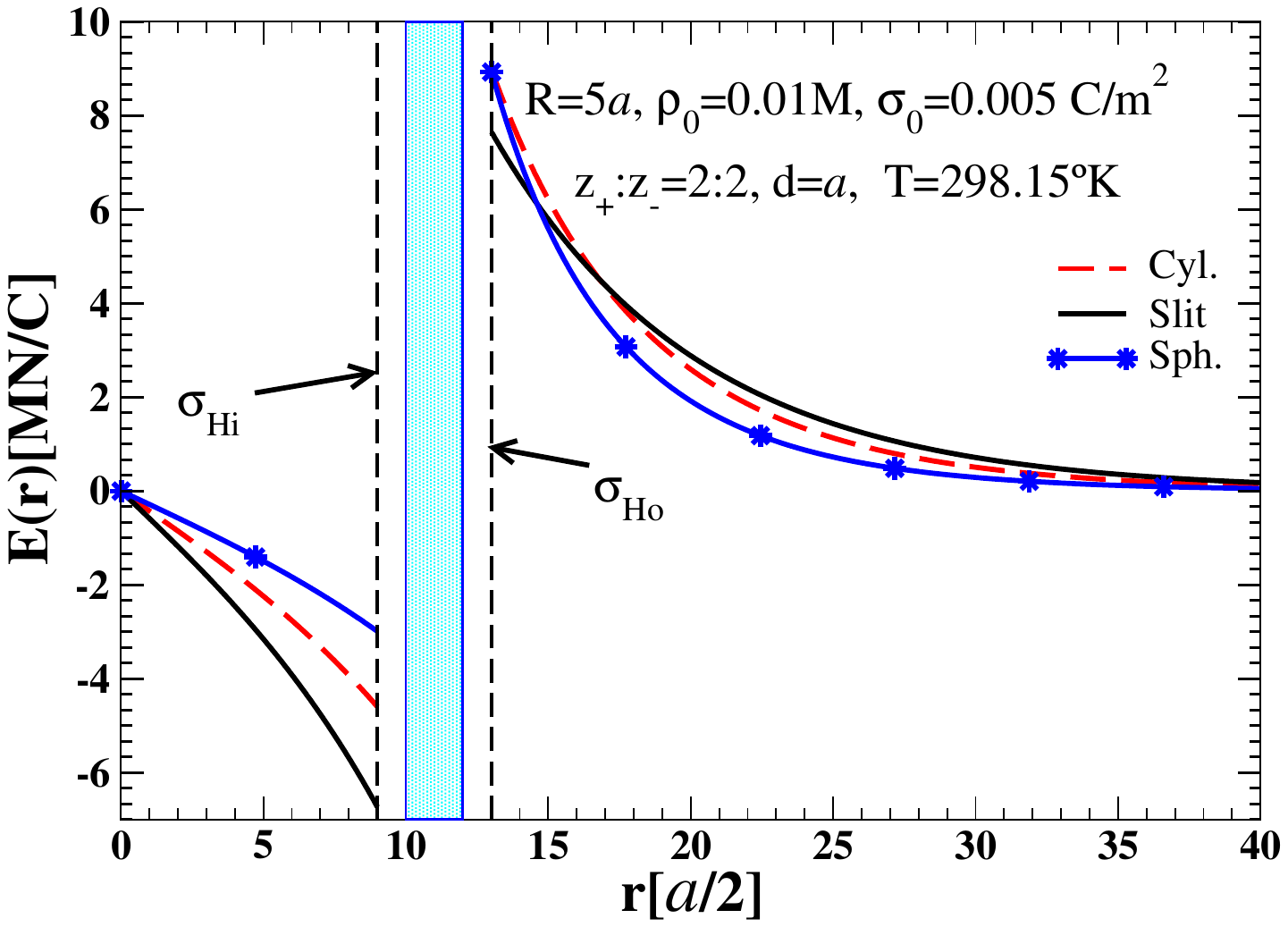}
		\caption{Electric field profile, $E(r)$, for planar,\\ cylindrical, and spherical hollow nanoparticles,\\ immersed in a $2{:}2$ electrolyte.}
		\label{Fig.E(r)_R5_d1a_rho0p01_s0p005_z2}
	\end{subfigure}
	\caption{Electric field profiles \( E(r) \) for planar, cylindrical, and spherical hollow nanoparticles with inner radius \( R \) and wall thickness \( d \), immersed in a symmetric 1:1 or 2:2 electrolyte of bulk concentration \( \rho_{\scriptscriptstyle{0}} = 0.01\,\mathrm{M} \). The surface charge density on both inner and outer walls is fixed at \( \sigma_{\scriptscriptstyle{0}} = 0.005\,\mathrm{C/m^2} \). 
		Figure~\ref{Fig.E(r)_R5a_d1a_rho0p01_s0p005_z1} shows the case of \( R = 5a \), \( d = a \), and a 1:1 electrolyte. Figures~\ref{Fig.E(r)_R10a_d1a_rho0p01_s0p005_z1}, \ref{Fig.E(r)_R5_d5a_rho0p01_s0p005_z1}, and~\ref{Fig.E(r)_R5_d1a_rho0p01_s0p005_z2} explore the effects of increasing the cavity radius, wall thickness, and salt valence \( z = 2 \), respectively.}
	\label{Fig.E(r)_all}
\end{figure}

In \cref{Fig.E(r)_all}, the electric field profile, \( E(r) \), inside and outside the hollow nanoparticles is shown for various model parameters. Since \( E(r) = \sigma(r) / (\varepsilon_{\scriptscriptstyle{0}}\varepsilon) \), its values at \( r = R - a/2 \) and \( r = R + d + a/2 \) determine the induced surface charge densities on the inner and outer shell surfaces, \( \sigma_{\scriptscriptstyle{Hi}} \) and \( \sigma_{\scriptscriptstyle{Ho}} \), respectively. These values also provide a measure of the thickness of the external electric double layers (EDLs). The internal EDLs, in contrast, are additionally modulated by the cavity size \( R \), due to the geometric confinement of the electrolyte.

\Cref{Fig.E(r)_R5a_d1a_rho0p01_s0p005_z1,Fig.E(r)_R10a_d1a_rho0p01_s0p005_z1} illustrate the effect of increasing the cavity radius, from \( R = 5a \) to \( R = 10a \), on \( E(r) \) for slit, cylindrical, and spherical geometries. Comparison of \cref{Fig.E(r)_R5a_d1a_rho0p01_s0p005_z1} with \cref{Fig.E(r)_R5_d5a_rho0p01_s0p005_z1,Fig.E(r)_R5_d1a_rho0p01_s0p005_z2} reveals the influence of shell thickness \( d \) and electrolyte valence. In all cases, \( \rho_{\scriptscriptstyle{0}} = 0.01\,\mathrm{M} \) and \( \sigma_{\scriptscriptstyle{0}} = 0.005\,\mathrm{C/m^2} \).

Within the cavity, \( E(r) \) decreases monotonically and reaches \( E(R - a/2) = \sigma_{\scriptscriptstyle{Hi}} / (\varepsilon_{\scriptscriptstyle{0}}\varepsilon) \), as expected from electrostatics (see \cref{Induced-charge-density-in}). For both radii, the field is strongest in the planar geometry, followed by the cylindrical, and weakest in the spherical geometry. This ordering arises purely from geometry: for a fixed \( \sigma_{\scriptscriptstyle{0}} \), the effective area is smallest for the planar case and largest for the spherical one. As \( R \to \infty \), all geometries approach the same limiting value \( \sigma_{\scriptscriptstyle{0}} \) (see \cref{Induced-charge-density-in}).

Outside the shell, \( E(r) \) decays most rapidly for the spherical geometry and most slowly for the slit, consistent with the geometry-dependent decay of the Coulomb potential. This trend is observed in all cases shown in \cref{Fig.E(r)_all}.

One might expect that a stronger internal induced field, \( E(R - a/2) \), would correspond to a weaker external field, \( E(R + d + a/2) \). However, in \cref{Fig.E(r)_R5a_d1a_rho0p01_s0p005_z1}, the cylindrical shell shows a larger \( E(R + d + a/2) \) than the spherical one, while in \cref{Fig.E(r)_R10a_d1a_rho0p01_s0p005_z1} this relation is reversed. These differences reflect the nonlinear dependence of \( E(r) \) on \( r \), as well as variations in the total surface charge \( Q_{\scriptscriptstyle{0}} = A_{\scriptscriptstyle{\gamma}} \sigma_{\scriptscriptstyle{0}} \), which depends on geometry, electrolyte concentration, and valence.

Note that the bare electric field produced by the imposed surface charge \( \sigma_{\scriptscriptstyle{0}} \) at the inner and outer shell walls is given by \( E(R) = E(R+d) = \sigma_{\scriptscriptstyle{0}} / (\varepsilon_{\scriptscriptstyle{0}}\varepsilon) = 5.6472\,\mathrm{MN/C} \). In all cases shown in \cref{Fig.E(r)_all}, the actual field satisfies \( E(R + d + a/2) > E(R + d) \), up to a certain cavity radius \( R \), depending on the system parameters. However, for some regimes, it is also observed that \( E(R - a/2) < E(R) \). These two phenomena correspond to \textit{Confinement Overcharging} (CO), when \( \sigma_{\scriptscriptstyle{Ho}} > \sigma_{\scriptscriptstyle{0}} \), and \textit{Confinement Charge Reversal} (CCR), when \( \sigma_{\scriptscriptstyle{Hi}} < \sigma_{\scriptscriptstyle{0}} \)~\cite{Lozada-Cassou_JML-2025}. In the slit geometry, CO is always present and CCR is never observed. In contrast, for cylindrical and spherical geometries, both CO and CCR may arise depending on the electrolyte parameters and shell dimensions.

CO and CCR originate from electrostatic correlations across the nanoparticle shell~\cite{Lozada-Cassou-PRE1997} and confinement effects, which lead to a violation of the local electroneutrality condition (see \cref{Eq.Violation_Local_Elec_Cond}). However, the global electroneutrality is still satisfied by the joint structure of the internal and external EDLs (see \cref{Electrical-field-balance,Electroneutrality-condition-general2}). Historically, overcharging, charge reversal, and charge inversion in inhomogeneous charged fluids and nanostructures were attributed to excluded-volume effects or configurational entropy~\cite{Attard_1996,Kjellander-charge-inversion-1998,Keshavarzi_2020,Keshavarzi_2022,Gonzalez-Calderon-EPJ-2021,Gonzalez-overcharging-cyl-macroions-2022,Boda-confinement-2024}.

Beyond their potential relevance in biological or soft-matter systems, these effects are fundamentally important for understanding the electrostatic energy–entropy balance in confined electrolytes. Notably, CO and CCR occur in the low-concentration regime, where configurational entropy is absent—except for the minimal excluded volume accounted for by the Stern correction. Their emergence, therefore, cannot be attributed to volume exclusion effects. In fact, they are enhanced as the ionic size tends to zero~\cite{Lozada-Cassou_JML-2025}, reinforcing their purely electrostatic origin. 

Moreover, while it is widely accepted in the literature that PB theory cannot account for such phenomena~\cite{Attard_1996,Kjellander-charge-inversion-1998}, we confirm that CO and CCR do not appear in planar electrolyte–electrode interfaces, but emerge solely as a consequence of confinement and curvature. Although these effects are predicted within the LPB approximation, their qualitative features are supported by more advanced approaches—such as density functional theory, integral equation methods, modified PB models, and computer simulations—in the low-concentration regime, as discussed in \cref{introduction}. A more detailed analysis of these phenomena lies beyond the scope of this work.

According to the capacitance expressions for the three geometries (see \cref{Hollow-nanoparticles-capacitance}), the capacitance depends solely on geometric factors and the Debye screening length \( \lambda_{\scriptscriptstyle{D}} \), which characterizes the EDL thickness. Thinner EDLs (i.e., shorter \( \lambda_{\scriptscriptstyle{D}} \) or larger \( \kappa \)) correspond to higher capacitance values. As shown in \cref{Fig.E(r)_all}, increasing either the cavity radius \( R \) or the wall thickness \( d \) leads to thicker EDLs, while increasing the electrolyte valence yields thinner EDLs. These trends are consistent with the capacitance behavior depicted in \cref{Cd_vs_rho_all}. Although not shown here due to space constraints, we have also verified that increasing the electrolyte concentration reduces the EDL thickness, in agreement with the trends observed in \cref{Cd_vs_R_all_bulk_concentrations}.
\begin{figure}[!hbt]
	\begin{subfigure}{.5\textwidth}
		\centering
		\includegraphics[width=1.1\linewidth]{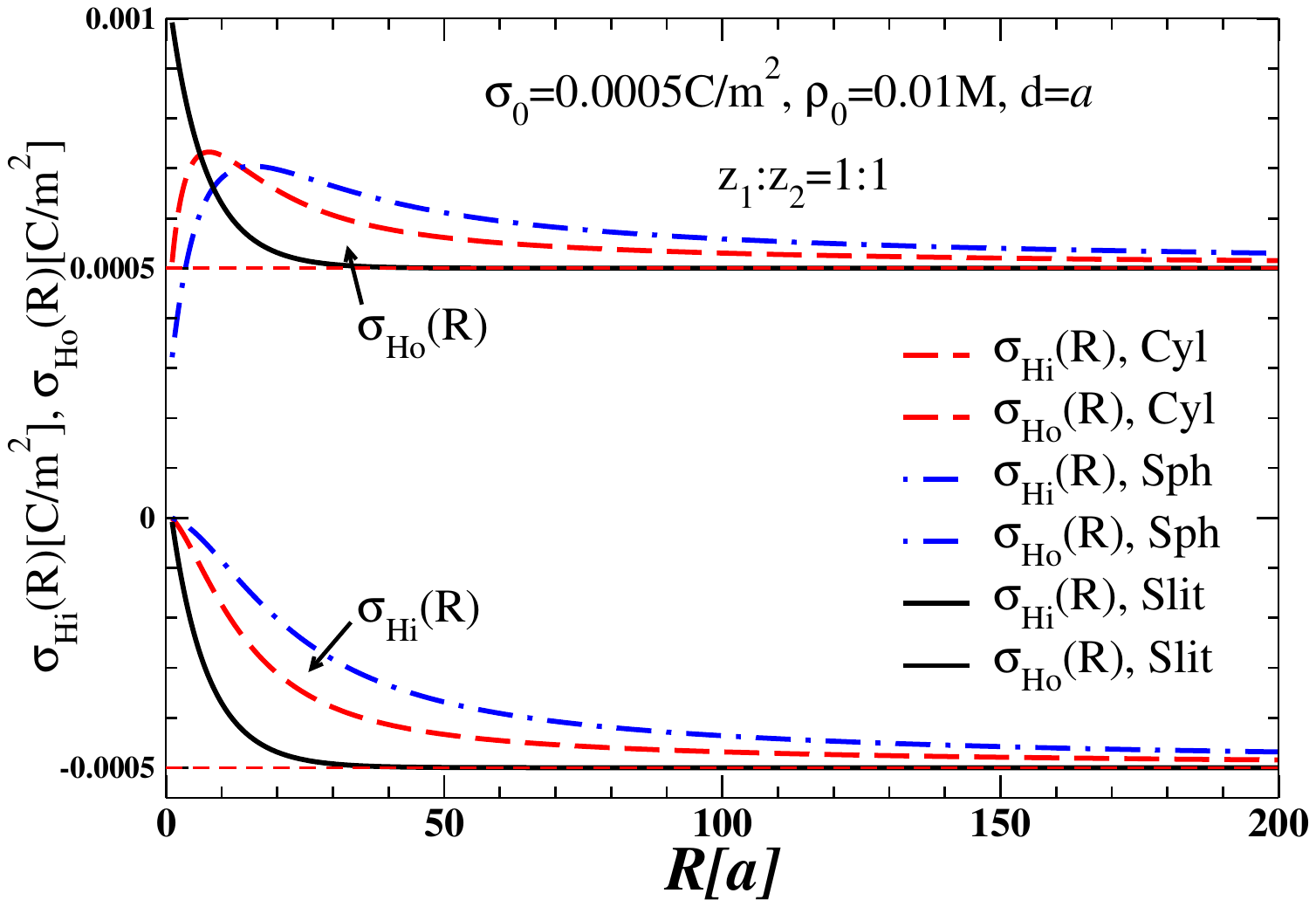}
		\caption{$\sigma_{\scriptscriptstyle{Hi}}(R)$ and $\sigma_{\scriptscriptstyle{Ho}}(R)$, at very low $\sigma_{\scriptscriptstyle{0}}$.}
		\label{Fig.sigma_Hi-sigmaHo-Cyl-Sphere-s0.0005_rho0.01}
	\end{subfigure}
	\begin{subfigure}{.5\textwidth}
		\centering
		\includegraphics[width=1.1\linewidth]{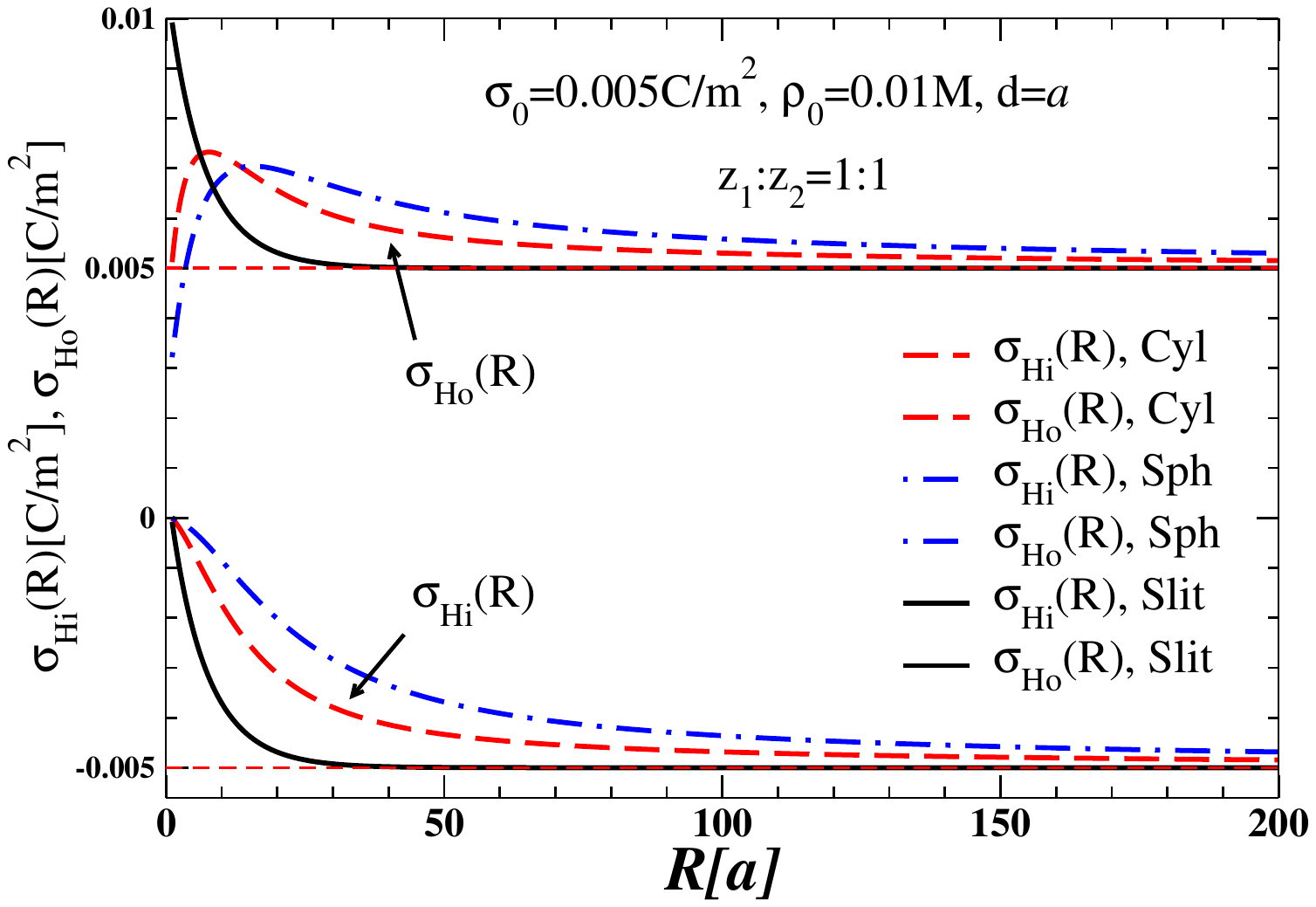}
		\caption{$\sigma_{\scriptscriptstyle{Hi}}(R)$ and $\sigma_{\scriptscriptstyle{Ho}}(R)$, at low $\sigma_{\scriptscriptstyle{0}}$.}
		\label{Fig.sigma_Hi-sigmaHo-Cyl-Sphere-s0.005_rho0.01}
	\end{subfigure}\\
	\begin{subfigure}{.5\textwidth}
		\centering
		\includegraphics[width=1.1\linewidth]{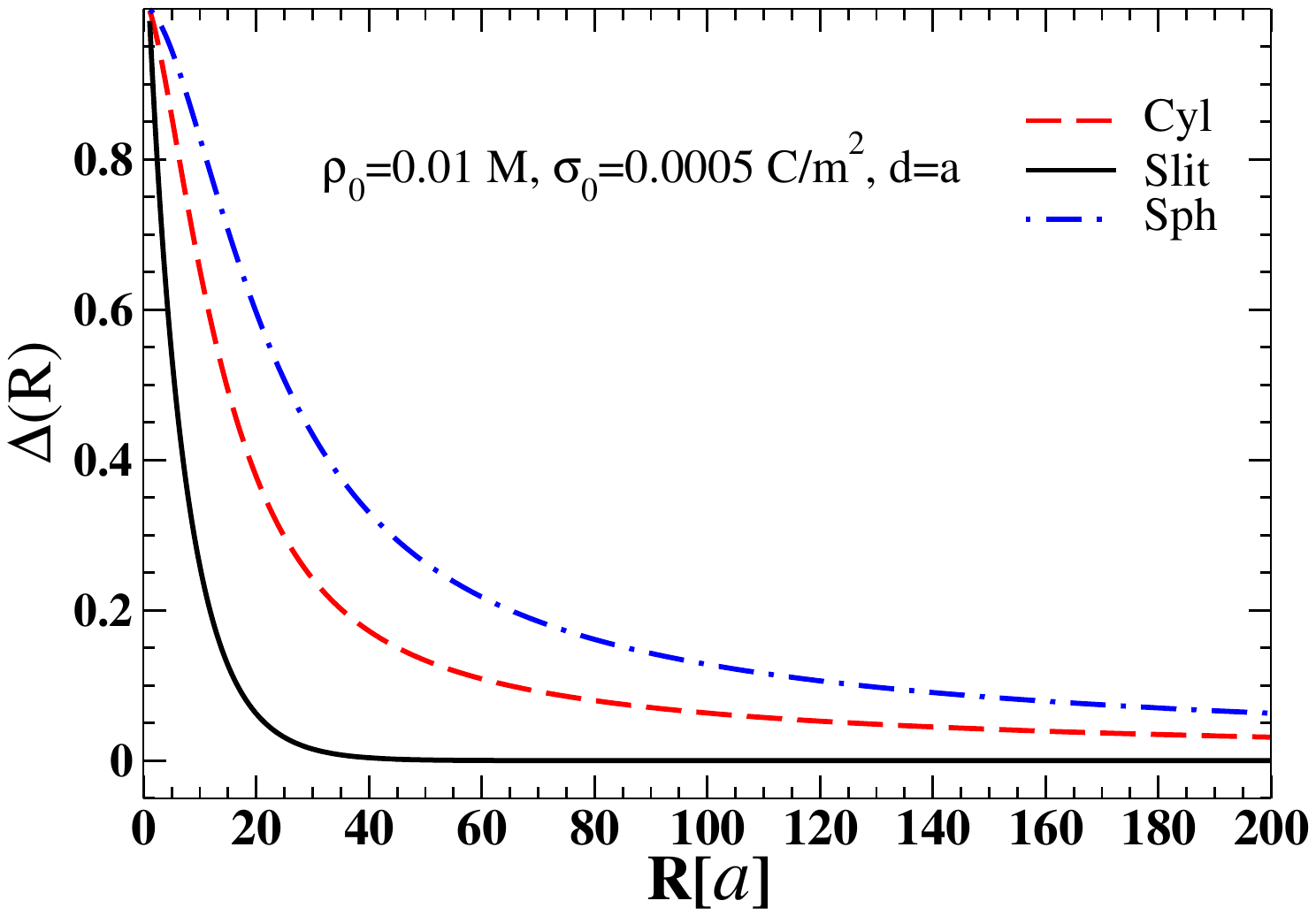}
		\caption{Electroneutrality deviation ratio as a function\\ of $R$, for very low $\sigma_{\scriptscriptstyle{0}}$.}
		\label{Delta-Sigma_Hi[R]-Sigma_0_z1_d1p0_s0p0005_rho0p01}
	\end{subfigure}
	\begin{subfigure}{.5\textwidth}
		\centering
		\includegraphics[width=1.1\linewidth]{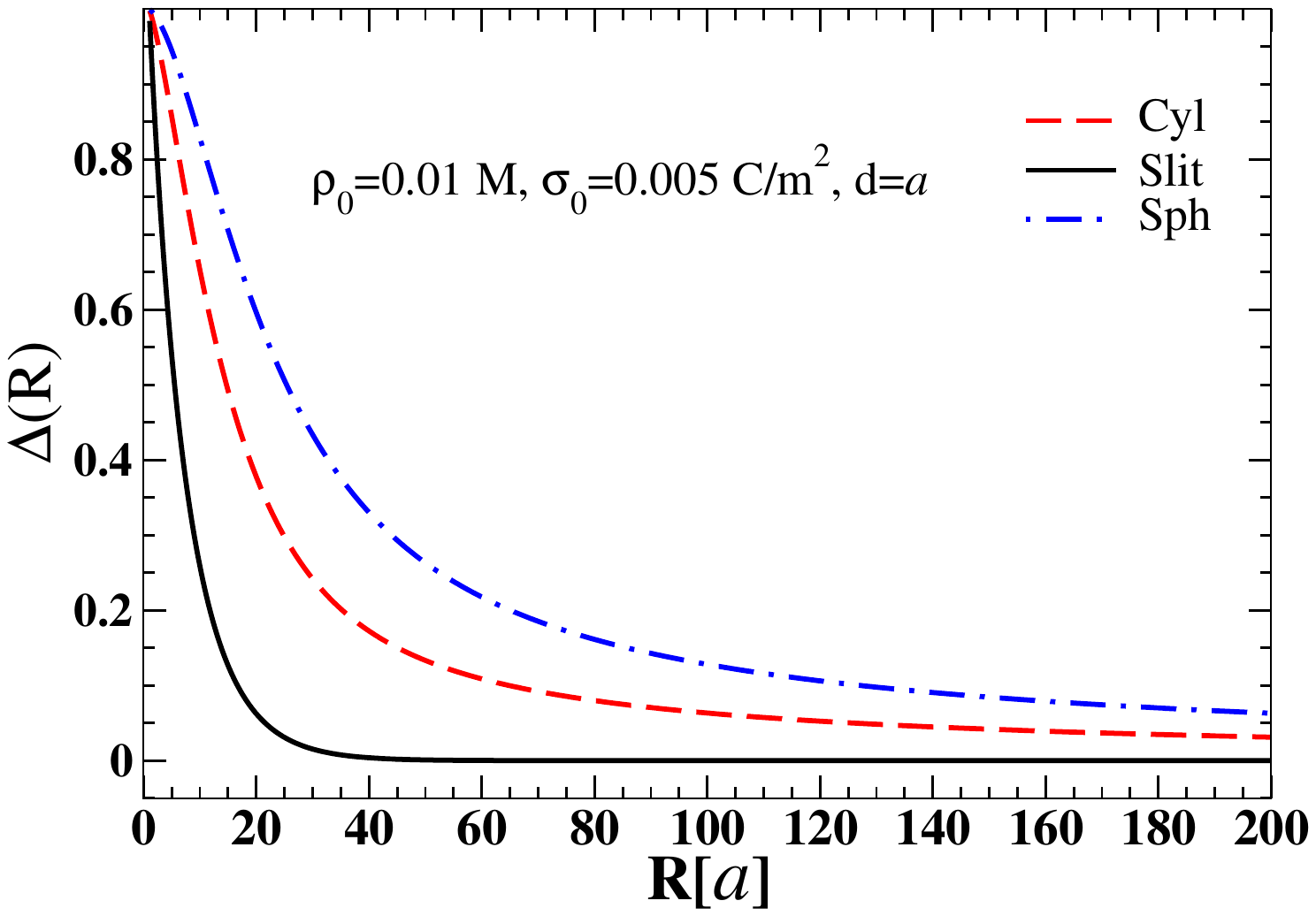}
		\caption{Electroneutrality deviation ratio as a function\\ of $R$, for higher $\sigma_{\scriptscriptstyle{0}}$.}
		\label{Delta-Sigma_Hi[R]-Sigma_0_z1_d1p0_s0p005_rho0p01}
	\end{subfigure}\\
	\caption{Violation of the local electroneutrality condition (VLEC) for hollow nanoparticles. In 
		\cref{Fig.sigma_Hi-sigmaHo-Cyl-Sphere-s0.0005_rho0.01,Fig.sigma_Hi-sigmaHo-Cyl-Sphere-s0.005_rho0.01}, 
		the effective surface charge densities on the inner and outer walls, \( \sigma_{\scriptscriptstyle{Hi}}(R) \) and \( \sigma_{\scriptscriptstyle{Ho}}(R) \), are plotted as functions of the cavity radius \( R \) for planar, cylindrical, and spherical geometries. Dashed horizontal lines mark the imposed values \( \sigma_{\scriptscriptstyle{Ho}} = \sigma_{\scriptscriptstyle{0}} \) and \( \sigma_{\scriptscriptstyle{Hi}} = -\sigma_{\scriptscriptstyle{0}} \). In \cref{Delta-Sigma_Hi[R]-Sigma_0_z1_d1p0_s0p0005_rho0p01,Delta-Sigma_Hi[R]-Sigma_0_z1_d1p0_s0p005_rho0p01}, the electroneutrality deviation ratio \( \Delta \) is plotted versus \( R \) for two different surface charge densities \( \sigma_{\scriptscriptstyle{0}} \).}
	\label{Fig.sigma-Hi_and_Ho_Delta}
\end{figure}

While the EDL thickness is influenced by the surface charge density \( \sigma_{\scriptscriptstyle{0}} \), the capacitance remains unaffected. Why, then, is the capacitance independent of \( \sigma_{\scriptscriptstyle{0}} \), despite its impact on the EDL structure? \Cref{Fig.sigma_Hi-sigmaHo-Cyl-Sphere-s0.0005_rho0.01,Fig.sigma_Hi-sigmaHo-Cyl-Sphere-s0.005_rho0.01} display the induced surface charge densities \( \sigma_{\scriptscriptstyle{Hi}} \) and \( \sigma_{\scriptscriptstyle{Ho}} \) as functions of \( R \) for two values of \( \sigma_{\scriptscriptstyle{0}} \). As expected, \( \sigma_{\scriptscriptstyle{Hi}} \) and \( \sigma_{\scriptscriptstyle{Ho}} \) scale proportionally with their corresponding electric fields at \( R - a/2 \) and \( R + d + a/2 \), and higher values of \( \sigma_{\scriptscriptstyle{0}} \) lead to thinner external EDLs. Nevertheless, the capacitance remains unchanged.


\noindent\textbf{Charge symmetry-breaking.}

As shown in \cref{Fig.sigma_Hi-sigmaHo-Cyl-Sphere-s0.0005_rho0.01,Fig.sigma_Hi-sigmaHo-Cyl-Sphere-s0.005_rho0.01}, the induced surface charge outside the shell, \( \sigma_{\scriptscriptstyle{Ho}} \), increases less steeply with \( R \) in spherical geometries compared to planar and cylindrical ones. This reinforces the asymmetry in the electric field response across the shell. Notably, despite these local asymmetries in the EDL structure, the overall electrostatic response remains symmetric under variations in \( \sigma_{\scriptscriptstyle{0}} \), why? We refer to this phenomenon as \emph{charge symmetry-breaking}.

The electrolyte inside and outside the cavity remains at the same chemical potential, as both regions are electrostatically correlated through the shell~\cite{Lozada-Cassou-PRE1997}. However, as discussed in \cref{electroneutrality condition}, local electroneutrality is violated when \( \sigma_{\scriptscriptstyle{Hi}} + \sigma_{\scriptscriptstyle{0}} \neq 0 \). To quantify this deviation, we define the \emph{electroneutrality deviation ratio} (EDR), \( \Delta \), as
\begin{equation}
	\Delta = \frac{\sigma_{\scriptscriptstyle{Hi}} - \sigma_{\scriptscriptstyle{0}}}{\sigma_{\scriptscriptstyle{0}}}.
\end{equation}

In \cref{Delta-Sigma_Hi[R]-Sigma_0_z1_d1p0_s0p0005_rho0p01,Delta-Sigma_Hi[R]-Sigma_0_z1_d1p0_s0p005_rho0p01}, we plot \( \Delta \) as a function of the cavity radius \( R \) for two different values of \( \sigma_{\scriptscriptstyle{0}} \). Remarkably, the resulting curves coincide, indicating that despite significant differences in the internal and external EDL profiles, the degree of electroneutrality violation~\cite{Lozada-Cassou-Yamada-1988,Lozada_1984,Lozada1996} depends solely on confinement geometry, and not on the absolute value of the surface charge.

We conclude, therefore, that although the capacitance depends only on bulk properties and geometry, the electrostatic profiles exhibit an implicit, nontrivial asymmetry with respect to the surface charge. This \emph{charge symmetry-breaking} arises from the combined influence of curvature, confinement, and shell thickness. That is, even when the surface charge densities on the inner and outer walls are equal in magnitude, the resulting electrostatic response is intrinsically asymmetric. This is a geometric and thermodynamic effect rather than one arising from compositional asymmetry or fluctuations. It is conceptually reminiscent of \emph{spontaneous symmetry breaking} in soft-matter systems~\cite[Ch.~6]{chaikin1995principles}, where symmetric governing equations produce asymmetric outcomes due to boundary constraints or geometric frustration. Similar effects were insightfully analyzed by Podgornik and coworkers~\cite{podgornik1989electrostatic,podgornik1998charge}, who showed that symmetric electrostatic boundary conditions can nonetheless yield asymmetric fields and forces when ionic correlations and confinement are properly accounted for. In our system, this asymmetry arises within linearized Poisson–Boltzmann theory purely from geometry and dielectric continuity, highlighting the subtle role of spatial constraints in driving effective symmetry breaking—even in weak-coupling regimes.

\noindent\textbf{Chemical potential symmetry-breaking.}

An equally striking and conceptually distinct phenomenon arises when comparing electrolytes of different ionic valence and concentration but identical Debye length, \( \lambda_{\scriptscriptstyle{D}} \). While the capacitance of hollow nanoparticle systems is strongly dependent on the Debye screening length—determined by the electrolyte’s concentration, temperature, and valence—we find that different electrolytes sharing the same \( \lambda_{\scriptscriptstyle{D}} \) lead to indistinguishable macroscopic electrostatic responses. Moreover, systems as distinct as a 1:1 electrolyte at high concentration and a 2:2 electrolyte at lower concentration, with the same \( \lambda_{\scriptscriptstyle{D}} \), exhibit identical reduced mean electrostatic potentials \( ez\beta \psi(r) \) and induced charge density profiles \( \rho_{\scriptscriptstyle{\mathrm{el}}}(r) \), and thus also share indistinguishable microscopic electrostatic responses.

\begin{figure}[!hbt]
	\begin{subfigure}{.5\textwidth}
		\centering
		\includegraphics[width=1.1\linewidth]{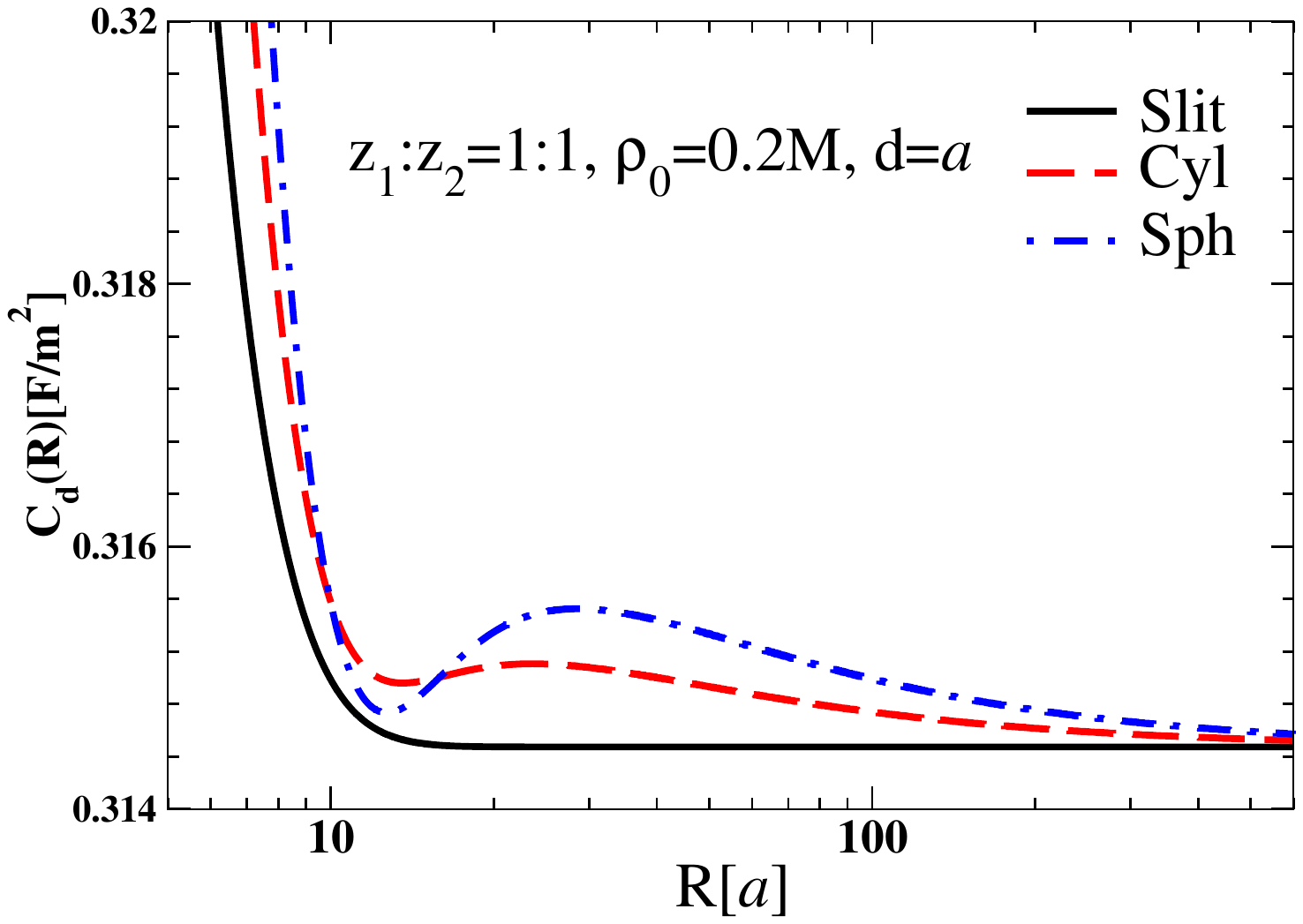}
		\caption{Capacitance: $1{:}1$, $0.2$\,M electrolyte.}
		\label{Fig.Cd_vs_R_rho0p2_d1_z1}
	\end{subfigure}
	\begin{subfigure}{.5\textwidth}
		\centering
		\includegraphics[width=1.1\linewidth]{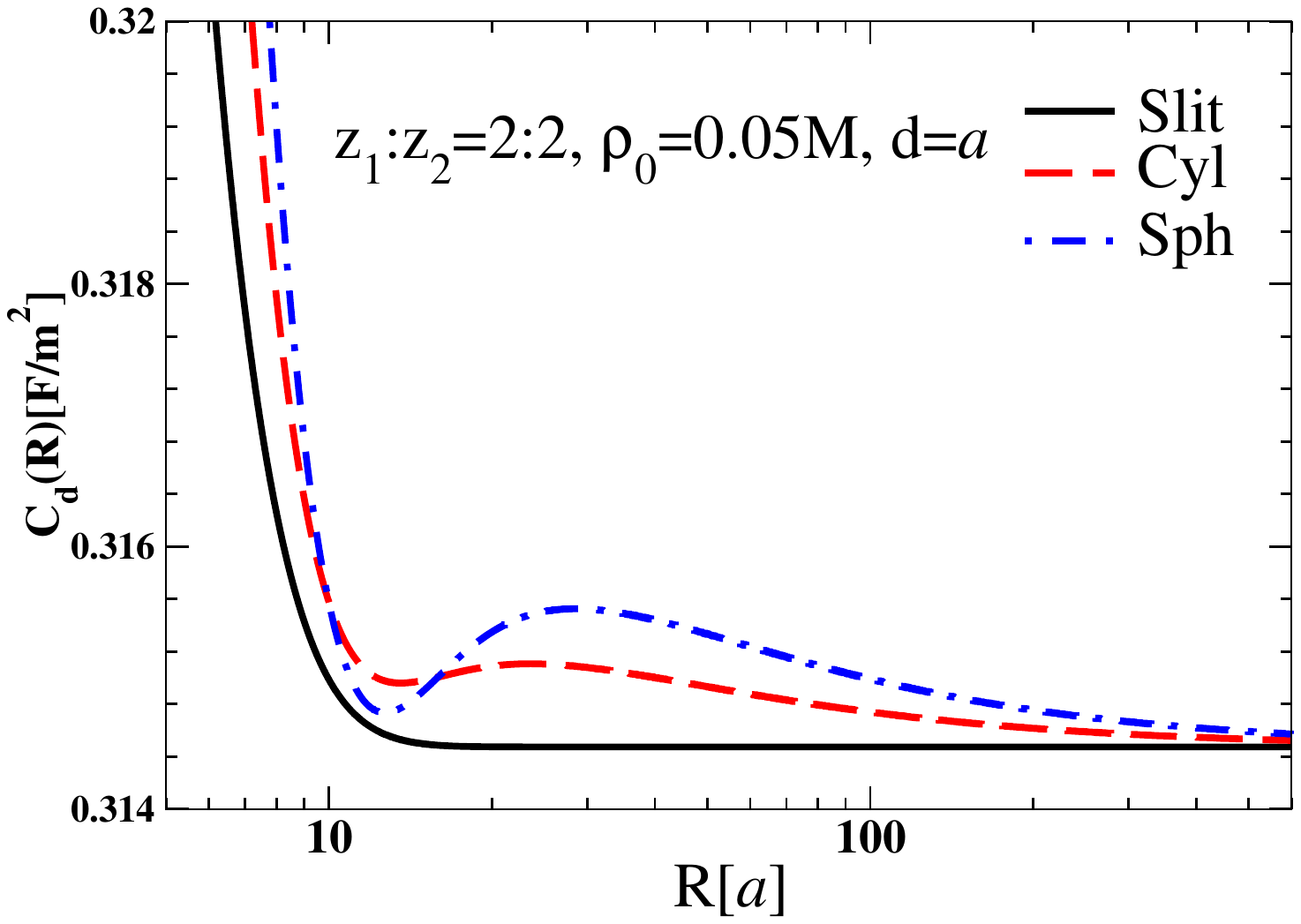}
		\caption{Capacitance: $2{:}2$, $0.05$\,M electrolyte.}
		\label{Fig.Cd_vs_R_rho0p05_d1_z2}
	\end{subfigure}\\
	\begin{subfigure}{.5\textwidth}
		\centering
		\includegraphics[width=1.1\linewidth]{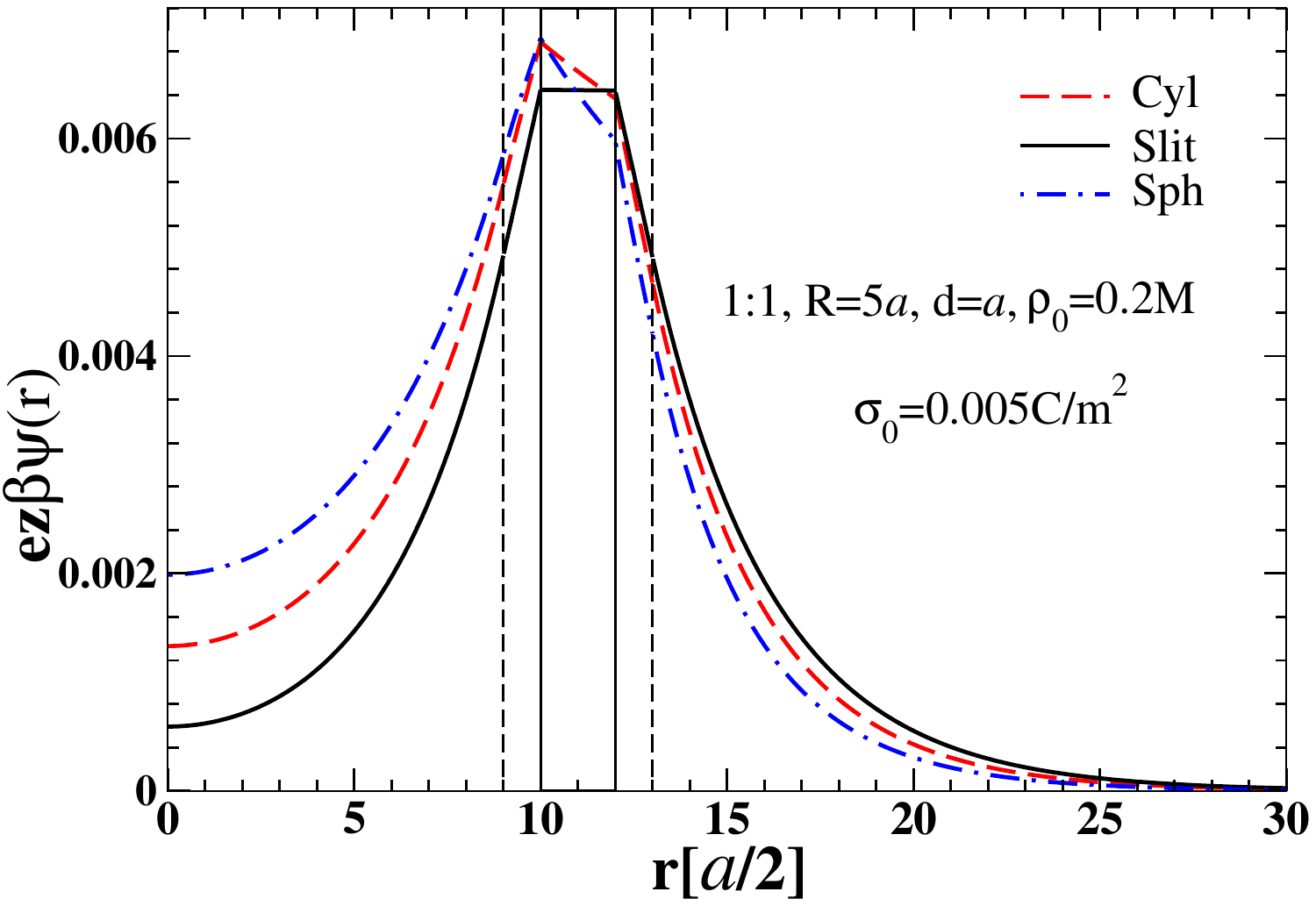}
		\caption{Reduced MEP: $1{:}1$, $0.2$\,M.}
		\label{Fig.psi[r]_R5p0_z1_d1_s_0p005_rho0p2}
	\end{subfigure}
	\begin{subfigure}{.5\textwidth}
		\centering
		\includegraphics[width=1.1\linewidth]{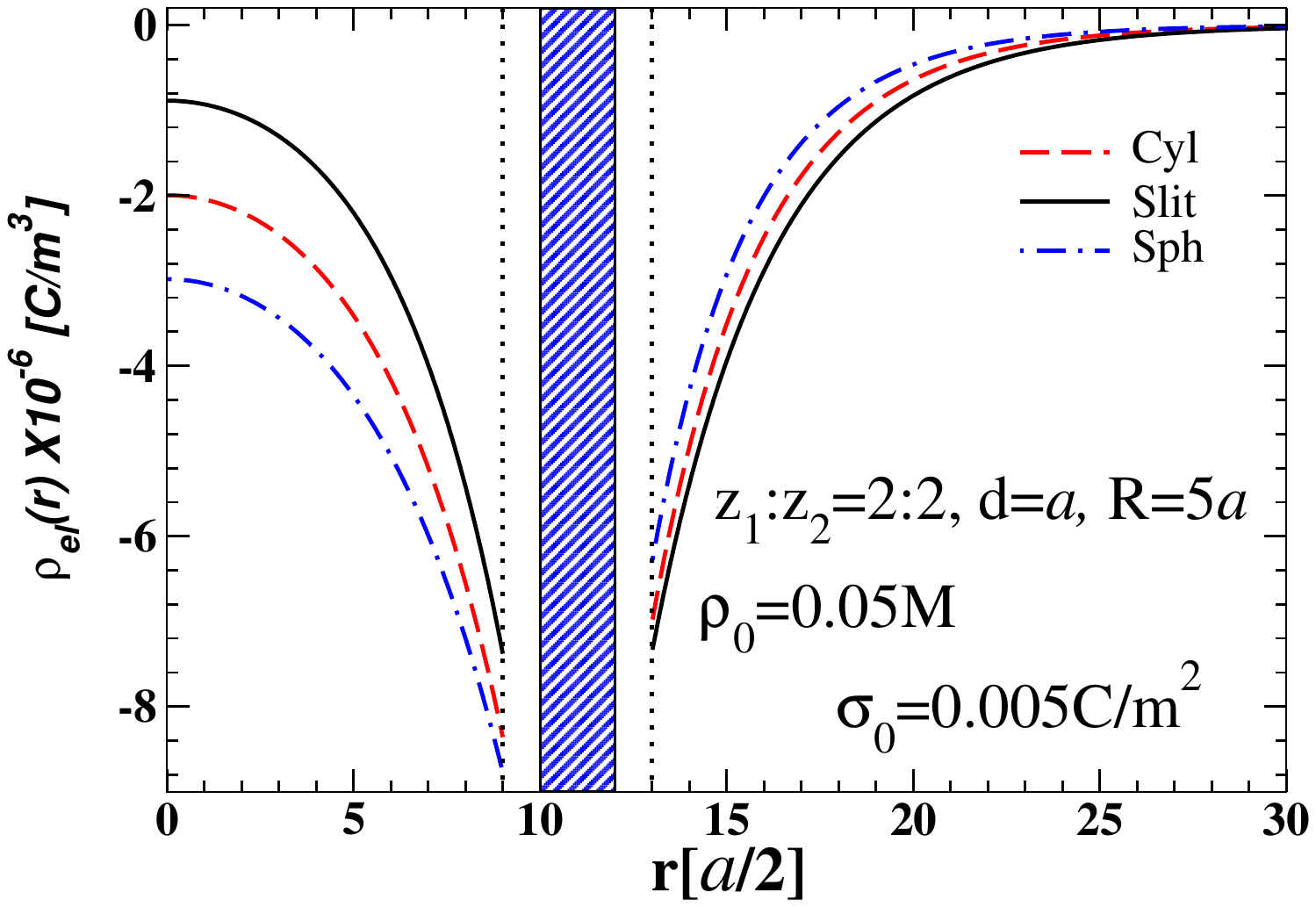}
		\caption{Charge density: $2{:}2$, $0.05$\,M.}
		\label{PCS_rhoel[r]_R5p0_z2_d1p0_s0p005_rho0p05}
	\end{subfigure}
	\caption{Capacitance and electrostatic field observables are identical for 1:1 and 2:2 electrolytes with equal Debye length \( \lambda_{\scriptscriptstyle{D}} \), despite differences in ion valence and bulk concentration.}
	\label{Fig.Chem_Pot_1}
\end{figure}

In \cref{Fig.Cd_vs_R_rho0p2_d1_z1,Fig.Cd_vs_R_rho0p05_d1_z2}, the specific capacitance for hollow nanoparticles immersed in a 1:1 electrolyte at \( \rho_0 = 0.2\,\mathrm{M} \) and a 2:2 electrolyte at \( \rho_0 = 0.05\,\mathrm{M} \) is shown. These salts have the same Debye screening length \( \lambda_{\scriptscriptstyle{D}} \), and hence exhibit identical capacitance. The same applies to their reduced mean electrostatic potential (MEP) and charge density profiles (\cref{Fig.psi[r]_R5p0_z1_d1_s_0p005_rho0p2,PCS_rhoel[r]_R5p0_z2_d1p0_s0p005_rho0p05}).

\begin{figure}[!htb]
	\begin{subfigure}{.5\textwidth}
		\centering
		\includegraphics[width=0.99\linewidth]{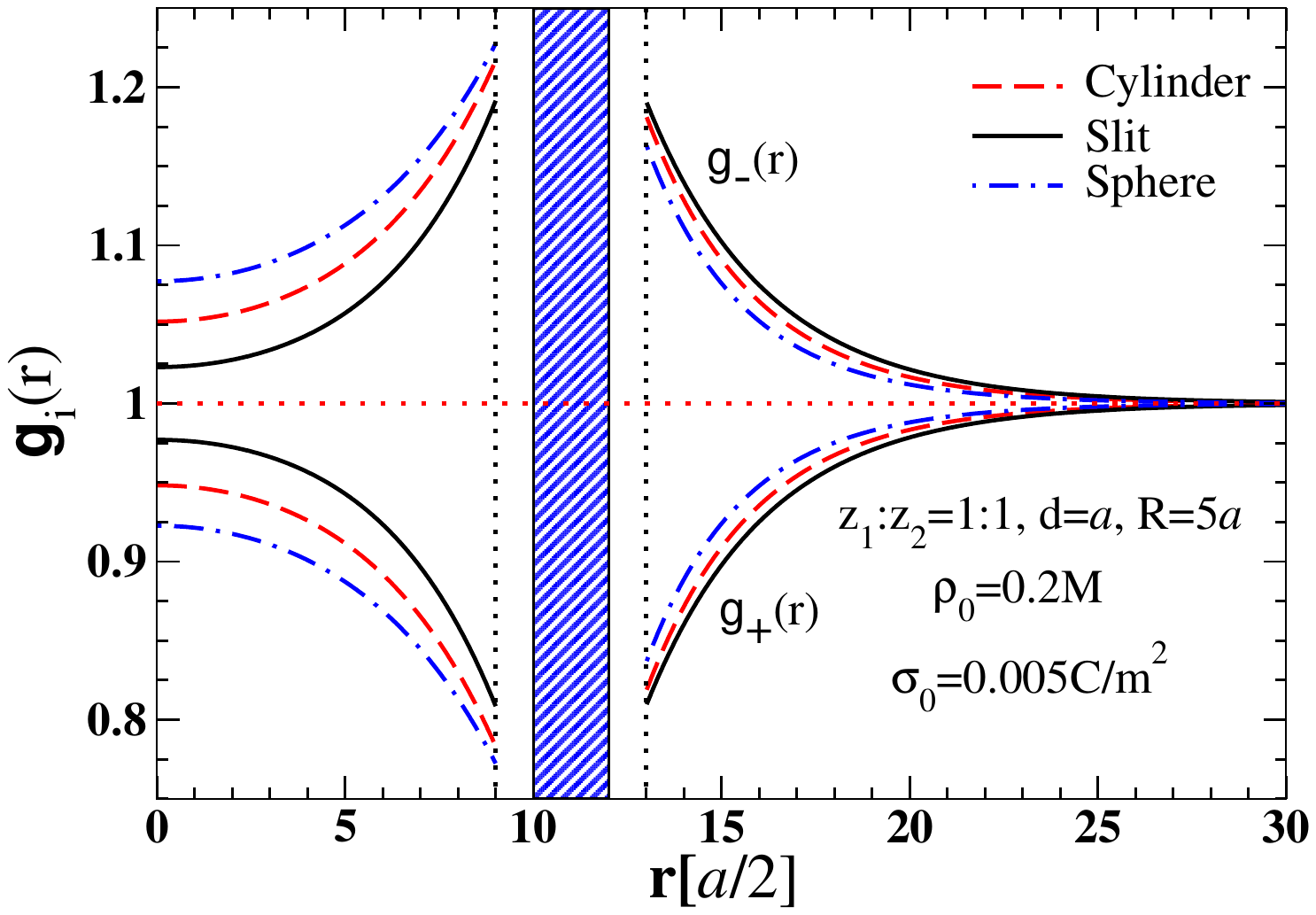}
		\caption{Reduced $g_i(r)$, $1{:}1$, $0.2$\,M.}
		\label{Fig.g[r]_R5p0_z1_d1_s0p005_rho0p2}
	\end{subfigure}
	\begin{subfigure}{.5\textwidth}
		\centering
		\includegraphics[width=0.99\linewidth]{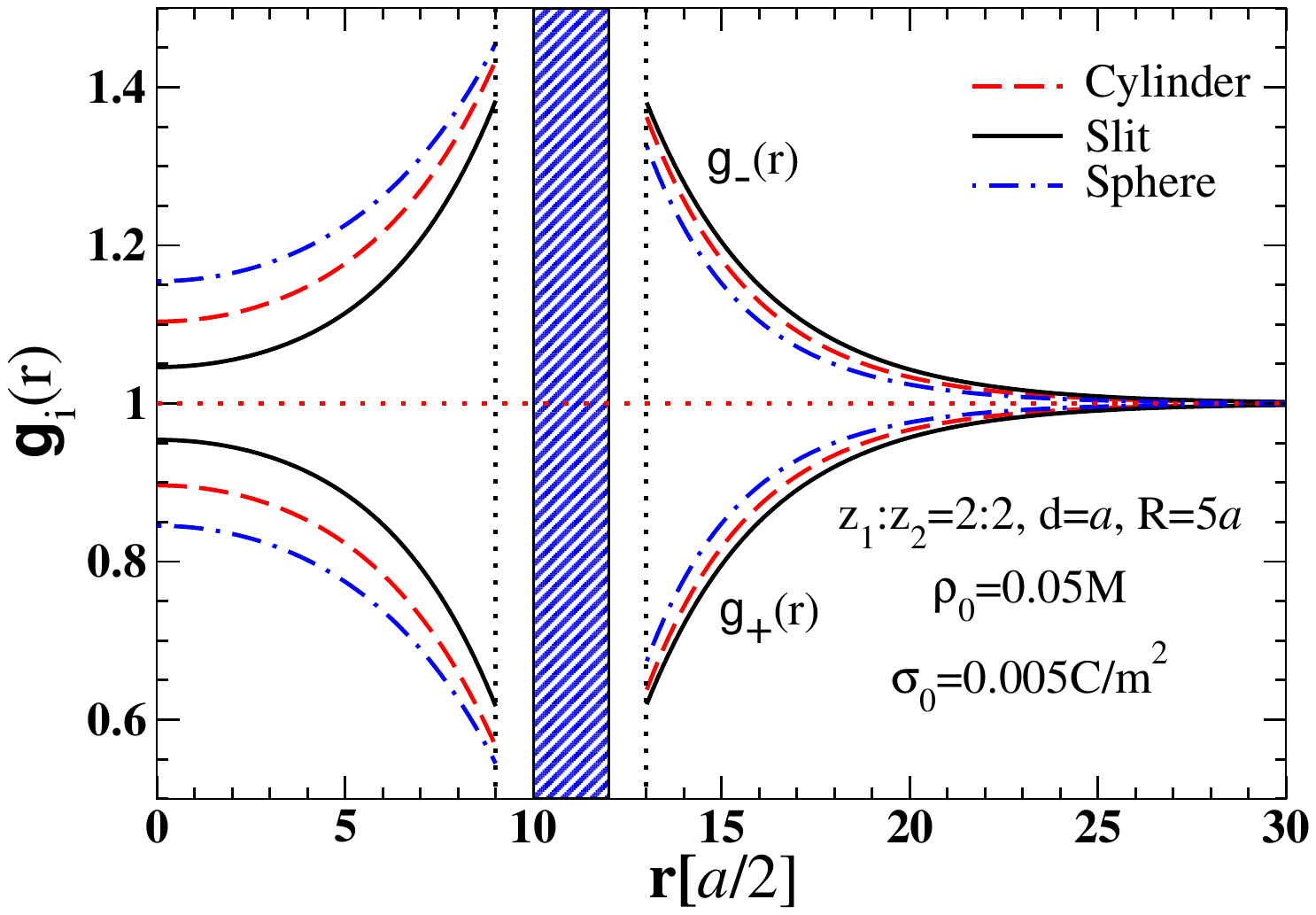}
		\caption{Reduced $g_i(r)$, $2{:}2$, $0.05$\,M.}
		\label{Fig.g[r]_R5p0_z2_d1_s0p005_rho0p05}
	\end{subfigure}
	\caption{Reduced ionic density profiles \( g_i(r) \) for 1:1 and 2:2 electrolytes differ significantly, even though they yield identical electrostatic observables when their Debye lengths \( \lambda_{\scriptscriptstyle{D}} \) are equal.}
	\label{Fig.Chem_pot_g[r]}
\end{figure}

In an inhomogeneous electrolyte near charged interfaces, such as inside or outside a hollow nanoparticle, the ionic chemical potential becomes spatially dependent. For a symmetric \( z:z \) electrolyte:

\begin{equation}
	\mu_i(r) = \mu_i^{\text{bulk}} + kT \ln\left[ \frac{\rho_i(r)}{\rho_0} \right],
	\label{eq:mu-local}
\end{equation}

\noindent or equivalently, in the Poisson–Boltzmann approximation:

\begin{equation}
	\mu_i(r) = \mu_i^{\text{bulk}} - e z_i \psi(r).
	\label{eq:mu-local-PB}
\end{equation}

Although \( \mu_i(r) \) varies locally with potential, the total electrochemical potential remains pinned at the reservoir value under grand canonical conditions:

\begin{equation}
	\sum_i \mu_i(r) = \sum_i \mu_i^{\text{bulk}} = \mu^{\text{bulk}}.
	\label{eq:mu_local_sum}
\end{equation}

Thus, even as \( \psi(r) \), \( \rho_i(r) \), and \( C_d \) remain unchanged for fixed \( \lambda_{\scriptscriptstyle{D}} \), the local chemical potentials \( \mu_i(r) \) differ due to differing bulk values. This reveals a form of \emph{chemical potential symmetry-breaking}: macroscopic electrostatic observables remain symmetric, even as bulk thermodynamic quantities differ.

This result is particularly striking in light of the many biological and chemical processes highly sensitive to ion valence. Divalent salts such as MgSO\(_4\) often cause stronger binding, distinct adsorption, or biological activity than monovalent salts like NaCl. The inconsistency between this chemically distinct behavior and their identical capacitance and electrostatic profiles (at equal \( \lambda_{\scriptscriptstyle{D}} \)) becomes understandable upon inspection of their distinct concentration profiles, shown in \cref{Fig.Chem_pot_g[r]}.

As an example, within the Debye–Hückel framework for bulk electrolytes, \cref{tab:mu_bulk_final_Text} quantifies the chemical potential components of MgSO\(_4\) and NaCl (see Appendix for derivation):

\begin{table}[ht]
	\centering
	\caption{Components of the bulk chemical potential (in kJ/mol per salt molecule) for NaCl (1:1, 0.2\,M) and MgSO$_4$ (2:2, 0.05\,M) at $T = 298.15\,\mathrm{K}$ in water ($\varepsilon = 78.5$), assuming a common Debye length ($\kappa = 1.4689 \times 10^9\, \mathrm{m}^{-1}$).}
	\begin{tabular}{lcc}
		\toprule
		\textbf{Contribution} & \textbf{NaCl (1:1, 0.2\,M)} & \textbf{MgSO$_4$ (2:2, 0.05\,M)} \\
		\midrule
		Kinetic term \( \mu^{\text{kin}} \)        & 39.36 & 38.65 \\
		Mixing (entropy) \( \mu^{\text{mix}} \)    & $-7.98$ & $-14.86$ \\
		Electrostatic \( \mu^{\text{el}} \)        & $-23.26$ & $-93.00$ \\
		\midrule
		\textbf{Total \( \mu^{\text{bulk}} \)}     & \textbf{8.12} & \textbf{$-69.21$} \\
		\bottomrule
	\end{tabular}
	\label{tab:mu_bulk_final_Text}
\end{table}

Despite matching \( \lambda_{\scriptscriptstyle{D}} \), the total bulk chemical potentials differ by nearly 77\,kJ/mol. This discrepancy arises from the stronger electrostatic self-energy of divalent ions and their lower mixing entropy.

Our results thus reveal a fundamental decoupling: identical electrostatic responses may emerge from distinct ionic environments. Chemical potential symmetry-breaking is not a violation of equilibrium, but a manifestation of the fact that \emph{distinct microscopic ion structures can yield the same macroscopic and microscopic electrostatic field response}. This contrasts with \emph{charge symmetry-breaking}, where variations in surface charge density \( \sigma_{\scriptscriptstyle{0}} \) yield different electric fields despite identical bulk electrolyte conditions.

\noindent\textbf{Topological classification of confining geometries}

Here, let us point out on some generality of our finding, which beyond the planar, cylindrical and spherical geometry. Even though the analysis of our results is based on the electric field properties of the nanocapacitors, the main rule parameter that explains our findings is the confinement of the electrolyte, with its concomitant violation of the local electroneutrality condition. Hence, our results go beyond the particularities of the local electric fields, since, to some extend, this confinement effect applies to nanocapacitors with topologies equal to the aforementioned geometries.

The three principal geometries used in our study—spherical, cylindrical, and planar—can be rigorously distinguished by their topological structure. Each defines a different type of spatial confinement and boundary, which can be described mathematically as follows:

\begin{table}[ht]
	\centering
	\caption{Topological classification of the confining geometries considered in this work. For clarity, full descriptions are condensed in the table and expanded in the main text.}
	\begin{tabular}{lll}
		\toprule
		\textbf{Geometry} & \textbf{Description} & \textbf{Mathematical Form} \\
		\midrule
		Spherical shell & Closed, compact & \( S^2 \times [0, \delta] \) \\
		Cylindrical tube & Open, axially symmetric & \( S^1 \times [0,L] \times [0,\delta] \) \\
		Planar slit & Parallel planes & \( \mathbb{R}^2 \times [0,\delta] \) \\
		\bottomrule
	\end{tabular}
	\label{tab:topologies}
\end{table}

\noindent
Here, \( S^2 \) represents the 2-sphere (surface of a 3D ball), \( S^1 \) the 1-sphere (a circle), and \( \delta \) the finite wall thickness of the shell. The cylindrical geometry includes an axial extent \( L \), while the planar geometry is translationally invariant in two spatial directions. These topologies are crucial in defining the boundary conditions and curvature effects that lead to the charge and chemical potential symmetry-breaking phenomena explored in this work.  



\section{Conclusions}\label{Conclusions}

In this article, we have investigated the specific differential capacitance, \( C_d \), of hollow nanoparticles with three distinct confining geometries: planar, cylindrical, and spherical. These nanocapacitors are characterized by cavity radius \( R \), wall thickness \( d \), and surface charge density \( \sigma_{\scriptscriptstyle{0}} \), and are immersed in a symmetric \( z:z \) electrolyte of bulk concentration \( \rho_{\scriptscriptstyle{0}} \).

We analyzed the dependence of \( C_d \) on the system’s geometric and electrostatic parameters. In general, \( C_d \) increases with increasing electrolyte concentration \( \rho_{\scriptscriptstyle{0}} \), ionic valence \( z \), and surface charge density \( \sigma_{\scriptscriptstyle{0}} \), while it decreases with increasing cavity radius \( R \) and shell thickness \( d \). Among the three geometries, spherical cavities typically exhibit the highest capacitance and planar slits the lowest, especially at small \( R \). As a function of \( R \), planar systems show a smooth monotonic decay of \( C_d \), whereas cylindrical and spherical geometries display strongly nonlinear, oscillatory behavior, including phenomena of \emph{confinement overcharging} (CO) and \emph{confinement charge reversal} (CCR), both of which are direct manifestations of the \emph{violation of local electroneutrality} (VLEC).

We obtained these results by solving the linearized Poisson–Boltzmann (LPB) equation, which we previously validated through detailed comparisons with density functional theory, integral equations, modified PB theories, and computer simulations—demonstrating its high accuracy at low concentrations and moderate surface charge densities.

\medskip
\noindent\textbf{Two Forms of Electrostatic Symmetry Breaking}

\smallskip
Our findings reveal two distinct yet complementary symmetry-breaking phenomena in the electrostatic response of confined systems:

\begin{itemize}
	\item \textbf{Charge symmetry breaking:} Varying \( \sigma_{\scriptscriptstyle{0}} \) leads to asymmetric electrostatic potentials and EDL profiles on inner and outer walls, even when surface charge magnitudes are equal. Remarkably, the differential capacitance remains invariant. This insensitivity arises because the degree of VLEC remains constant for a fixed geometry and screening length..
	
	\item \textbf{Chemical potential symmetry breaking:} Bulk electrolytes with different ionic valences and concentrations—such as 1:1 NaCl and 2:2 MgSO\(_4\)—can exhibit identical electrostatic responses, including \( \psi(r) \), \( \rho_{\scriptscriptstyle{\mathrm{el}}}(r) \), and \( C_d \), provided their Debye lengths are equal. However, these systems differ significantly in their bulk chemical potentials and local ion concentration profiles, reflecting broken thermodynamic symmetry despite electrostatic indistinguishability.
\end{itemize}

\medskip
\noindent\textbf{Geometric and Topological Influences}

\smallskip
We also show that topology plays a central role in shaping the electrostatic behavior of confined electrolytes. In cylindrical and spherical geometries, the capacitance exhibits a non-monotonic dependence on \( R \), with maxima at intermediate confinement. These effects stem from curvature-modulated electric fields and are absent in planar systems. At large \( R \), all geometries converge, recovering the planar limit.

\medskip
\noindent\textbf{A Unified Perspective}

\smallskip
Taken together, these observations support the existence of a deeper organizing principle in confined electrostatics:

\begin{quote}
	\emph{Capacitance is governed by the geometry, topology, and screening length—not by the absolute values of surface charge or bulk chemical potential.}
\end{quote}

This principle points to a broader class of electrostatic invariants, where systems with distinct microscopic or thermodynamic parameters yield identical macroscopic observables. We believe that the framework of symmetry-breaking—whether by charge, confinement, or chemical potential—provides a powerful lens for understanding nonlinear and topological effects in soft condensed matter and nanofluidic systems.

\medskip
\noindent\textbf{Beyond Linearized Theory: Future Directions}

\smallskip
While our analysis is based on the LPB theory, the underlying symmetry-breaking mechanisms are expected to persist beyond the linear regime. In particular, low-density expansions and integral equations approaches suggest that both CO and CCR phenomena survive in systems with excluded-volume effects and ionic correlations. Extensions to models with finite ion size and short-range interactions may reveal additional structural transitions or bistability, particularly under strong confinement. These generalizations are currently under development and will be reported elsewhere.

\medskip

\newpage
 \appendix
\section{Detailed Calculation of the Bulk Chemical Potential}

In this appendix, we provide the detailed steps used to compute the bulk chemical potential components for NaCl (1:1) at \( \rho_0 = 0.2\,\mathrm{M} \) and MgSO$_4$ (2:2) at \( \rho_0 = 0.05\,\mathrm{M} \), both at \( T = 298.15\,\mathrm{K} \), in water with dielectric constant \( \varepsilon = 78.5 \). The inverse Debye length \( \kappa \) is the same for both salts and is taken as:

\begin{equation}
	\kappa = 1.4689 \times 10^9\, \mathrm{m}^{-1}.
\end{equation}

The total bulk chemical potential per salt molecule is given by:
\begin{equation}
	\mu^{\text{bulk}} = \mu^{\text{kin}} + \mu^{\text{mix}} + \mu^{\text{el}},
\end{equation}
where:

\begin{itemize}
	\item \( \mu^{\text{kin}} = \sum_{i} kT \ln\left[ \left( \frac{2\pi m_i kT}{h^2} \right)^{3/2} \right] \)
	\item \( \mu^{\text{mix}} = \sum_{i} kT \ln \rho_0 \)
	\item \( \mu^{\text{el}} = -\frac{z^2 e^2 \kappa}{8\pi \varepsilon_0 \varepsilon} \sum_i 1 \) (one term per ion)
\end{itemize}

\vspace{1em}
\subsection*{Constants used:}

\begin{align*}
	k &= 1.380649 \times 10^{-23}\, \mathrm{J/K} \\
	T &= 298.15\, \mathrm{K} \\
	h &= 6.62607015 \times 10^{-34}\, \mathrm{J\,s} \\
	e &= 1.602176634 \times 10^{-19}\, \mathrm{C} \\
	\varepsilon_0 &= 8.8541878128 \times 10^{-12}\, \mathrm{F/m} \\
	\varepsilon &= 78.5 \\
	N_A &= 6.02214076 \times 10^{23}\, \mathrm{mol}^{-1}
\end{align*}

\vspace{1em}
\subsection*{Molar masses used (kg/mol):}

\begin{align*}
	M_{\mathrm{Na}^+} &= 22.99 \times 10^{-3} \\
	M_{\mathrm{Cl}^-} &= 35.45 \times 10^{-3} \\
	M_{\mathrm{Mg}^{2+}} &= 24.31 \times 10^{-3} \\
	M_{\mathrm{SO}_4^{2-}} &= 96.06 \times 10^{-3}
\end{align*}

\vspace{1em}
\subsection*{Kinetic contribution \(\mu^{\text{kin}}\):}

\textbf{NaCl (1:1):}
\begin{align*}
	\mu_{\mathrm{Na}^+}^{\text{kin}} &= kT \ln\left[ \left( \frac{2\pi m_{\mathrm{Na}} kT}{h^2} \right)^{3/2} \right] = 19.68\, \mathrm{kJ/mol} \\
	\mu_{\mathrm{Cl}^-}^{\text{kin}} &= kT \ln\left[ \left( \frac{2\pi m_{\mathrm{Cl}} kT}{h^2} \right)^{3/2} \right] = 19.68\, \mathrm{kJ/mol} \\
	\Rightarrow \mu^{\text{kin}} &= 39.36\, \mathrm{kJ/mol}
\end{align*}

\textbf{MgSO\(_4\) (2:2):}
\begin{align*}
	\mu_{\mathrm{Mg}^{2+}}^{\text{kin}} &= 18.68\, \mathrm{kJ/mol}, \quad
	\mu_{\mathrm{SO}_4^{2-}}^{\text{kin}} = 19.97\, \mathrm{kJ/mol} \\
	\Rightarrow \mu^{\text{kin}} &= 38.65\, \mathrm{kJ/mol}
\end{align*}

\vspace{1em}
\subsection*{Mixing (entropic) term \(\mu^{\text{mix}}\):}

\textbf{NaCl:}
\[
\mu^{\text{mix}} = 2kT \ln(0.2) = 2 \times kT \times (-1.6094) = -7.98\, \mathrm{kJ/mol}
\]

\textbf{MgSO\(_4\):}
\[
\mu^{\text{mix}} = 2kT \ln(0.05) = 2 \times kT \times (-2.9957) = -14.86\, \mathrm{kJ/mol}
\]

\vspace{1em}
\subsection*{Electrostatic term \(\mu^{\text{el}}\):}

\[
\mu^{\text{el}} = -\frac{z^2 e^2 \kappa}{4\pi \varepsilon_0 \varepsilon} \cdot \frac{1}{1000 \cdot N_A}
\]

\textbf{NaCl (1:1, \(z=1\)):}
\[
\mu^{\text{el}} = -\frac{(1)^2 \cdot (1.602 \times 10^{-19})^2 \cdot 1.4689 \times 10^9}{4\pi \cdot 8.854 \times 10^{-12} \cdot 78.5} \cdot \frac{1}{1000 \cdot 6.022 \times 10^{23}} \approx -23.26\, \mathrm{kJ/mol}
\]

\textbf{MgSO\(_4\) (2:2, \(z=2\)):}
\[
\mu^{\text{el}} = -\frac{(2)^2 \cdot (1.602 \times 10^{-19})^2 \cdot 1.4689 \times 10^9}{4\pi \cdot 8.854 \times 10^{-12} \cdot 78.5} \cdot \frac{1}{1000 \cdot 6.022 \times 10^{23}} \approx -93.00\, \mathrm{kJ/mol}
\]

\vspace{1em}

\subsection*{Final values:}

\begin{table}[htb!]
	\centering
	\caption{Components of the bulk chemical potential (in kJ/mol per salt molecule) for NaCl (1:1, 0.2\,M) and MgSO$_4$ (2:2, 0.05\,M) at \( T = 298.15\,\mathrm{K} \) in water (\( \varepsilon = 78.5 \)), assuming a common Debye length \( \lambda_{\scriptscriptstyle{D}} \) with \( \kappa = 1.4689 \times 10^9\, \mathrm{m}^{-1} \). Each total value corresponds to the sum over both cation and anion.}
	\begin{tabular}{lcc}
		\toprule
		\textbf{Contribution} & \textbf{NaCl (1:1, 0.2\,M)} & \textbf{MgSO$_4$ (2:2, 0.05\,M)} \\
		\midrule
		Kinetic term \( \mu^{\text{kin}} \)        & 39.36 & 38.65 \\
		Mixing (entropy) \( \mu^{\text{mix}} \)    & $-7.98$ & $-14.86$ \\
		Electrostatic \( \mu^{\text{el}} \)        & $-23.26$ & $-93.00$ \\
		\midrule
		\textbf{Total \( \mu^{\text{bulk}} \)}     & \textbf{8.12} & \textbf{$-69.21$} \\
		\bottomrule
	\end{tabular}
	\label{tab:mu_bulk_appendix}
\end{table}

\paragraph{Note on Standard vs Model-Based Chemical Potentials.}
The values reported in \cref{tab:mu_bulk_final_Text} are obtained from first-principles statistical mechanical considerations, including kinetic, mixing (entropic), and electrostatic (Debye–Hückel) contributions. These values differ significantly from standard tabulated chemical potentials, such as \( \mu^\ominus_{\mathrm{NaCl}} = -384.024\,\mathrm{kJ/mol} \), which are based on experimental data and include solvation effects, reference state conventions, and activity corrections. In the standard thermodynamic framework, the bulk chemical potential is given by
\[
\mu = \mu^\ominus + RT \ln a \approx \mu^\ominus + RT \ln c,
\]
where \( a \) is the activity and \( c \) the molar concentration. For example, using \( R = 8.314\,\mathrm{J/(mol\cdot K)} \), \( T = 298.15\,\mathrm{K} \), and \( c = 0.2\,\mathrm{M} \), we obtain
\[
RT \ln(0.2) \approx 2.4788\,\mathrm{kJ/mol} \cdot (-1.609) \approx -3.99\,\mathrm{kJ/mol},
\]
yielding \( \mu_{\mathrm{NaCl}} \approx -388.01\,\mathrm{kJ/mol} \). This contrasts with our model-based value of \( \mu^{\mathrm{bulk}} = 8.12\,\mathrm{kJ/mol} \), which is intended for consistent theoretical comparisons across electrolytes with equal Debye length \( \lambda_{\scriptscriptstyle{D}} \). The model values are internally consistent and appropriate for analyzing symmetry breaking within the Poisson–Boltzmann framework, even though they omit complex real-world effects such as ion solvation and non-ideal activity behavior.


\section*{Acknowledgment}

The support of UNAM (PAPIIT Clave: IN108023) is acknowledged.


\end{document}